# Analysis of the performance of U-Net neural networks for the segmentation of living cells

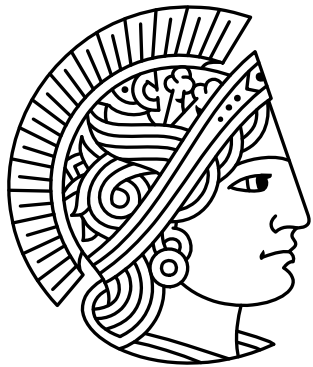

André Oliveira Françani, B.Sc.

M.Sc. Thesis

Supervisors:
Prof. Dr. techn. Heinz Koeppl
Tim Prangemeier, M.Sc.

Bioinspired Communication Systems
Technische Universität Darmstadt

December 2019

German title:

# Analyse der Leistungsfähigkeit von U-Net convolutional neural networks zur Segmentierung von lebenden Zellen

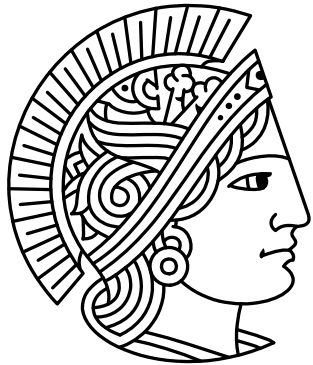

**Erklärung zur Abschlussarbeit gemäß § 23 Abs. 7 APB der TU Darmstadt**

Hiermit versichere ich, André Oliveira Françani, die vorliegende Master-Thesis ohne Hilfe Dritter und nur mit den angegebenen Quellen und Hilfsmitteln angefertigt zu haben. Alle Stellen, die Quellen entnommen wurden, sind als solche kenntlich gemacht worden. Diese Arbeit hat in gleicher oder ähnlicher Form noch keiner Prüfungsbehörde vorgelegen. Bei der abgegebenen Thesis stimmen die schriftliche und die zur Archivierung eingereichte elektronische Fassung überein.

Darmstadt, den 02.12.2019

______________________
(André Oliveira Françani)

**Thesis Statement pursuant to §22 paragraph 7 of APB TU Darmstadt**

I herewith formally declare that I have written the submitted thesis independently. I did not use any outside support except for the quoted literature and other sources mentioned in the paper. I clearly marked and separately listed all of the literature and all of the other sources which I employed when producing this academic work, either literally or in content. This thesis has not been handed in or published before in the same or similar form. In the submitted thesis the written copies and the electronic version are identical in content.

Darmstadt, den 02.12.2019

______________________
(André Oliveira Françani)

# Abstract

The automated analysis of microscopy images is a challenge in the context of single-cell tracking and quantification. This work has as goals the analysis of the performance of deep learning for segmenting microscopy images and the improvement of the previous available pipeline for tracking single cells. Deep learning techniques, mainly convolutional neural networks, have been applied to cell segmentation problems and they have shown high accuracy and fast performance. To perform the image segmentation, an analysis of hyperparameters was done in order to implement a convolutional neural network with U-Net architecture. Furthermore, different models were built in order to optimize the size of network and number of learnable parameters. The trained network is then used in the pipeline that localizes the traps in a microfluidic device, performs the image segmentation on trap images, and evaluates the fluorescence intensity and the area of single cells over time. The tracking of the cells during an experiment is performed by image processing algorithms, such as centroid estimation and watershed. Finally, with all improvements in the neural network to segment single cells and in the pipeline, a quasi real time image analysis was enabled, where 6.20GB of data were processed in 4 minutes.

# Contents

# List of Figures

# List of Tables

# Chapter 1

# Introduction

## 1.1 Overview

This study is about the performance analysis of convolutional neural networks for the segmentation of individual cells. In addition, a pipeline has been developed in order to evaluate the fluorescence intensity and area of each cell. This thesis begins with a small motivation in chapter 1, introducing the application of convolutional neural networks in machine learning context. Next, the state-of-the-art of convolutional neural network layers and image processing algorithms are described in chapter 2. Chapter 3 presents the methods used to obtain the architecture of the used neural network, as well as the methods for tracking and calculating the fluorescence of cells. The results are addressed in chapter 4, where the influence of some hyperparameters and the application of the pipeline in a microfluidic experiments with yeast cells are discussed. Finally, chapter 5 contains the conclusions throughout this work as well as an outlook with ideas for future projects.

## 1.2 Motivation

Machine learning (ML) techniques have attracted attention of the scientific community in the recent decades. One important field of ML is the deep learning (DL), since it is universal in its application domain, it is robust once it does not require previous feature extraction, and it is an approach easily generalizable for different data types and applications [1]. One approach of DL is the Neural Network (NN), and there is a subfield of NN called Convolutional Neural Networks (CNN), which is a method that has shown to be very accurate in many tasks. The most common applications of DL are object localization [2], object detection [3], face detection [4], face recognition [5], speech recognition [6], art and texture transfer [7], semantic segmentation [8], instance segmentation [9], among others [10, 11, 12]. Figure 1.1 shows some of the classical computer vision problems, for example image classification, object localization, semantic segmentation, and instance segmentation.

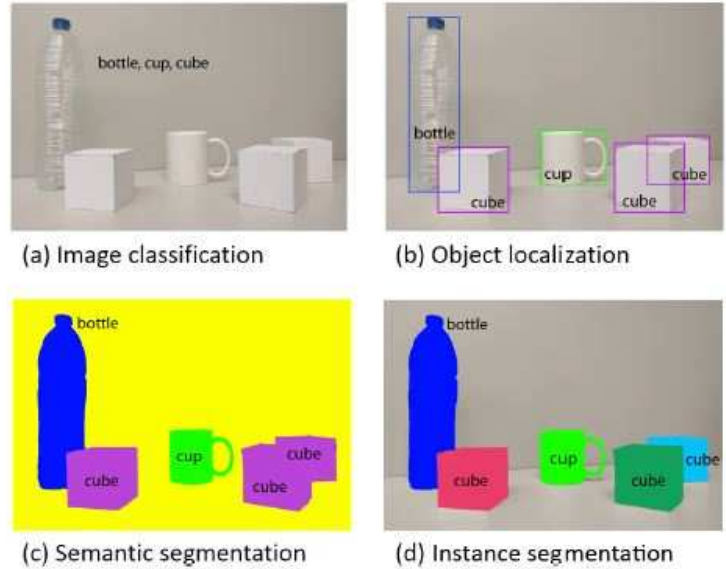

Figure 1.1: Classical computer vision tasks. Extracted from [13].

Image segmentation [14, 13] is one of the most challenging application of DL in computer vision tasks. It gives to every pixel of an image a class label. The most widespread image segmentation applications are human-machine interaction [15], biomedical image processing [16, 17, 18, 19, 20], and autonomous driving [21, 22, 23].

This thesis is inserted in the biological context, more specifically in cell segmentation to localize them and obtain their contours in microscope imagery. This detection allows the analysis of individual cells to extract morphological features and fluorescence intensity, which is a field of interest in cell and molecular biology [24]. Brightfield images from different focal planes can be used to complement the detection and localization of the cells, helping the fluorescent intensity measurement of single cells. Furthermore, brightfield channel segmentation allows measurement of zero fluorescence. The main motivation of this thesis is to understand deeply how the neural network is extracting the features to do the cell segmentation, as well as to analyze the behavior of the network for different environment conditions and for quasi real time experiments.

## 1.3 Objectives

This thesis has three main topics, namely the analysis of the performance of the convolutional neural network in the segmentation of yeast cell images, improvement of the previous pipeline available in the Bioinspired Communication Systems Lab (BCS), and tracking of cells to trace the fluorescence intensity curve over time for a given experiment. Among these topics, some core goals were established for the development of this thesis, which are the main approaches of this thesis:

- Analysis of hyperparameter influence, such as batch size, number of epochs, optimal input image size for the available dataset, and number of samples required for good image segmentation;

- Performance analysis of different U-Net models, varying the number of parameters to be learned in order to reduce the size of the original U-Net;

- Tracking of single cells in time and measurement of fluorescence intensity emitted by the cells;

- Build a simple and useful pipeline, with availability in a software development platform, such as GitHub, for the use of future researchers in this biomedical area of single cell tracking;

- Analysis of the neural network performance for a possible real-time experiment.

Therefore, the analyses and goals of this work lead to the following main questions:

- What is the influence of hyperparameters in the performance of the network?

- Can the original U-Net architecture be reduced to a smaller architecture for this application?

- Is it feasible to process an experiment in real time in such a way that the fluorescence intensity of individual cells is measured during an experiment over time?

# Chapter 2

# Background

## 2.1 Convolutional neural network

The first CNN architecture is considered the "*Neocognitron*", presented by Kunihiko Fukushima in 1980 [25]. This neural network model has an architecture inspired by the hierarchy model of the visual nervous proposed by Hubel and Wiesel in [26], and it was built for a pattern recognition problem. The great advantageous of this network, compared to the other models proposed before, is that it is shift invariant. This means that the response is not affected by shift in position of the input patterns [25].

The most recent architectures and layers arose from *LeNet-5* in 1998 [27], which is trained with a backpropagation algorithm and has a key sequence of layers: convolution, pooling and non-linearity activation function. With the development of technology, the graphics processing unit (GPU) gained interest from the neural network community in 2010 [28] with faster computational by integrating GPU and central processing unit (CPU). Using multiple GPU's, the architecture called *AlexNet* [29] won a ImageNet competition in 2012, increasing the interest of using deep convolutional neural networks in classification problems. This network is a wider version of *LeNet-5*, but with more layers and, as a consequence, more parameters to be computed (60 million parameters). However, the training was made faster by using an efficient GPU implementation and a non-saturating activation function in its neurons [29]. After *AlexNet*, the history of deep learning with CNN as approach changed course and attracted even more attention from researchers in this area [1]. The networks began to become deeper and new state-of-the-art architectures have emerged, such as *VGG-16* [30], *GoogLeNet* [31], *U-net* [17], *Residual Networks* (ResNets) [32], *DenseNet* [33] and others [34]. Figure 2.1 illustrates a CNN in a semantic segmentation problem.

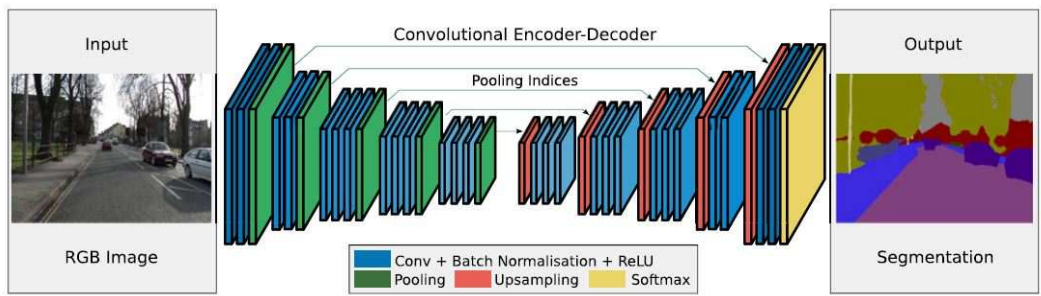

Figure 2.1: Convolutional neural network applied in a semantic segmentation task. Extracted from [34].

The layers showed in Figure 2.1 are detailed in Section 2.2.

## 2.2 Layers

This section contains all the useful layers, which are necessary to understand how a CNN works. It is important to mention that there are also other layers beyond these, but they will not be described in the present section because they are not used in the convolutional neural network architecture selected for this project.

### 2.2.1 Convolution

In signal processing and image processing area, the convolution is a fundamental operation. Let an image be denoted by $\mathbf{s}(i,j)$, where $(i,j)$ is the coordinate of an element in this image. Given the images $\mathbf{f}(i,j) \in \mathbb{R}^{M \times N}$ and $\mathbf{w}(i,j) \in \mathbb{R}^{M \times N}$, the 2D discrete convolution is given by [35]

$$\mathbf{g}(i,j) = \mathbf{w}(i,j) * \mathbf{f}(i,j) = \frac{1}{MN} \sum_{k=0}^{M-1} \sum_{l=0}^{N-1} \mathbf{w}(k,l)\mathbf{f}(i-k, j-l), \qquad (2.1)$$

where $i = 0, 1, \cdots, M-1$ and $j = 0, 1, \cdots, N-1$. Many machine learning libraries implement the convolution function as a cross-correlation function, since we deal with real functions (images) and the libraries do not flip the kernel relative to the input image, which is done in a convolution operation [36]. Furthermore, the normalization factor is not implemented and they use square sized images [36]. The cross-correlation function can be interpreted as a linear spatial filtering of digital image processing [35], and the function $\mathbf{w}(i,j)$ is commonly called *convolution kernel* [35]. With unity stride and no padding, this operation throw away information from the edges and it shrinks the output, also known as downsampling. As a consequence of the convolution operation, for an input of size $N \times N$ and a $F \times F$ filter, an output image $\mathbf{g}(i,j) \in \mathbb{R}^{M \times M}$ will have dimension [35]

$$M = N - F + 1. \qquad (2.2)$$

Figure 2.2 shows an example of edge detection using a linear $3 \times 3$ spatial filtering, where the operations are very similar to what happens in a convolutional layer for an input image of size $256 \times 256$.

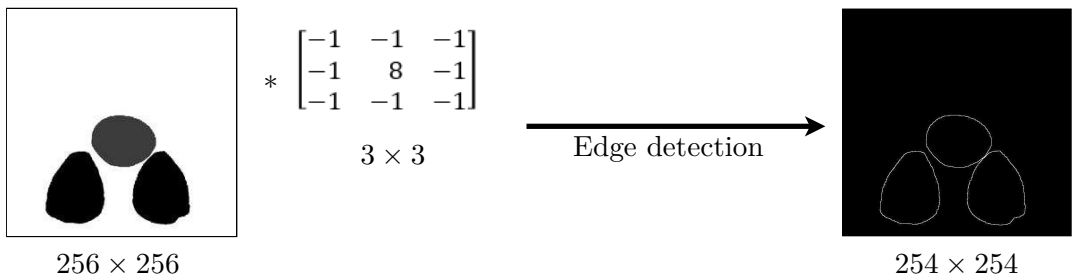

Figure 2.2: Visual edge detection filter an $256 \times 256$ input image, $3 \times 3$ filter kernel, resulting in a $254 \times 254$ output image

It can be seen in Figure 2.2 that a $3 \times 3$ filter applied to a $256 \times 256$ input image results in an $254 \times 254$ output image. One way to avoid this downsampling is by doing padding.

Padding the image means adding rows and columns of zeros, or replicating rows or columns on the borders [35]. This allows the output to have the same size as the input of the filtering process. Normally, the libraries have the terminology "*same*", if the output size should be the same as the input size, or "*valid*" if the output does not need to have the same size of the input [36]. In the first case, the output image $\mathbf{g}(i,j) \in \mathbb{R}^{M \times M}$ will have dimension [35]

$$M = N + 2P - F + 1, \qquad (2.3)$$

where $P$ is amount of padding [1, 36]. This means that, in the example of Figure 2.2, the amount of padding should be $P = 1$ so that $M = 256$. This means adding a border of zeros around the image. In the other case ('*valid*'), the output dimension will follow Eq. (2.2).

Another parameter that plays an important role in the downsampling is the stride. The stride $S$ represents the number of pixels skipped between two consecutive filter positions [37]. Therefore, the bigger the stride, the bigger the downsampling. The complete equation of the output dimension after convolution, with padding and stride, is then given by [37]

$$M = \left\lfloor \frac{N + 2P - F}{S} + 1 \right\rfloor. \qquad (2.4)$$

**Multiple filters**

An image can have different channels, for example the Red, Green and Blue channels (RGB). This means that an image can have a generic dimension $N_h \times N_w \times N_c$, respectively *height*, *width* and *number of channels* or *depth*. In the convolutional neural network context, the output of the convolution is called *feature map*, and it is common to have multiple feature maps stacked one onto another [36, 37]. Thus, it is possible to have filters of size $F \times F \times N_c$, where this last dimension represents the number of kernels of the previous convolutional layer. For example, for a RGB input image of size $10 \times 10 \times 3$ and a convolutional layer with 3 kernels of size $3 \times 3 \times 3$ with no padding and unity stride, the output feature map is a volume of dimension $8 \times 8 \times 3$, since the 3 kernels were stacked together and the output of the RGB channels are summed up into one output. This example is shown in Figure 2.3, which can also be called convolution over a volume due to the last dimension of the kernels.

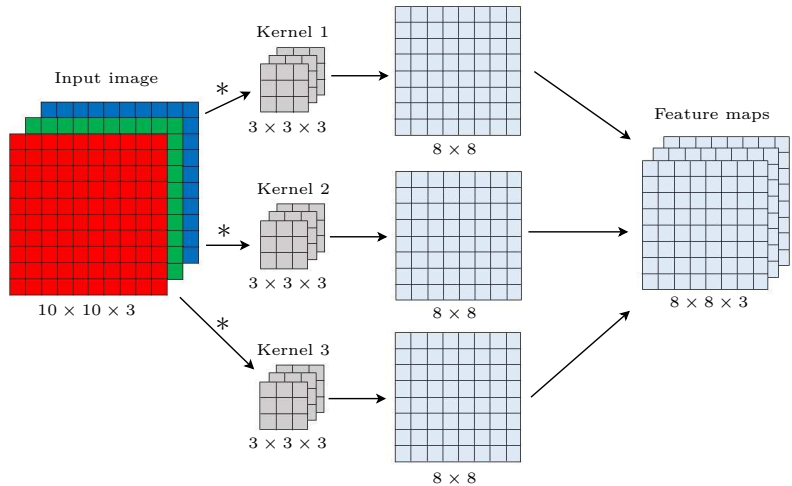

Figure 2.3: Multiple filter example.

Continuing the example of Figure 2.3, if a following convolutional layer has 20 kernels of size 5 × 5, the output feature map of this layer will have size 4 × 4 × 20 considering unity stride and no padding.

Unlike the spatial linear filtering of image signal processing, the elements of the kernel are not fixed in the machine learning context. The elements of the kernel are the learned parameters of a neural network [36]. This means that the network is learning different filters to extract information from the image, that is why the convolutional layer is the most important layer in a CNN.

### 2.2.2 Activation functions

The activation functions are considered as part of the convolution layer and are used right after it. After the convolution, the learned parameters of the kernels go through a non-linear activation function $\varphi(\cdot)$, applied element-wise, resulting the output feature map [36]. The most common functions are *sigmoid*, *hyperbolic tangent*, *softmax* and *rectified linear units* (ReLU), shown in Figure 2.4.

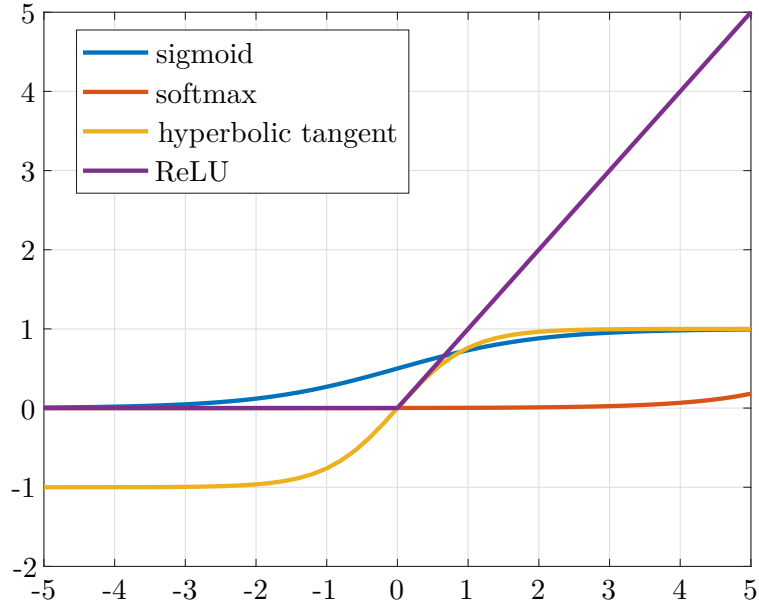

Figure 2.4: Curves of common activation functions.

In this thesis, only the ReLU and *softmax* functions are defined because they are the only ones used.

The ReLU function is given by

$$\varphi(w_{ij}) = \max(0, w_{ij}), \qquad (2.5)$$

where $w_{ij}$ is the element $(i, j)$ of the output feature map $\mathbf{w} \in \mathbb{R}^{F \times F}$, $i = j = 0, 1, \cdots, F-1$. This function does not saturate, does not have a normalization factor and it is easy to compute. Then it is commonly used as activation function in the hidden layers of a network.

The *softmax* function is expressed as

$$\varphi(w_{ij}) = \frac{\exp(w_{ij})}{\sum_i \sum_j \exp(w_{ij})}. \qquad (2.6)$$

It has a normalization factor and can be interpreted as a probability, where the high values will have the higher probability, and it is mostly used at the final layer of the neural network as a classification function in multiple classification problems.

### 2.2.3 Batch normalization

The batch normalization layer was introduced in [38] to speed up training. The so called *mini-batch* is related to the number of training examples in one training step (iteration). This means that, for a mini-batch of size *b*, in every training iteration are used *b* training examples. With this layer, every activation layer output is normalized after the training mini-batch, enabling higher learning rates and reducing overfitting [38]. By using this normalization layer, the authors achieved a better result than the *AlexNET* in the ImageNet competition mentioned in Section 2.1.

### 2.2.4 Pooling

Pooling is an essential layer of subsampling, reducing the size of the output feature maps. It has no parameters to learn and can be applied after the activation function of a convolutional layer. The most used pooling functions are *max* and *average*, and they summarize certain regions of the output [36]. This layer also helps the shift invariance property because the presence rather than position is important. The output size of this layer depends on 2 hyperparameters: filter size $F_p$, which determines the region in the input image, and the stride $S_p$. Therefore, given an $N \times N$ input, the $M \times M$ output dimension can be calculated as [37]

$$M = \left\lfloor \frac{N - F_p}{S_p} + 1 \right\rfloor. \qquad (2.7)$$

Figure 2.5 shows a numerical example of the output of a pooling layer for different pooling functions.

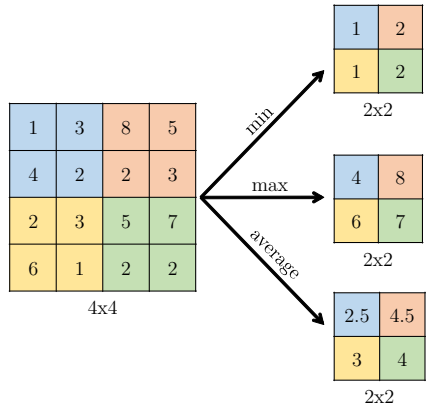

Figure 2.5: Pooling with filter size $F_p = 2$ and stride $S_p = 2$ for the *min*, *max* and *average* pooling functions, which outputs the minimum, maximum and the average of the filter region, respectively. With these hyperparameters, the output dimension of each operation is reduced by half.

### 2.2.5 Convolution transpose

In a semantic segmentation problem, if the output of the CNN should have the same size of the input, it should be applied an upsampling method. One of the existent methods is the *transposed convolution*, also called *deconvolution*, which goes to the opposite direction of a convolution [37]. The parameters of this transposed convolution layer are learned during training. Therefore, there is no need of defining a fixed interpolation method to perform the upsampling [8].

## 2.3 U-Net architecture

The use of deep learning in image segmentation tasks is challenging. The pixel-wise classification is really important in some applications, such as biomedical image processing, because the information of "what" and "where" can be extracted from the output image, which is also a visual task. The segmentation of neuronal membranes and cell segmentation have shown to be feasible with CNN [18, 17]. For this reason, the *U-Net* architecture [17], illustrated in Figure 2.6, was chosen as a model for this work.

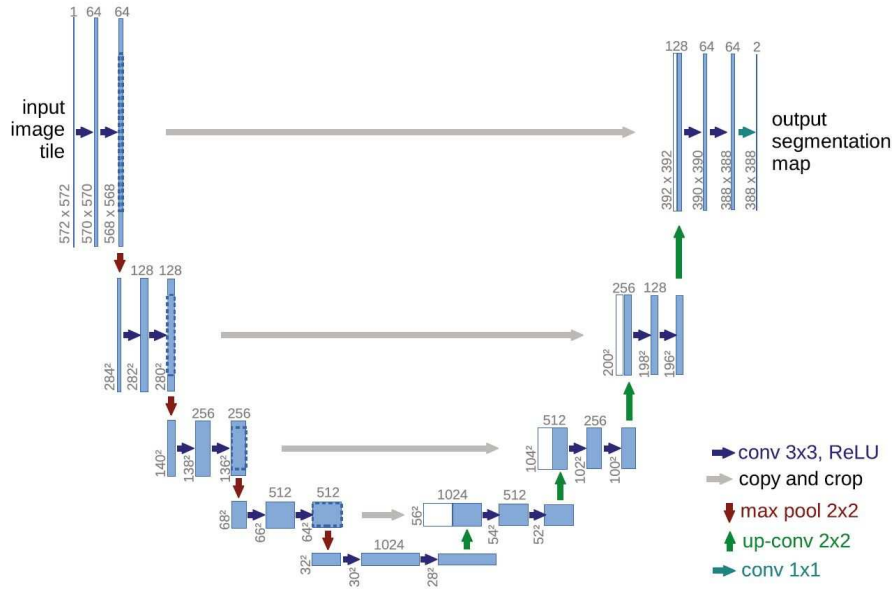

Figure 2.6: U-Net architecture. Extracted from [17].

This network consists of a contracting path (encoder) and an expansive path (decoder). It can be seen that, after every $2 \times 2$ max pooling with stride 2, the number of feature channels is doubled (denoted on the top of each blue box), and the dimension of the blue box is halved (lower left edge of the box). Furthermore, in the expansive path, the transpose convolution also halves the number of feature channels, but there is a concatenation with a cropped feature map from the encoder. This cropping and concatenation allows a better localization of the features due to the combination of features from the decoder and encoder [17], so that the loss of border pixels is reduced. The final layer is a $1 \times 1$ convolution to output a map with the desired number of classes. More detailed information about this architecture can be seen in [17].

## 2.4 Loss functions

The correct choice of a loss function plays an important role in the training of a neural network [39]. The optimization process requires a loss function such that the weights or parameters of the model are updated using backpropagation of the prediction error. This is commonly done by using the stochastic gradient descent as optimization algorithm. In this work, a custom function using both the categorical cross-entropy and the dice loss function were chosen as loss function.

### 2.4.1 Categorical cross-entropy loss

Let $\hat{\mathbf{Y}} \in \mathbb{R}^{M \times N \times C}$ be the output of the last layer after the classification function, with elements $\hat{y}_{m,n,c}$, and let $\mathbf{Y} \in \mathbb{N}^{M \times N \times C}$ be the corresponding expected output labels with elements $y_{m,n,c}$, where $C$ is the number of classes in the classification problem. The categorical cross-entropy loss $L_{cce}$ is given by [40]

$$L_{cce} = \frac{1}{mn} \sum_{m=1}^{M} \sum_{n=1}^{N} \sum_{c=1}^{C} -y_{m,n,c} \log(\hat{y}_{m,n,c}). \tag{2.8}$$

### 2.4.2 Dice loss

The dice loss $L_{dice}$ [41, 19, 42] is defined as

$$L_{dice} = 1 - DICE = 1 - \sum_{c=1}^{C} \frac{2 \sum_{m=1}^{M} \sum_{n=1}^{N} \hat{y}_{m,n,c} y_{m,n,c}}{\sum_{m=1}^{M} \sum_{n=1}^{N} \hat{y}_{m,n,c} + \sum_{m=1}^{M} \sum_{n=1}^{N} y_{m,n,c}}, \tag{2.9}$$

where $DICE$ is called dice coefficient and can be used as an evaluation metric.

## 2.5 Evaluation metrics

The metrics are important measurements to compare models and judge their performances. The functions used in this project are *accuracy*, *IoU*, weighted *IoU* and *Dice coefficient*, which are described in the following.

### 2.5.1 Accuracy

The accuracy is basically the percent of pixels that are correctly classified. A high accuracy means that the most of the pixels were predicted correctly. However, this metric is not robust for the case of a dataset with imbalanced classes. Figure 2.7 illustrates two different cases, where the metric is reliable and where it is not.

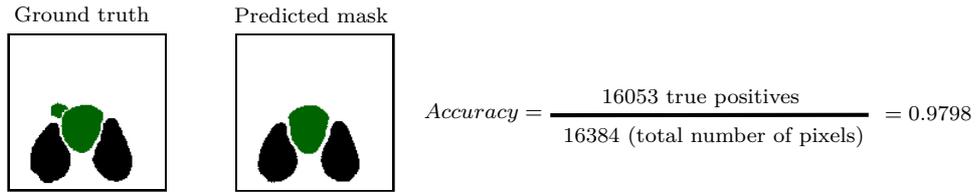
(a) Accuracy for a good prediction case.

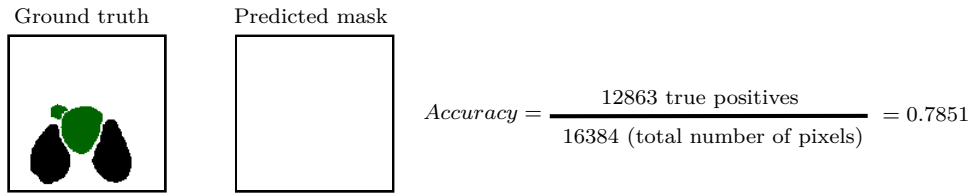
(b) Accuracy for a bad prediction case: only background.

Figure 2.7: Illustration of the accuracy for different prediction masks.

In Figures 2.7a and 2.7b can be observed that, only the accuracy is not enough to validate the model for the imbalanced classes case, as it can be seen from Figure 2.7b, where only a background prediction has an accuracy of 78.51%. This accuracy can be considered a high value in a dataset with balanced classes. Therefore, another metrics should also be applied in order to analyze the performance and quality of a model.

### 2.5.2 Intersection-over-Union

The Intersection-over-Union ($IoU$), also known as Jaccard Index, is a metric commonly used in semantic segmentation problems. As the name suggest, it is calculated by getting the intersection between the predicted image and its ground truth, that is, the number of pixels that the prediction and ground truth overlap, divided by the union between the two images. It can be described mathematically as

$$IoU(A, B) = \frac{|A \bigcap B|}{|A \bigcup B|}, \qquad (2.10)$$

where $A$ and $B$ are the original and predicted image, respectively, $\bigcap$ is the mathematic intersection, $\bigcup$ is the union, and the operator $|\cdot|$ represents the sum of all pixels of the resulting image. Equation 2.10 is applied for every class and then averaged, which is also called *mean IoU*. Figure 2.8 shows graphically the computation of the *mean IoU* for every class for the same example of Figure 2.7.

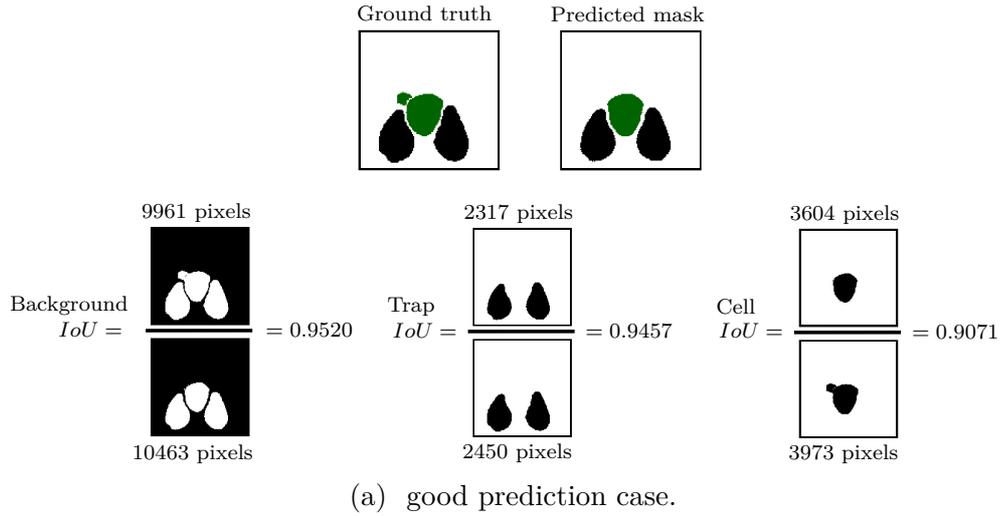

(a) good prediction case.

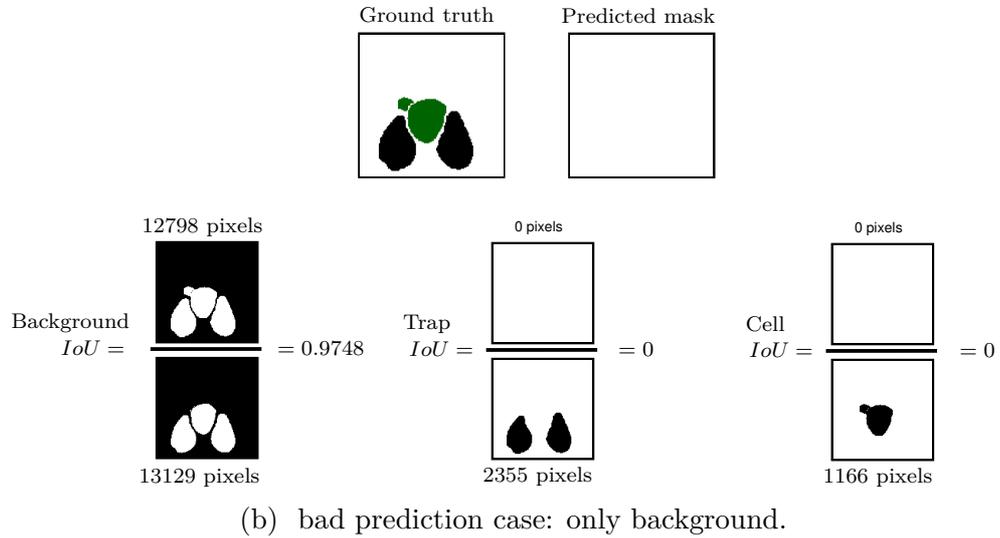

(b) bad prediction case: only background.

Figure 2.8: Mean IoU with graphical intersections and unions for every class. The images are one hot-encoded and the pixel values 0 and 1 are, respectively, white and black.

Figure 2.8a has a *mean IoU* given by the average of the *IoU* of every class, that is,

$$IoU = \frac{0.9520 + 0.9457 + 0.9071}{3} = 0.9349.$$

By doing the same procedure for Figure 2.8b, the *mean IoU* is

$$IoU = \frac{0.9748}{3} = 0.3249.$$

Therefore, *mean IoU* is a good metric for an imbalanced classes case, since it is able to output a low value for a bad prediction.

### 2.5.3 Weighted Intersection-over-Union

Once the *mean IoU* is computed by averaging all the *IoU* classes, it is also possible to apply a weighted arithmetic mean. Sometimes one class is more important than the others, that is, the prediction of this class is more relevant than another one, and the metric should put more weight on this class in order to output a low value if this class is not properly predicted. Therefore, the so called weighted *IoU* in this thesis is the weighted mean of the *IoU* of the individual classes.

### 2.5.4 Dice coefficient

The Dice coefficient (*DICE*), also known as F1 Score or overlap index, is defined as

$$DICE(A, B) = \frac{2|A \bigcap B|}{|A| + |B|}. \tag{2.11}$$

Its other mathematical formulation was defined in Section 2.4.2, and it has a similar interpretation to *IoU*.

## 2.6 Image augmentation

Data augmentation helps the network to be invariant to a characteristic. For example, if the network should be robust to rotated images, a rotation in the original dataset to generate more data should be an interesting approach. Besides, since new images and new situations are generated, it is commonly used when a small dataset is available to train the network model. Depending on the desired characteristic to be invariant, the augmentation is done by translating, rotating, flipping, and deforming the original images.

### 2.6.1 Elastic deformation

Elastic deformation [43, 17] is one of the functions used as augmentation technique. To generate the deformations, a random displacement vectors on a coarse $3 \times 3$ grid was created and a Gaussian distribution with standard deviation of $\sigma = 4$ pixels was used to sample displacements on the grid. The grid is then interpolated using a bicubic interpolation to compute a displacement for each pixel in the input image in 4 points of the deformation grid. Figure 2.9 shows an example of elastic deformation.

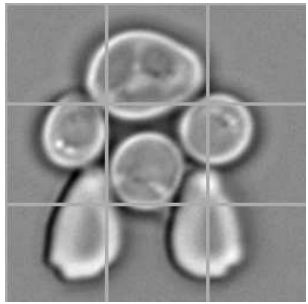 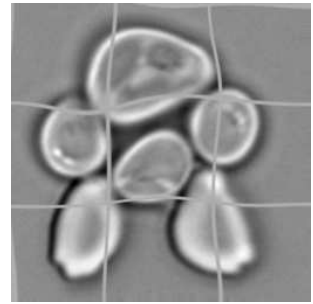

(a) Original image with $3 \times 3$ grid.    (b) Deformed image with $3 \times 3$ deformed grid.

Figure 2.9: Elastic deformation using a $3 \times 3$ grid and $\sigma = 4$ to augment the training set.

The elastic deformation as augmentation was chosen because the network should be invariant to cell shapes. Therefore, by doing those deformations, more cell shapes are generated such that the network acquire more experience and can be invariant to it.

## 2.7 Weight map

The map of weights is a mathematical formulation that highlights the region between objects by assigning high pixel values in it. The smaller the distance between two objects, the greater the pixel value in the region between these objects. This map, described in [17], helps the model to learn how to separate nearby regions of two different objects. In this study the objects are cells and traps. It was introduced into the loss function such that the loss near the boundaries regions were highly penalized. The weight map [17] is defined as

$$w(x_{ij}) = w_c(x_{ij}) + w_0 \cdot \exp\left(-\frac{(d_1(x_{ij}) + d_2(x_{ij}))^2}{2\sigma^2}\right), \qquad (2.12)$$

where $x_{ij}$ is the element $(i,j)$ of the mask image $\mathbf{X} \in \mathbb{N}^{N \times N}$, $i = j = 0, 1, \cdots, N-1$, $w_c \in \mathbb{R}^{N \times N}$ is the class probability map, $d_1(\cdot)$ and $d_2(\cdot)$ are, respectively, the distance to the border of the nearest cell and the distance to the border of the second nearest cell, $w_0$ is a border weight parameter, and $\sigma$ is a border width parameter related to the distance in pixels. These two last parameters are explored in Figure 2.10.

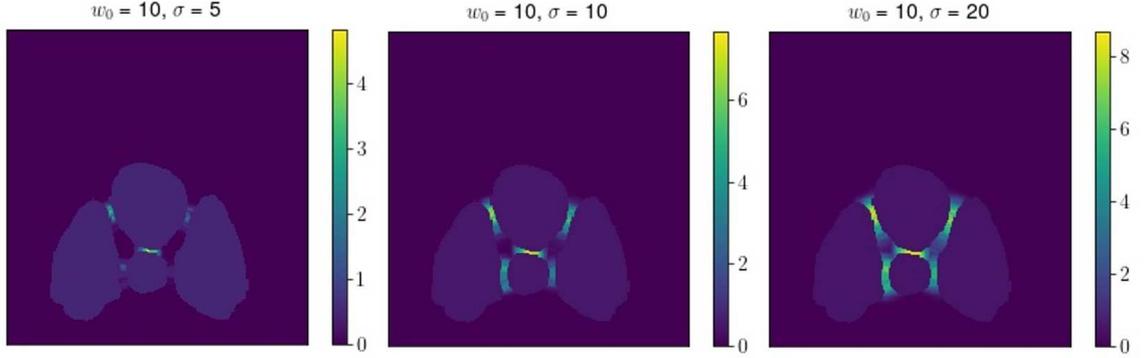

(a) Influence of $\sigma$ on the weight map.

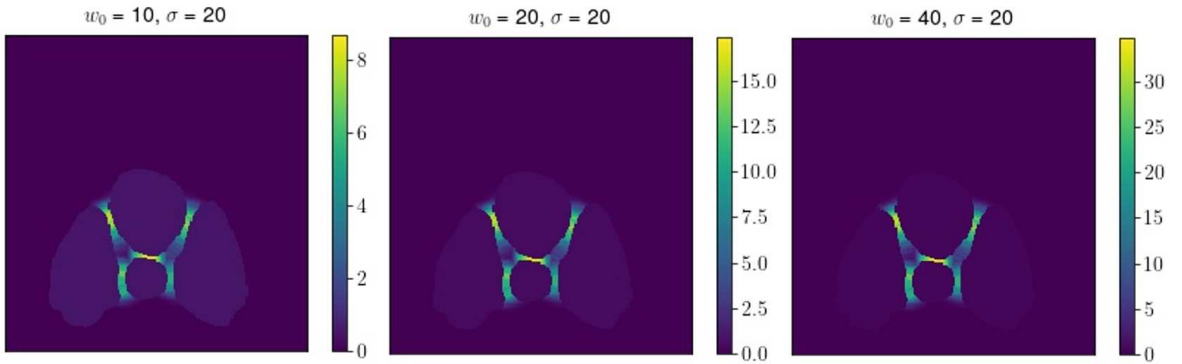

(b) Influence of $w_0$ on the weight map.

Figure 2.10: Influence of $w_0$ and $\sigma$ parameters on the weight map.

The higher $w_0$, the higher is the pixel value between the cells, and the higher $\sigma$, more pixels are selected to have high amplitude. In other words, the regions between the objects are better defined with high values.

## 2.8 Image moments and centroid

The centroid estimation of one image can be calculated by using image moments [44]. Let an digital image be denoted by $I(x, y) \in \mathbb{N}^{P \times Q}$, where $(x, y)$ is the coordinate of an element in this image. The two-dimensional $(m + n)$th order moments $M_{mn}$ of this digital image $I(x, y)$ is defined as

$$M_{mn} = \sum_{x=1}^{P} \sum_{y=1}^{Q} x^m y^n I(x, y), \quad (2.13)$$

where $x^m y^n$ is the basis function.

For a binary image, the elements $(x, y)$ have values 0 or 1. Therefore, the zeroth moment $M_{00}$ represents the sum of all pixel values 1, which can be interpreted as mass of the image or area of the object. The first order moments $M_{10}$ and $M_{01}$ can be interpreted, respectively, as the mean coordinates $x$ and $y$ of this image where the pixels are 1. Therefore, the centroid with coordinates $(x_c, y_c)$ can be defined by the moments as follows:

$$\begin{aligned} x_c &= \frac{M_{10}}{M_{00}}, \\ y_c &= \frac{M_{01}}{M_{00}}. \end{aligned} \quad (2.14)$$

## 2.9 Mathematical morphology

Mathematical morphology can be understood as any technique or algorithm that allows the processing of images to extract morphological features such as shape and boundaries [35]. In this thesis, the *opening* and *watershed* transform algorithms were used to extract morphological features.

### 2.9.1 Opening

The *opening* is constituted by two fundamental morphological algorithms, namely *dilation* and *erosion*. In a few words, *dilation* expands a binary image and *erosion* shrinks it, and the *opening* consists of an erosion followed by a dilation [35]. The opening usually smoothes the outline of a binary object and eliminates fine bumps and noise [35].

### 2.9.2 Watershed

The *watershed* transform is a technique widely used in image segmentation by mathematical morphology. This algorithm allows to segment objects based on regions and pixel similarities [45]. To do this, the objects to be separated have markers, called object markers. These marker objects are found with other morphological transformations. In this work, the markers are found with erosion followed by component connection.

In this way, it separates regions that characterize the objects with internal markers and the background as external markers [35]. Using the markers that indicate the objects separately with centers and the original images to be segmented as masks, the watershed can be applied to define the objects separately. Using an water analogy, as the name suggests, it can be thought that the markers of the objects are dams and the water floods starting from these dams delimited by the masks.

# Chapter 3

# Experimental setup and methods

## 3.1 Definition of traps and cells

The dataset employed in this study were acquired in the BCS Lab and consists of microscope images of an experiment using a microfluidic device. In this study, an experiment means obtaining microscopic images within a certain period of time. This experiment can last for hours, and each time period of approximately 10 minutes is called a timepoint, that is, the microscope runs every 10 minutes through the microfluidic device generating images in brightfield and fluorescent channels. The brightfield image obtained from the microscope at a given timepoint and position is presented in Figure 3.1.

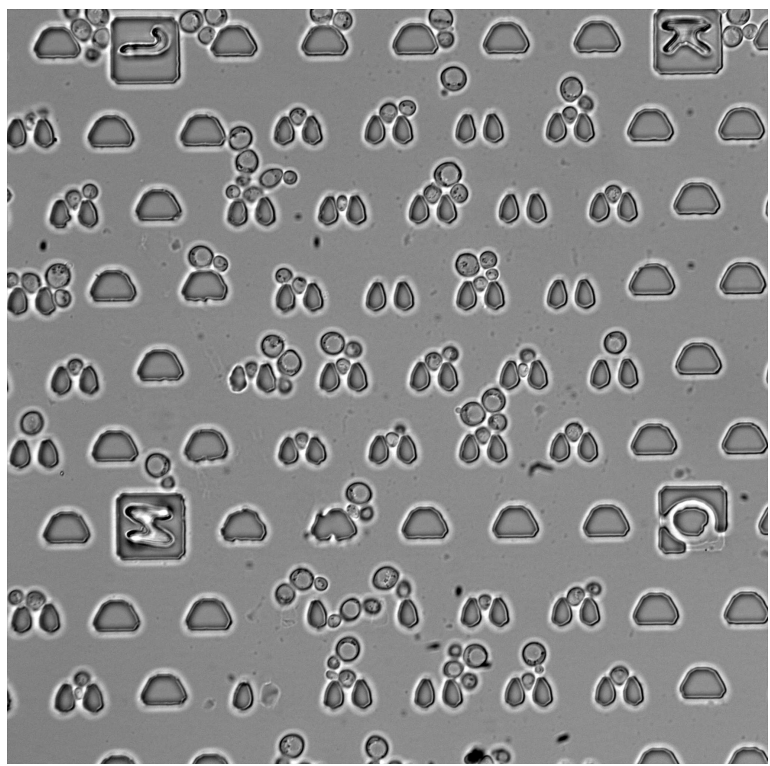

Figure 3.1: Example of a microscope image obtained in an experiment.

The fluorescent channel used in this project is green fluorescent protein channel (GFP). However, all procedures will work with other colours. Figure 3.2 shows an GFP image obtained at one specific timepoint and position of the microfluidic device.

Figure 3.2: GFP image captured at a particular time point and position of an experiment.

In the microscope images, the cells are immobilized in microscopic trap structures. Figure 3.3 shows a microscopic image with cells and traps as well as the label for one specific trap image.

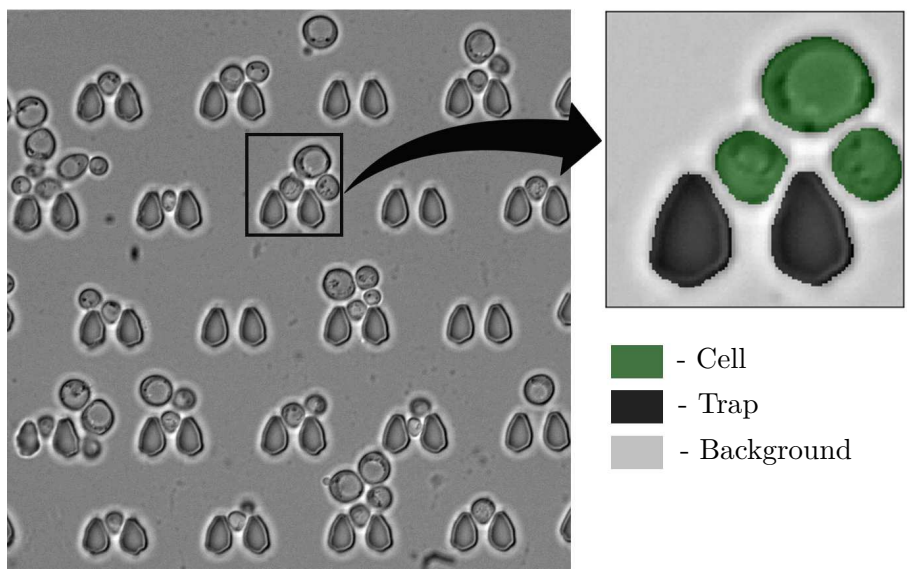

Figure 3.3: Microscope image with example of yeast cells and trap structures.

The trap structures may have different shapes depending on the microfluidic device, as depicted in Figure 3.4.

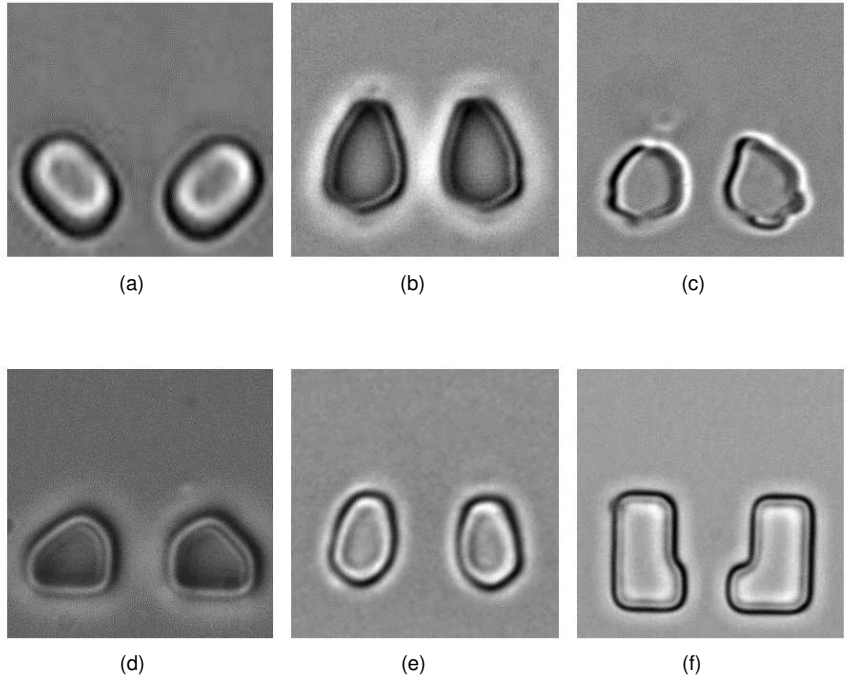

Figure 3.4: Different trap shapes employed in BCS lab.

In this study, the labeled images have traps shape of type (b) or similar.

## 3.2 Overview of training dataset

The training dataset consists of small trap images cropped from a big image such as Figure 3.1. The small images contain the traps and cells of some experiment. Figure 3.5 shows an example of 5 training samples used as training dataset.

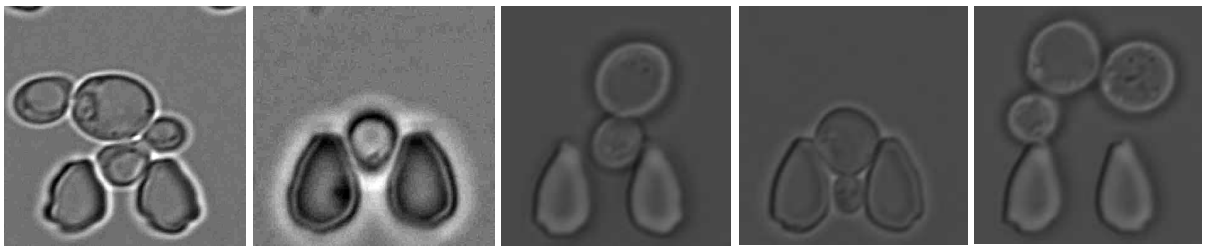

Figure 3.5: Example of 5 training samples.

Training images have different cell arrangements and cell sizes. For example, there are samples with only one trapped cell, several cells around a trapped cell, two trapped cells, or cells around an empty trap. Although not present in Figure 3.5, there is also the case that the trap is empty without the presence of any cells in the trap image.

These trap images were then labeled with four classes:

- background: pixel value 0;

- trap: pixel value 1;

- secondary cell: pixel value 2;

- target cell (cells between the traps): pixel value 3.

Figure 3.6 shows the labeled image with corresponding pixel values depending on the class.

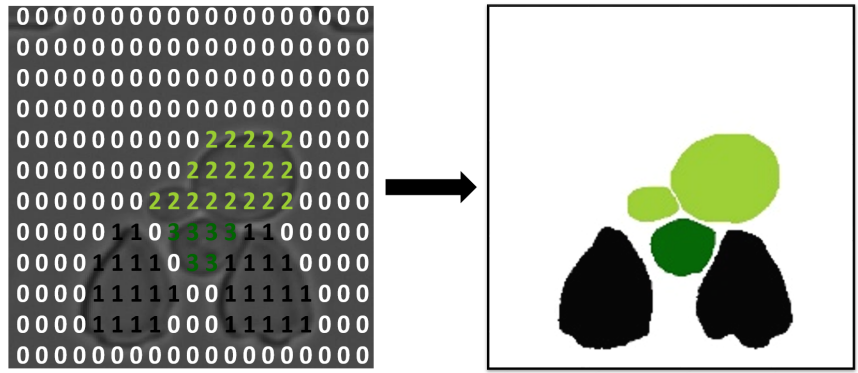

Figure 3.6: Schematic four classes labeled image with corresponding pixel values depending on the class.

Despite having the images with four classes, they can be easily reduced to three classes just by changing the value of the target cell to cell. Figure 3.7 shows one trap image and its corresponding labeled image with four and three classes.

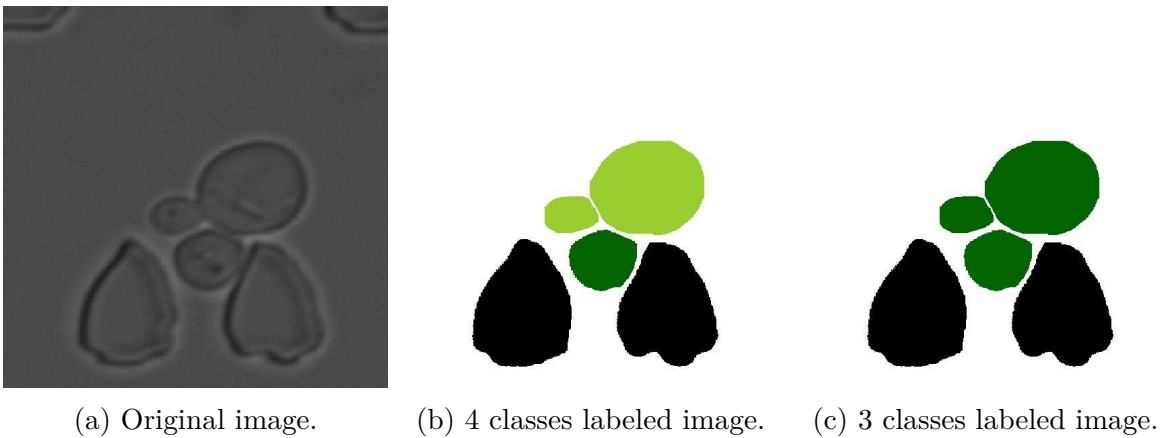

(a) Original image. (b) 4 classes labeled image. (c) 3 classes labeled image.

Figure 3.7: Example of three and four classes labeled image.

## 3.3 Focal plane

Microscopy images can be obtained from different focal planes. The focus outlines the borders of the cell, which is an essential measure for the segmentation problem. The brightfield microscopy available in the laboratory provides 1, 5 or 9 different focal planes in a *z-stack* format. All focal planes have been used, however in this study 5 focal planes are often used. The *z-stack* format means that the measures are taken from different profiles according to the position of the plane in the *z-dimension*, as depicted in Figure 3.8.

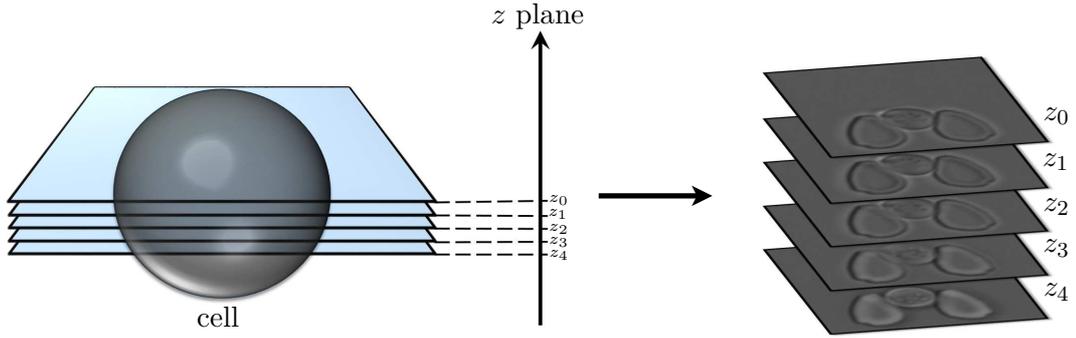

Figure 3.8: On the left of the picture, there is a sideview of a cell in different planes, in blue. On the right, the measured z-stack brightfield images.

Depending on the measured focal plane, the image has different morphological characteristics such as size, shape, brightness and halos around the objects in the image. Figure 3.9 shows the brightfield images obtained by 5 different focal planes.

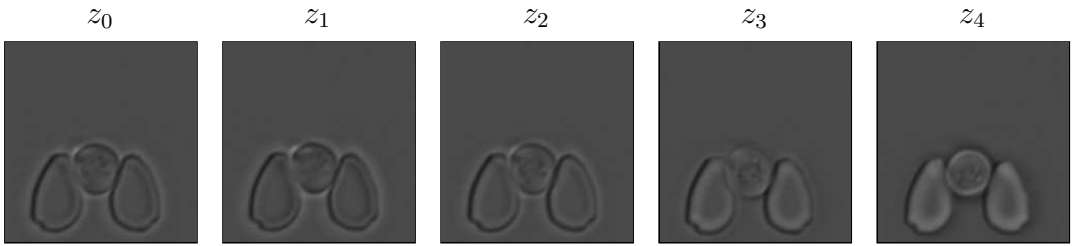

Figure 3.9: Brightfield images measured by different focal planes.

It can be observed that the planes $z_0$, $z_1$ and $z_2$ are quite similar. They have a bright halo around traps and cells, and the contour of both can be clearly observed. However, it can be seen that $z_2$ is slightly brighter than $z_0$ and $z_1$. The image obtained by the plane $z_3$ starts getting different from the previous ones. The halos can barely be seen and inside of the traps is getting whiter. Finally, $z_4$ shows a beginning of a dark halo around the cell and traps, and inside of them becomes to be brighter. The $z$-position, however, are not always the same for different experiments. The examples in Figure 3.9 are only a selection of halos, but other halos are also possible.

## 3.4 Computational setup

Both central processing unit and graphical processing unit are used in machine learning tasks. The GPU has a capability of parallel computing, which is fast for repetitive computations. That is why the GPU is normally used to train neural networks once the repetitive task can be parallelized. Furthermore, there are available for the machine learning community parallel computing platforms, such as CUDA, and optimized libraries such as cuDNN, both developed by NVIDIA. Therefore, the GPU is highly optimized to implement and train convolutional neural networks [46].

In this thesis, the CPU Intel Xeon Processor (Skylake) with 2.340 GHz and GPU GeForce RTX 208 with 1.575 GHz were used to perform the machine learning image segmentation task. Both belong to the Euler server of the BCS group. The servers CPU was used to perform computations such as reading and processing database, while the GPU was used to train the convolutional neural networks and perform the prediction labels of the image segmentation.

## 3.5 Machine learning libraries

The open source machine learning library used was TensorFlow 2.0 and the high-level library Keras built on top of TensorFlow. Since the project language is Python, other core data science and engineering libraries were used, such as NumPy, SciPy, Matplotlib, and SciKit-Learn as additional machine learning packages.

## 3.6 Image preprocessing

Preprocessing is an essential step in computer vision problems. It is really important that the data have the same range of pixel values or the same size, since some architectures work for some specific data shapes. Therefore, the following steps were applied as preprocessing step in order to standardize the input data:

1. resize;

2. rescale;

3. standard normalization;

4. one-hot encoding.

The first step allows that all the images have the same shape. That is, even if the dataset has images with different initial sizes, after this step they will have the same number of pixels. For the input data, a bilinear interpolation is used in this project to downscale or upscale the images. However, for the output labeled data, the interpolation method is the nearest-neighbor interpolation in order to preserve the integer values of the pixels.

The *rescale* step is only applied to the input data. It makes the pixel values in the range [0, 1]. Normally, the input images have type unsigned 8 bit integer (*uint8*). This means that the pixel values belong to the interval [0, 255], since $2^8 = 256$ integer values are available with 8 bits. Therefore, the rescale is easily done by dividing every pixel of the images by 255, for the *uint8* type.

The standard normalization makes the pixel values of the normalized image have mean $\bar{\mu} = 0$ and standard deviation $\bar{\sigma}_N = 1$. This step is only applied to the input data. The mean $\mu$ and the standard deviation $\sigma$ of an image $\mathbf{X} \in \mathbb{R}^{M \times N}$ with elements $x_{ij}$, $i = 0, 1, \cdots, M-1, j = 0, 1, \cdots, N-1$ are calculated by

$$\mu = \frac{\sum\limits_{i=0}^{M-1} \sum\limits_{j=0}^{N-1} x_{ij}}{M + N}, \ \sigma = \sqrt{\frac{\sum\limits_{i=0}^{M-1} \sum\limits_{j=0}^{N-1} (x_{ij} - \mu)^2}{M + N}}. \tag{3.1}$$

Therefore, the normalized image $\overline{\mathbf{X}}$ is given by

$$\overline{\mathbf{X}} = \frac{\mathbf{X} - \mu}{\sigma}. \tag{3.2}$$

In other words, subtraction of the mean and division by the standard deviation were performed for every pixel.

Lastly, one-hot encoding was applied to the labeled data. Since the image segmentation is a multi-class classification problem, whose classes were defined in Section 3.1, the classes can be written individually using only binary values. For the image case, the one-hot encoding makes an $N_c$ classes image of size $N \times N$ have shape $N \times N \times N_c$, only with binary pixel values, where 1 means that the pixel belongs to the class and 0 means that the pixel does not belong to the class. Figure 3.10 illustrates an one-hot encoding for a 4 classes labeled image with size $128 \times 128$.

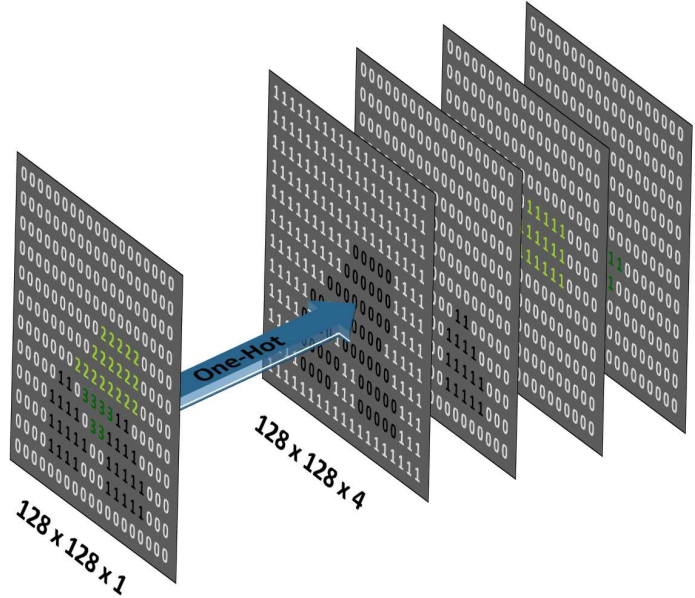

Figure 3.10: Visual example of one-hot encoding for a 4 classes image.

## 3.7 Choice of batch size

The batch size is a hyperparameter that controls the number of samples that go through the network to update the model parameters. This means that the training dataset is divided in batches and the error and model parameters are only computed and updated after those batches. Commonly, the batch size is a group of power of 2 samples. In order to analyze the influence of the batch size on training time, number of epochs to achieve a certain accuracy, neural networks were trained only by varying the batch size. In addition, the batch size also influences the use of GPU memory. A small batch size needs to store fewer images in memory at one iteration of the neural network in one epoch. However, because the iteration is done with fewer images, the number of iterations to complete one epoch is greater than when using a larger batch size. Finally, the batch size limit will be analyzed in the range of 2 to 256, so that the neural network can be trained on the Euler server's GPU.

## 3.8 Number of samples

The number of samples contained in the training dataset is crucial for the quality and especially robustness of predictions, once the dataset must be representative. A small dataset may not contain the information needed or not provide the desired variability for the neural network to extract features and learn different characteristics to the point of becoming as generic as possible. Therefore, since focal planes also influence the quality of predictions, tests were performed on significant samples of each focal plane in order to identify the minimum number of samples required so that the accuracy of predictions is satisfactory.

Following the same procedure as with batch size, identical neural networks were trained only by varying the training dataset, but with the same validation dataset for subsequent comparison and analysis of results. To do so, two images of each focal plane were added to each dataset test, i.e. a dataset that has five focal planes and five samples per focal plane, actually contains 25 images.

## 3.9 Trained models

An experiment was performed to analyze the behavior of the neural network according to some quantitative factors of its architecture, such as the depth $N$ of the neural network and the number of initial filters in the first convolutional layer $N_f$ . In this case, depth means the number of blocks in the encoder and decoder path of the neural network. For example, a model with depth 4 has 4 encoder and 4 decoder blocks. The initial number of filters means the number of filters at the beginning of each encoder/decoder block. In addition, the initial number of filters will also influence the number of parameters and structure of the CNN, since after each block in the encoder, the number of filters in the next block is doubled. There are three main blocks, which are decoder block, encoder block and center block, as depicted in Figure 3.11, which depicts a general simplified model.

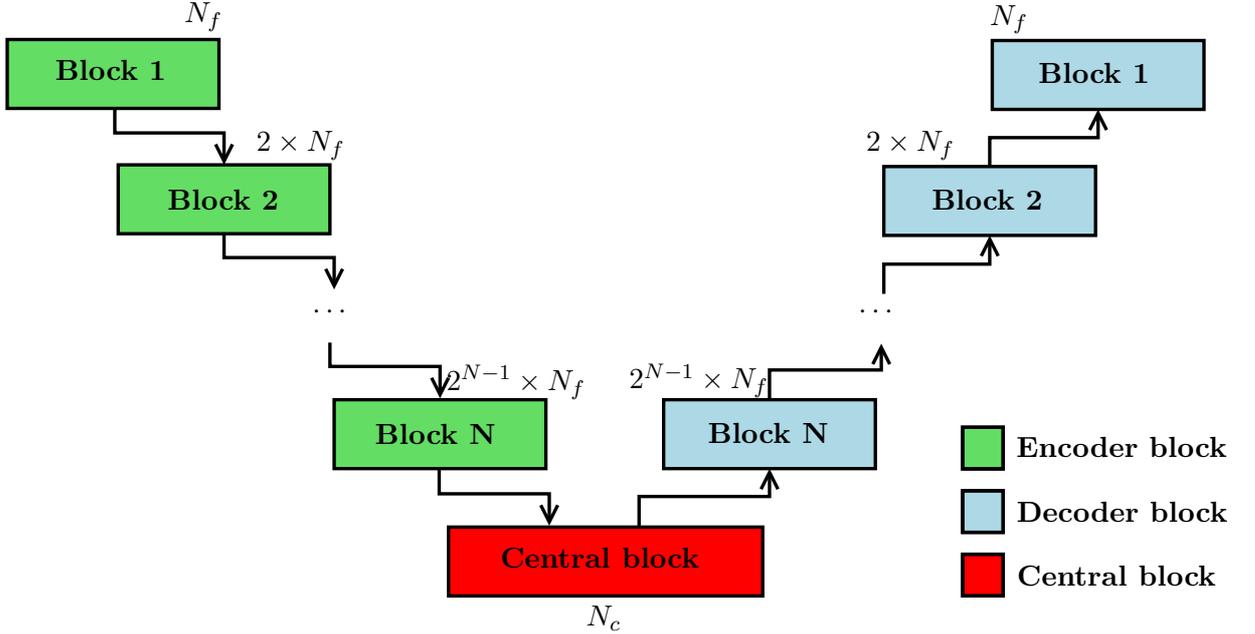

Figure 3.11: General simplified structure of trained models. $N_f$ corresponds to the number of filters in the initial convolutional layer, and $N_c$ is the number of filters in the central block.

The number of filters in the central block $N_c$ is normally set either as twice the number of filters in the previous encoder block, or as the same number of filters in this previous block. The encoder block consists of 2 convolutional layers followed by max pooling. The decoder block consists of 1 transposed convolution layer followed by a concatenation with the corresponding encoder block and 2 convolutional layers. The central block consists of only 2 convolutional layers. The layers and internal structures of the blocks can be seen in more detail in Section 2.3. By using the generic model of Figure 3.11, 15 different models were trained by varying the depth and both the number of filters $N_f$ and $N_c$. The models are summarized in Table 3.1, which also shows the number of parameters of each neural network, that is, the number of trainable parameters or weights.

Table 3.1: Trained U-Net models with different depth and number of filters.

| Model | Depth N | Nf | Nc | Number of parameters |
|---|---|---|---|---|
| 1 | 4 | 32 | 256 | 5.137.635 |
| 2 | 4 | 16 | 256 | 1.940.851 |
| 3 | 4 | 32 | 512 | 7.759.587 |
| 4 | 4 | 64 | 1024 | 31.030.723 |
| 5 | 3 | 16 | 128 | 481.779 |
| 6 | 3 | 32 | 256 | 1.925.091 |
| 7 | 3 | 64 | 512 | 7.696.323 |
| 8 | 3 | 128 | 1024 | 30.777.219 |
| 9 | 2 | 16 | 64 | 116.787 |
| 10 | 2 | 32 | 128 | 466.019 |
| 11 | 2 | 64 | 256 | 1.861.827 |
| 12 | 2 | 128 | 512 | 7.442.819 |
| 13 | 1 | 32 | 64 | 101.027 |
| 14 | 1 | 64 | 128 | 402.755 |
| 15 | 1 | 32 | 256 | 734.243 |

It is worth noting that the model type 4 is the closest one to the original U-Net of Figure 2.6, which has the most parameters among the other trained models.

## 3.10 Input image size

The size of the input image influences the computational cost during training, as the convolutional layers depend on the size of the input image. In addition, the image size can affect the performance of the neural network in terms of the accuracy of the outputs, because if the image is resized to a smaller or larger size than the original size, it loses resolution due to approximations done in the interpolation methods. Due to the Keras Tensorflow library, the image size must be square, which means that the input images must have shape $I_s \times I_s \times N_c$, where $I_s$ is the input size and $N_c$ is the number of channels. In addition, the U-Net architecture has concatenation layers, and each encoder block reduces the size of the images by half due to the max pooling operation. Therefore, the input size will depend on the depth of the neural network, defined in Section 3.9. In general, the input size depends on the depth of the chosen U-Net model. Taking into account the max pooling after each decoder block and that the concatenation layer does not accept be an odd shape, the input size is given by

$$I_s = 2N(2n), \qquad (3.3)$$

where $N$ is the depth of the model and $n \in \mathbb{Z}_{>0}$. In other words, the final size should be an even number after the input sized be reduced by half in $N$ blocks.

## 3.11 Looking inside the network

An interesting aspect to accomplish is to analyze the interior of the neural network [47], that is, to visualize the information extracted by the neural network by observing the outputs of the learned filters, especially after the convolutional layers. Due to the large amount of filters present in the U-Net architecture, only a few images obtained after the output of some convolutional layers are shown in Figure 3.12.

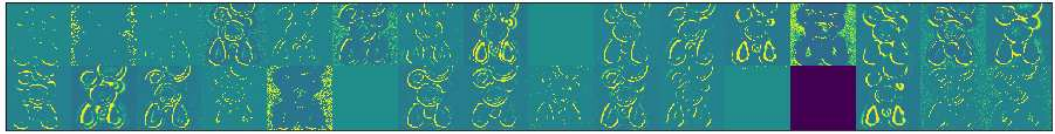

(a) Output images of the 32 filters after the second convolutional layer of the first encoder block.

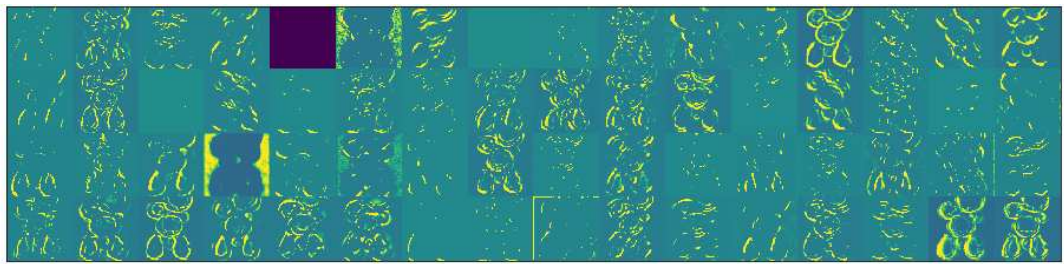

(b) Output images of the 64 filters after the second convolutional layer of the second encoder block.

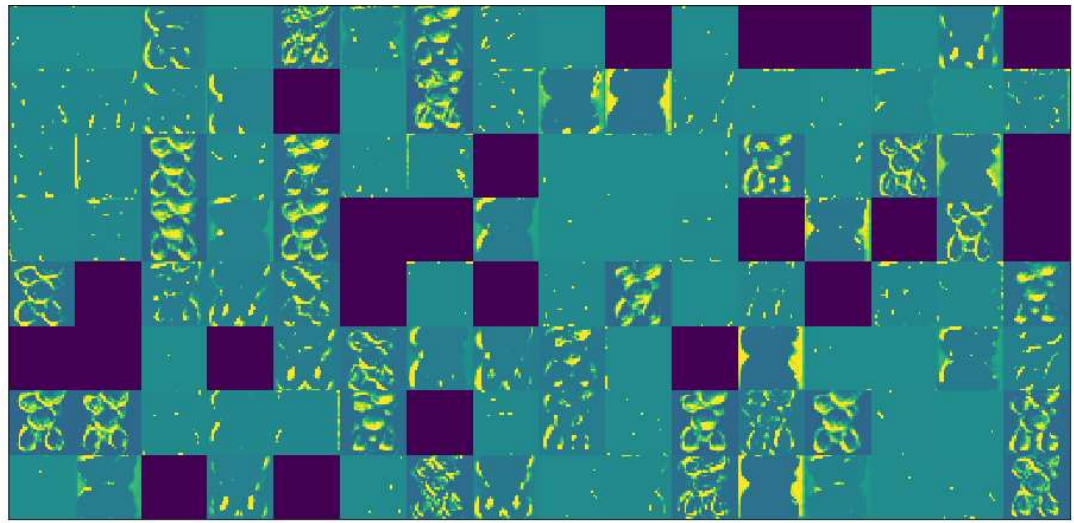

(c) Output images of the 128 filters after the second convolutional layer of the third encoder block.

Figure 3.12: Visualizing the interior of the U-Net. The yellow represents high activation value, and purple represents zero activation. The blue scale in between yellow and purple represents intermediate activation values.

It can be seen that the output images of the first layers are highly activated for the borders and edges of the cells and traps. Therefore, some low-level features are learned once details are identified, as depicted in Figure 3.12a. However, the deeper the network, the more it tends to extract information of entire objects, as it can be seen in Figure 3.12c. In this Figure, the entire cells or traps are highly activated, which means that they represent high-level features. This approach is interesting since it is also possible to notice that not all filters are activated, as in the case of null outputs represented by entire purple images. This may also indicate that some filters can be supplied in order to optimize computational processing. Output images from more convolutional layers are found in Appendix A.

## 3.12 Touching cells problem

The problem of touching cells occurs when two nearby cells are not segmented separately, that is, there is no background region between the edges of the cells. Since the prediction of the cells labels should be accurate, this problem can prejudice the subsequent tracking of individual cells, once cells are merged. Figure 3.13 illustrates the case of touching cells. Note that there is no defined background region between the cells, making it difficult to assign an ID to each individual cell.

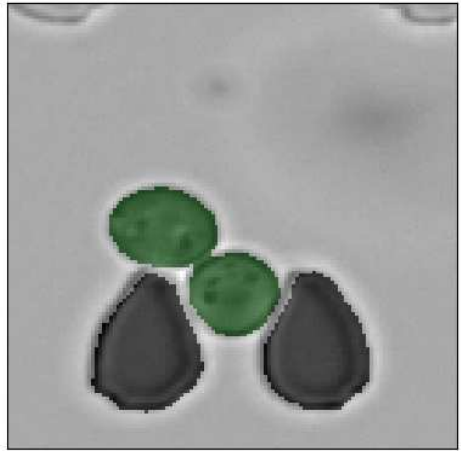

Figure 3.13: Example of the touching cells problem, where two cells are merged.

To solve this problem or reduce the number of cells that touch each other, two approaches were implemented, namely weighted cross-entropy loss and the concatenation of a new label class that represents the regions between the cells.

### 3.12.1 Weighted cross-entropy loss

The weighted cross-entropy loss approach was introduced in [17], and was proposed exactly for the task of separation of touching objects of the same class. The idea behind this approach is to apply large weight to the region between the cells so that in loss function these regions become more important during the training. Thus, this region is more penalized during training in case of a wrong prediction. These weights are obtained from the weight maps defined in Section 2.7. Then, the weight maps are integrated into the cost function by doing a pixel-wise multiplication by the logarithm of the output of the network. Therefore, the weighted cross-entropy loss $L_{wcee}$ is given by

$$L_{wcce} = \frac{1}{mn} \sum_{m=1}^{M} \sum_{n=1}^{N} w_{m,n} \sum_{c=1}^{C} -y_{m,n,c} \log(\hat{y}_{m,n,c}), \qquad (3.4)$$

where $\hat{y}_{m,n,c}$ is the element $(m, n, c)$ of the output predicted labels $\hat{\mathbf{Y}} \in \mathbb{R}^{M \times N \times C}$ with $C$ classes, $y_{m,n,c}$ is the element $(m, n, c)$ of the true labels $\mathbf{Y} \in \mathbb{N}^{M \times N \times C}$, and $w_{m,n}$ is the element $(m, n)$ of the weighted map $\mathbf{W} \in \mathbb{R}^{M \times N}$.

By doing this, the regions between cells are penalized in the loss function since these regions already have large pixel values due to the weight map, and it will remain high after the pixel-wise multiplication in the loss function.

### 3.12.2 Interface class

Besides the weighted cross-entropy loss, a new class called interface class was proposed in such a way that the neural network learns to identify these regions between cells. This means that the regions between objects must be a label different from the labels of the true labels. To do this, the same weight maps defined in Section 2.7 were used to generate this new class. Figure 3.14 shows the weight map and the generated interface class.

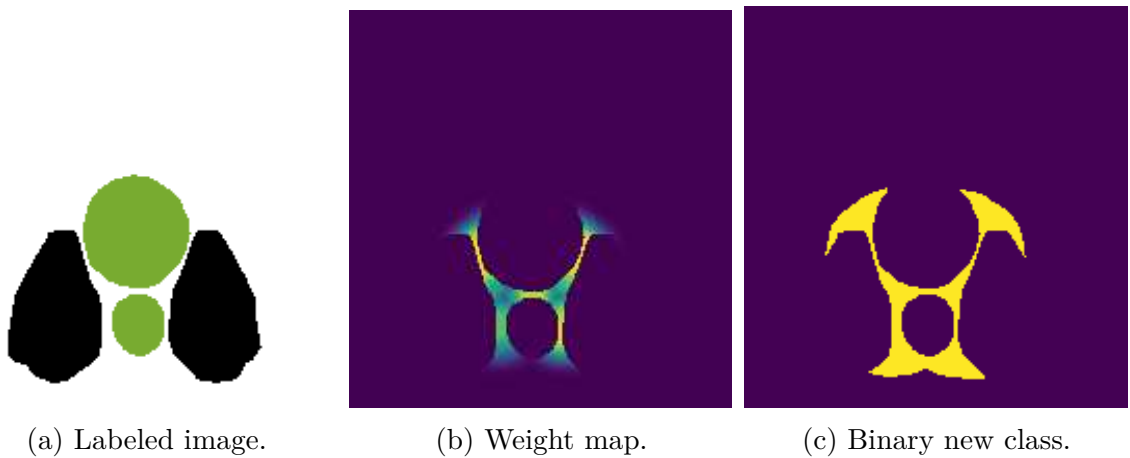

(a) Labeled image.      (b) Weight map.      (c) Binary new class.

Figure 3.14: Interface class generated from the weight map of the labeled image. The threshold here was set to 0.1, and the parameters $w_0$ and $\sigma$ of the weight map are both equal to 50, such that a big gap between cells can be captured by weight map.

Starting from the weight map, the new interface class was created by only assigning the value 1 for weights bigger than a defined threshold, and 0 for the others. That is, a binarization was made using a threshold. After that, the interface class is then concatenated with the other labels.

## 3.13 Experiment pipeline

This section describes the processes performed during an analysis of an experiment. The processes include the localization of traps and cropping into smaller images, image segmentation by the trained neural network, postprocessing of predicted labels, calculation of fluorescence and tracking of each cell.

### 3.13.1 Trap localization

The localization of traps in the device is done by the cross-correlation between the reference image, timepoint 0, and some representative templates. This method is known as *template matching*. Figure 3.15 shows three examples of templates used to perform the trap localization.

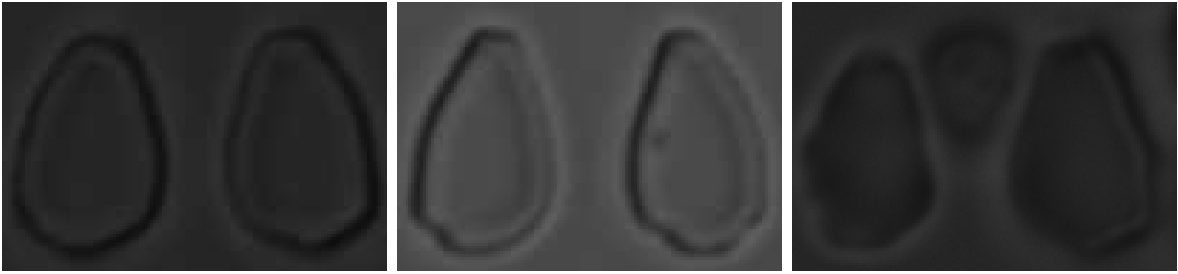

Figure 3.15: Three template examples used to localize traps via *template matching*.

After *template matching*, it is possible to find the positions in the big image where the template appears with high similarity. To get the positions of the templates in the image, a threshold is applied in order to get the highest values where the template occurs with highest similarity. The trap localization step of this work was done by using the previous pipeline available in the BCS group. Figure 3.16 shows the trap localization at timepoint 0.

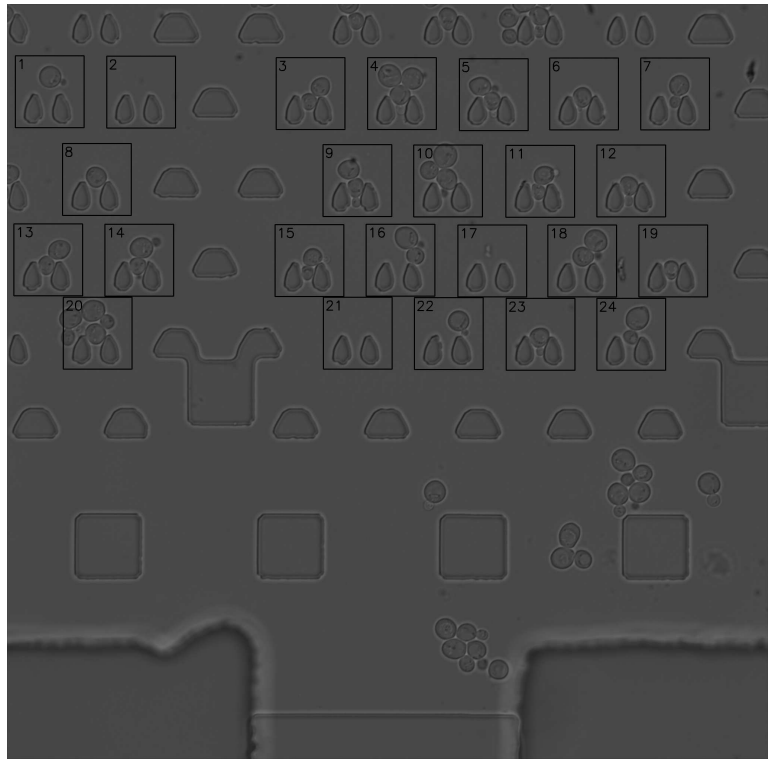

Figure 3.16: Trap localization at timepoint 0.

In Figure 3.16, 24 traps were localized, and the regions inside the black boxes are then used as small images to test the trained network.

### 3.13.2 Shift correction

First of all, the time point 0 is used as a reference for the other time instants. Images of the next time instants can occur in a slightly shifted position relative to the reference. This means that, without pre-processing, the same position of the first localized traps cannot be used in the other time instants because the locations are not exactly the same. With this shift correction, it is ensured that the same traps are located at all timepoints. In addition, the computational cost is also lower, since it is not necessary to localize the traps at every time point if this shift correction is performed.

To make this correction, a cross-correlation is made between the reference timepoint and the current timepoint. Once this is done, a maximum value is obtained that indicates the position where the images present the highest similarity. From the position of this maximum value and the center of the reference image, one can obtain the number of pixels in the $x$ and $y$ axes that the image is shifted related to the reference, thus being able to perform the correction. The *template matching* done at timepoint 0 allows to obtain the positions of the traps in the reference image. Therefore, it is possible to find the traps in the other timepoints by using the position of the traps found in the reference image combined with the correction pixels obtained from the shift correction.

In the previous pipeline, the cross-correlation was made between the two entire images. Each image has size $2048 \times 2048$, which means that the computational cost to find the amount of pixels to shift the current timepoint is high. However, this computation is not necessary, since all pixels are equally shifted from one timepoint to the next. That is, only a portion of the image can be used, because any portion of the image will suffer the same displacement of pixels in relation to the initial timepoint. Therefore, a cross-correlation using only a $750 \times 750$ was performed. This size is representative enough to detect the number of pixels to shift the current timepoint. From this improvement, the computational cost was significantly reduced, also contributing to a possible application of the pipeline almost in real time.

### 3.13.3 Segmentation step

The segmentation step is performed using the Keras Tensorflow machine learning library. Once the neural network is trained, it is necessary to build the model and load the learned weight parameters. With the built model, the small images of the traps obtained in the experiment after localization of the traps can then be segmented as test input. The output of the neural network will then be an array with the same size of the input images due to the built U-net model. It contains in the last dimension the class probabilities of each pixel. Therefore, to create a new image with integer labels, a threshold is used to assign the value of the corresponding class where the pixel probability is greater than this threshold. As a numerical example, knowing that the class order is background, trap and cell, an output $\hat{y}_{m,n} = [0.05, 0.06, 0.89]$ for the pixel position $(m, n)$ means that the pixel has high probability of being a cell, once the last index, which represents the cell class, contains the highest value. Extending this example for all pixels in the output image, it is possible to obtain the predicted label mask.

This step significantly influences the processing and application time of the pipeline, since the predicted label mask is performed after each segmentation of the trap images. In the previous pipeline, this step was not optimally implemented because there were many for loops that scanned the entire image comparing the probabilities of each class and assigning the value of the class with the highest probability. In the pipeline of this study, however, the label mask was implemented in an efficient way, using only a search that satisfies a condition, working with boolean arrays, additions and multiplications to reconstruct the predicted label mask. This step may seem irrelevant to the final processing of the pipeline, but it significantly influences the processing time if it is not properly implemented.

### 3.13.4 Postprocessing

The postprocessing of the neural network output is a very important step to complement its prediction of the classes. The focus of the postprocessing in this work is on eliminating small islands where false predictions have been made, and also to improve the results of the touching cells problem. To do this, morphological operations were applied as a tool to extract components of the image that are useful in representing the shape, edges and boundaries. A fundamental algorithm was used for this morphological processing, namely *opening*.

In the case of this project, the *opening* algorithm eliminates predictive noises, which could be counted as small cells, and can also complement the separation of touching cells due to erosion. Figure 3.17 shows the effect of *opening* as postprocessing of a cell prediction.

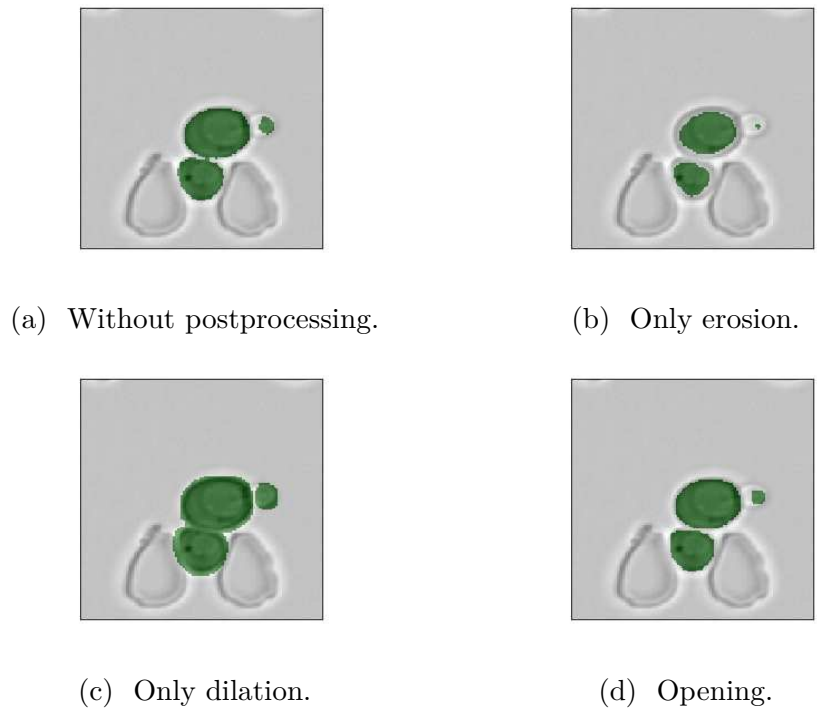

(a) Without postprocessing.    (b) Only erosion.

(c) Only dilation.    (d) Opening.

Figure 3.17: Example of postprocessing applied to a cell prediction.

It can be perceived that, compared to the prediction in Figure 3.17a without postprocessing, in Figure 3.17b the binary prediction of cells is shrunk, and in Figure 3.17c it is expanded. Furthermore, it is noted that in Figure 3.17d the opening allows to correct some isolated predictions, as well as to create a background region between the cells.

In addition to *opening*, it will be necessary to separate the label from the cells so that they can subsequently be tracked individually. This means that each individual cell will have a different label value. If the prediction was perfect, it would only be necessary to connect the nearby elements and set a different value. However, it may happen that the cells are still in contact with each other even after all the processing to avoid the problem of touching cells. In addition, it may happen that the prediction was not good or that there were still small prediction islands that prevent elements from only being connected as the same label. As a consequence of poor prediction, there might be the identification of a new cell without it actually being a cell. Therefore, the separation of cell labels is done as follows:

1. apply *erosion* to the prediction;
2. identify connected components and set a new pixel value to each connected element (cell);
3. apply watershed by using the previous connected elements and the original prediction images as mask.

The first step is performed in order to reduce the touching cells area. This time the erosion can be applied once the shape and area is not important, but the regions separately to be used as markers of the watershed algorithm. The shape will be recovered after the last step. Furthermore, step 1 also allows to get rid of small island regions where there was false alarm, that means, there was a cell prediction where it was actually background. The second step sets a different label to each element region by connecting components. This means that regions surrounded by background will have a different label each. The third step uses the previous labels and regions found on the previous steps as markers and reconstructs the prediction. Figure 3.18 illustrates the three steps for one specific prediction.

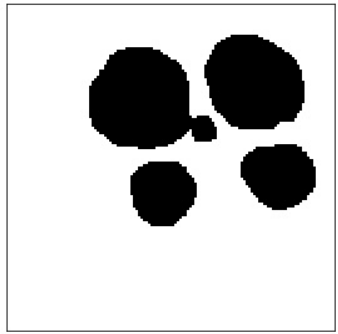
(a) Original cells prediction.

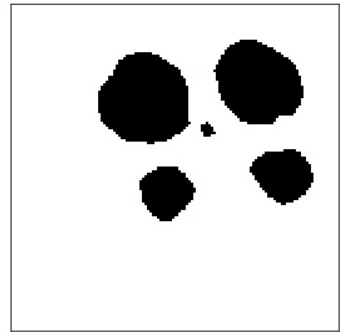
(b) Prediction after erosion.

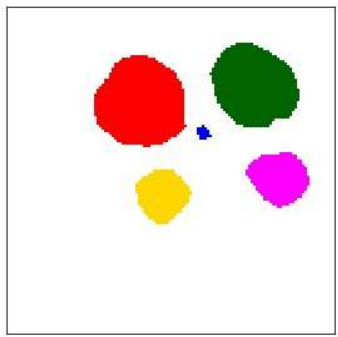
(c) Prediction after connected components.

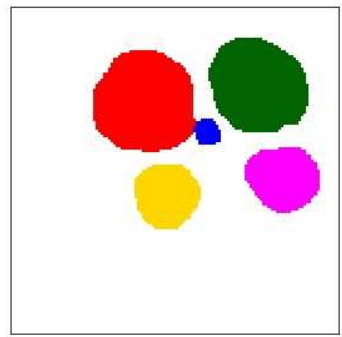
(d) Prediction after watershed.

Figure 3.18: Example of postprocessing applied to set a different label to every cell. In (a) and (b), the predictions are binary, where black is cell label and white background. In (c) and (d), every color represents a different label.

### 3.13.5 Fluorescence calculation

Once cell predictions are made, the GFP channel is used to measure the fluorescence of individual cells. For this, using the same positions of the traps obtained during shift correction and trap localization, the regions of the traps are selected on smaller images.

Then, with the predictions of the cells in the brightfield channel, it is possible to identify the pixels in which the cell is present, making the image binary. Thus, a pixel-wise multiplication is sufficient to obtain the region in which the cell is present in the GFP channel. Once this is done, the pixel values obtained are summed up, since what is not a cell has a value of 0 after the pixel-wise multiplication. Therefore, the measured fluorescence of each cell is equal to the sum of the pixel values in which the cell was present. Furthermore, it is important to make a background correction, that is, to eliminate the background fluorescence. To calculate the background fluorescence, a first method was done by averaging the GFP of the entire image. This was an approximation considering that the classes are unbalanced and the largest portion of the pixels is background. However, this method is not accurate for the case where lots of cells are present in the image. Then a second method was implemented by doing the same procedure used to calculate the cell fluorescence, but applied to the background class, i.e. pixel-wise multiplication between the background mask and the GFP trap image, followed by the sum of the pixel values resulting from the multiplication. This method is also valid for the case where many cells are present in the image, since the background class is well defined. Figure 3.19 shows the intermediate steps of calculating the fluorescence of one single cell.

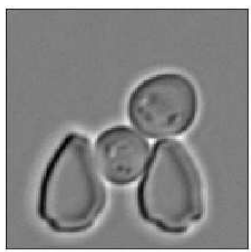
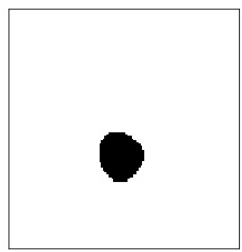

(a) Trap 2 brightfield image.    (b) Single cell predicted label.

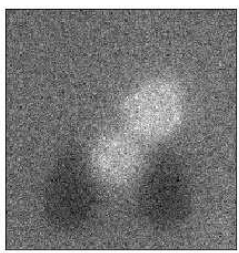
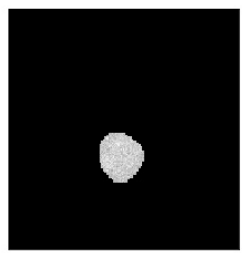

(c) Trap 2 GFP image.    (d) Single cell fluorescence.

Figure 3.19: Fluorescence calculation steps. Figures (a) and (b) are respectively the original trap brightfield image and its single cell prediction. Figure (c) shows the same trap in the GFP channel, and (d) its single cell GFP image after pixel-wise multiplication by the single cell prediction (b).

### 3.13.6 Cell tracking

The cell tracking is a challenging step of the pipeline. It consists of tracking individual cells in time, measuring their fluorescence so that in the end it is possible to analyze the intensity of fluorescence throughout the experiment for each individual cell. Therefore, all cells can be analyzed individually or together by traps or by position in the experiment.

The tracking of the cells is performed by estimating the centroid of every single cell for timepoint 0 as reference and by setting a cell ID and trap ID for every cell. The trap ID is important to facilitate the search for all the cells present in this trap, and the cell ID is used in the end to get the calculated information of every single cell by its ID. For the next timepoints, the tracking is done by the following steps:

1. estimate centroid of current single cell;

2. get centroids of the same trap ID at previous timepoint;

3. calculate Euclidean distance between the current centroid and all previous centroids of the trap;

4. set current cell ID as the same ID of the cell with the minimum Euclidean distance, if this distance is smaller than a threshold;

5. if the distance is higher than the threshold for all previous cells, repeat steps 1 to 4 for the previous 5 timepoints, if they exist. This step avoids the case where there was a poor cell prediction at one timepoint and not at others;

6. if the current centroid is not close enough to any of the previous centroids of the trap, set new cell ID as the maximum cell ID of the trap ID incremented by one;

### 3.13.7 Save data information

The data is saved in such a way that all the information needed to analyze the experiment can be read in only one file. To do this, the data structure adopted is the pickle *.p* file, which is similar to *.json* file (JavaScript Object Notation). The pickle protocols allow object serialization of python objects.

In python, the data is saved as a list of python dictionaries. Each cell is a python dictionary whose keys are cell ID, trap ID, fluorescence, area, timepoint, centroid and cell coordinates. Therefore, to analyze only cell 1 of trap 4, for example, it is sufficient to search for the key "trap ID" whose value is 4, then filter for the key "cell ID" whose value is 1. Once this is done, all the elements that satisfy this search are available as a list, being able to plot curves as the fluorescence over time.

It is also possible to convert the saved the data into a *.mat* file in MATLAB format. This is interesting since MATLAB is a powerful tool for analyzing data and working with matrices and vectors.

# Chapter 4

# Results and discussion

## 4.1 Influence of focal planes

Different data scenarios with different halos and contrasts can be built in the dataset depending on the focal plane of the measured images, as shown in Figure 3.9. The following experiment aims to analyze the influence of the focal planes during training the network. First, the networks have the same *U-Net* architecture and were trained on a small dataset with 43 training samples, 20 epochs, batch size of 2 and images of size $128 \times 128$. The visual results of the experiments are summarized in Table 4.1.

Table 4.1 shows that when the network is trained and tested on the same z-plane, it returns a good prediction. A good prediction result is when the accuracy is above 90% because this dataset is totally unbalanced, that is, around 80% of those images is background. The results indicate that it is possible to aggregate the dataset according to the brightness of the halo. For example, $z_0$, $z_1$ and $z_2$ can be grouped together in a $z_+$ group since they have bright halo and the predictions have high accuracy for them trained separately (over 98%). Furthermore, training only on $z_3$ has also a good result when tested on $z_4$. However, since training only on $z_4$ was not so accurate when tested on $z_3$, it is better to have two different groups of data, which are $z_0$ for no halo images and $z_-$ for dark halo images. It is important to mention that in some experiments the halos can be brighter or even darker than the images of Figure 3.9. Lastly, when the network is trained with all the $z's$ categories, the prediction accuracy is over 98% for all cases. Therefore, a good training dataset should have a balanced number of $z_+$, $z_0$ and $z_-$ images.

Table 4.1: Results of the U-Net trained on different focal planes. The values in the images represent the accuracy of the predictions, which best result is in bold.

| Trained on | Tested on $z_0$ | Tested on $z_1$ | Tested on $z_2$ | Tested on $z_3$ | Tested on $z_4$ |
|---|---|---|---|---|---|
| $z_0$ | 98.44% 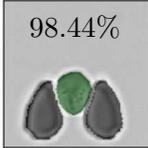 | **98.99%** 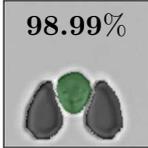 | 98.91% 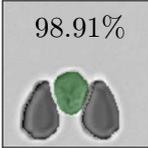 | 91.30% 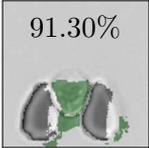 | 81.88% 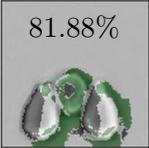 |
| $z_1$ | 98.30% 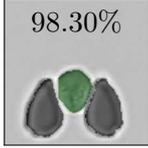 | **98.99%** 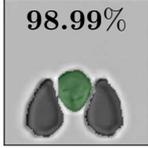 | 98.36% 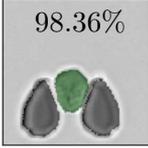 | 87.67% 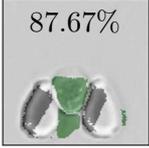 | 78.34% 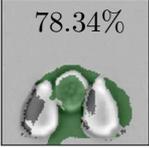 |
| $z_2$ | 98.37% 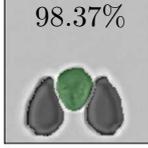 | **98.77%** 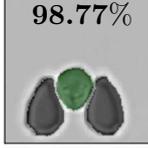 | 98.71% 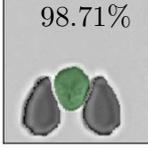 | 95.78% 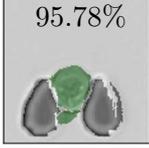 | 91.99% 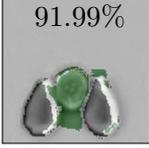 |
| $z_3$ | 86.41% 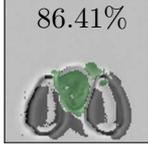 | 81.18% 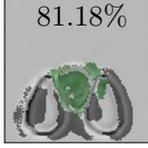 | 93.41% 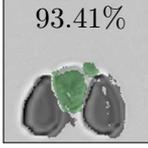 | 97.61% 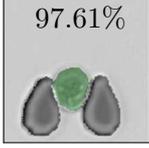 | **97.91%** 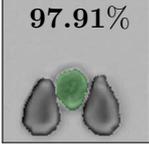 |
| $z_4$ | 69.45% 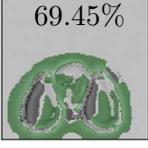 | 66.51% 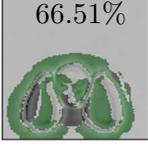 | 73.91% 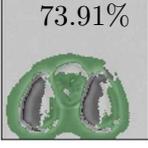 | 95.79% 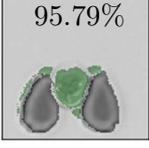 | **98.50%** 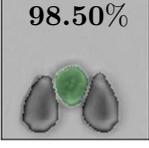 |
| All $z$'s | 98.51% 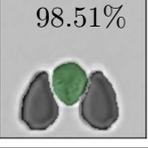 | 98.90% 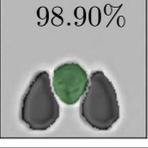 | **98.97%** 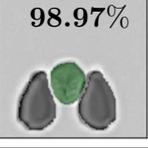 | 98.22% 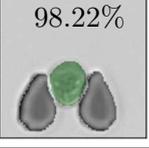 | 98.88% 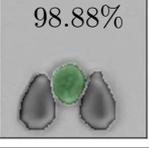 |

## 4.2 Influence of batch size

In this section different batch sizes were evaluated to analyze the accuracy, loss and training time curves. Figures 4.1 and 4.2 show the accuracy and the loss, respectively, of the same network trained with different batch sizes.

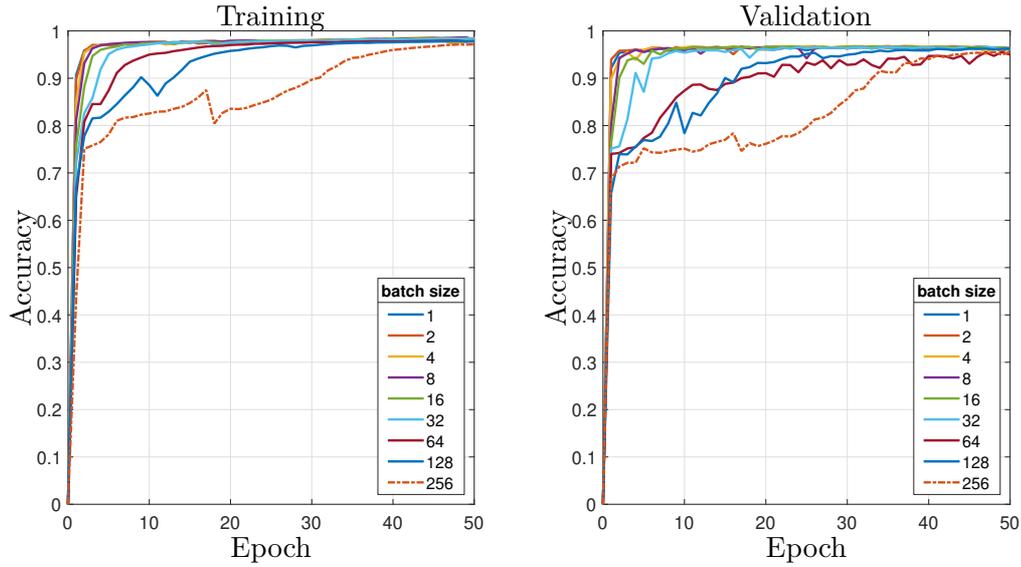

Figure 4.1: Comparison of the accuracy for different batch sizes during training and validation.

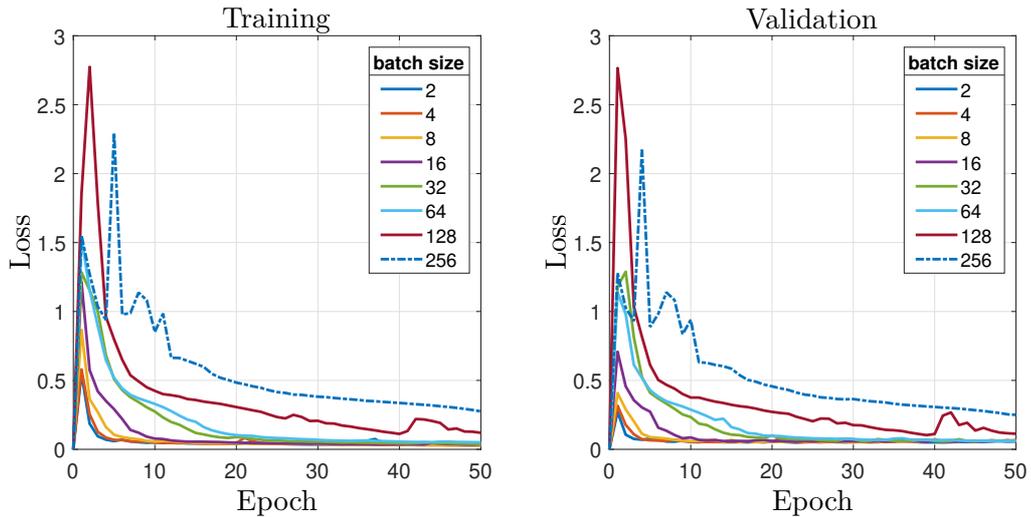

Figure 4.2: Comparison of the loss for different batch sizes during training and validation.

The networks of model 1 were trained with 927 samples of size $128 \times 128$, with image augmentation, and 50 epochs. It can be seen that a large batch size (larger than 64) requires more epochs to reach high accuracy, once with large batch size the network has less updates than with a lower batch size. Furthermore, the smaller the batch size, the faster decays the loss function, as depicted in Figure 4.2. Table 4.2 shows the mean time necessary to compute one epoch during training the network.

Table 4.2: Mean training computational time per epoch regarding the batch size using CPU and GPU.

| Batch size | CPU time/epoch (s) | GPU time/epoch (s) |
|---|---|---|
| 2 | 148.8658 | 10.2002 |
| 4 | 125.5211 | 6.5894 |
| 8 | 130.7373 | 5.6960 |
| 16 | 117.1715 | 5.0116 |
| 32 | 115.0120 | 4.4128 |
| 64 | 112.9666 | 4.4077 |
| 128 | 109.4667 | 4.3100 |
| 256 | 107.3481 | - |

The smaller the batch size, the larger the computational time. It is also related with the updates of the model parameters, that is, a small batch size has more parameter updates than a high batch size since it has more iterations per epoch to go through all samples. That is why a large batch size requires more epochs to reach high accuracy, once it has less iterations and less updates per epoch. Furthermore, a small batch size requires less memory, since the number of samples is reduced during one training iteration. Therefore, a small batch size also contributes to avoid the CPU or GPU to run out of memory, which was what happened for batch size larger than 128 on the GPU. For this application, a good batch size is deemed to be in the range of 4 and 64. A batch size of 8 was chosen to be used during training the networks in this work.

## 4.3 Influence of epochs

Since the number of epochs should be larger the larger the batch size, for the chosen batch size 8, it should be analyzed the number of epochs such that the neural network learns as much as possible with high accuracy without overfitting. Figures 4.3 and 4.4 show the influence of the number of epochs on the CNN trained with batch size 8 for accuracy and loss, respectively.

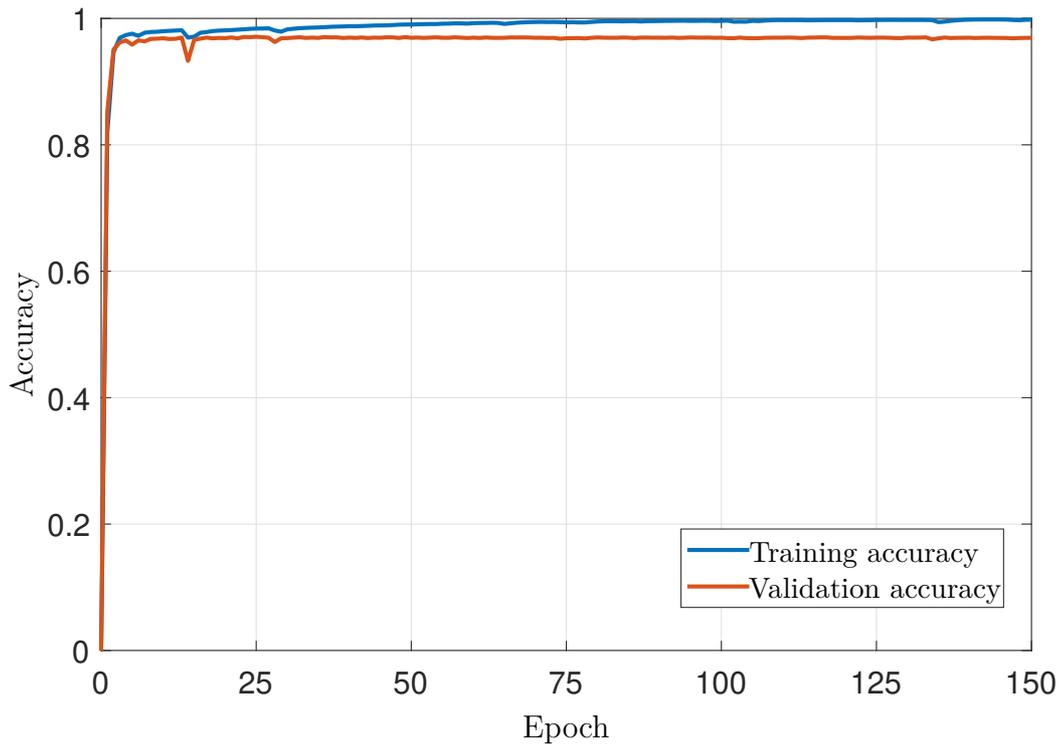

Figure 4.3: Accuracy in 150 epochs of the CNN trained with batch size 8.

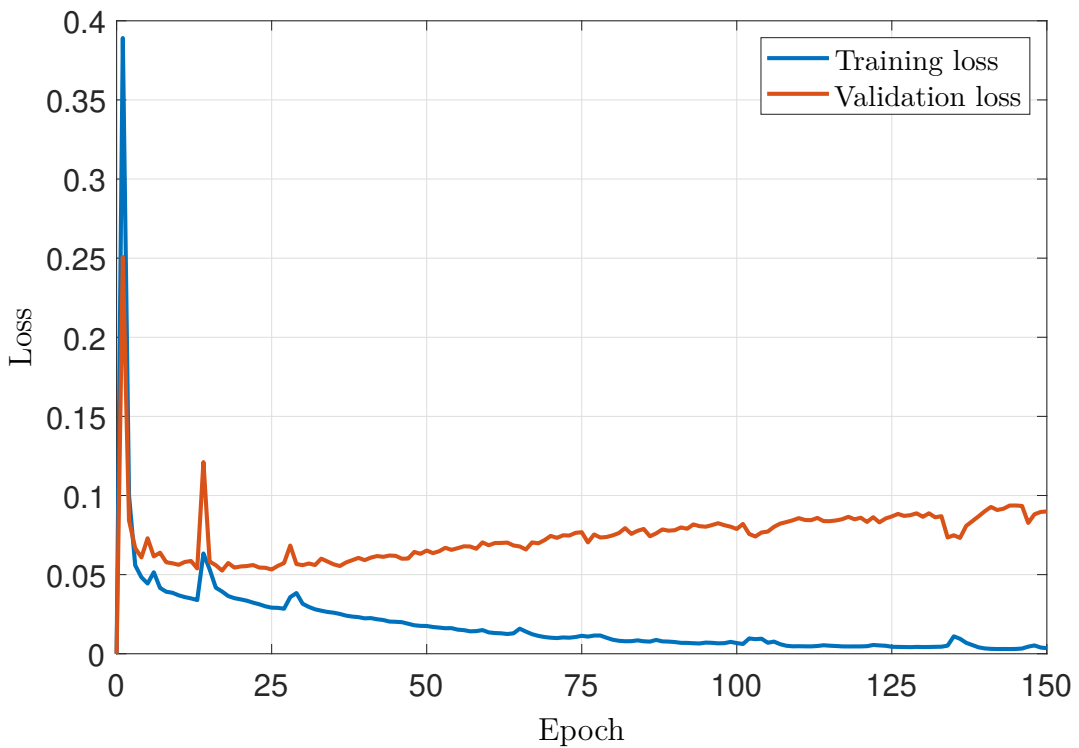

Figure 4.4: Loss in 150 epochs of the CNN trained with batch size 8.

Figure 4.3 shows that there is overfitting after approximately 30 epochs since the loss validation curve is no longer decreasing with the increase in the number of epochs. This means that the model is not able to segment unseen samples, and it is just learning the training dataset. In Figure 4.4, it is also possible to observe that, after epoch 30, the accuracy remains almost constant. Therefore, the neural network with batch size 8 will be trained in 30 epochs.

## 4.4 Comparison and analysis of trained models

The models defined in Section 3.9 were evaluated. To compare the models, the metrics *accuracy*, *DICE*, *IoU* and *weighted IoU* were computed. Table 4.3 shows all of these computations for all built models.

Table 4.3: Metrics results for the different trained models.

| Model | Number of parameters | Prediction time/sample (ms) | Acc | DICE | IoU | Weighted IoU |
|---|---|---|---|---|---|---|
| 1 | 5.137.635 | 8.9964 | **0.9723** | **0.9675** | **0.9376** | **0.8848** |
| 2 | 1.940.851 | 15.3873 | 0.9697 | 0.9642 | 0.9315 | 0.8599 |
| 3 | 7.759.587 | 8.6817 | 0.9692 | 0.9623 | 0.9281 | 0.8625 |
| 4 | 31.030.723 | 26.1425 | 0.9699 | 0.9639 | 0.9310 | 0.8643 |
| 5 | 481.779 | 4.7730 | 0.9645 | 0.9565 | 0.9177 | 0.8367 |
| 6 | 1.925.091 | 6.1760 | 0.9669 | 0.9607 | 0.9252 | 0.8536 |
| 7 | 7.696.323 | 7.4643 | 0.9693 | 0.9632 | 0.9298 | 0.8697 |
| 8 | 30.777.219 | 26.0819 | 0.9693 | 0.9643 | 0.9318 | 0.8660 |
| 9 | 116.787 | 4.8651 | 0.9636 | 0.9581 | 0.9206 | 0.8503 |
| 10 | 466.019 | 5.2996 | 0.9647 | 0.9577 | 0.9199 | 0.8443 |
| 11 | 1.861.827 | **4.3409** | 0.9664 | 0.9612 | 0.9262 | 0.8628 |
| 12 | 7.442.819 | 10.3818 | 0.9650 | 0.9588 | 0.9218 | 0.8514 |
| 13 | 101.027 | 4.4552 | 0.9529 | 0.9431 | 0.8939 | 0.7762 |
| 14 | 402.755 | 4.9630 | 0.9538 | 0.9456 | 0.8983 | 0.7838 |
| 15 | 734.243 | 5.3356 | 0.9521 | 0.9421 | 0.8923 | 0.7704 |

The number of parameters of each model was repeated from Table 3.1 for convenience, in order to compare how big the networks are. From Table 4.3 it can be concluded that there is a link between the number of parameters and the prediction time. In general, the larger the number of parameters in the neural network, the longer the prediction time, once there is more computation in the network to output the class probabilities. The number of parameters of each model can be consulted in Table 3.1. As an example, models 8 and 4 have the most parameters, and they also have the longest prediction time. On the other hand, models 14 and 13 have few parameters in their models and have relatively small prediction times.

The values of *accuracy*, *DICE*, and *IoU* are very close among the models, so the *weighted IoU* is the most relevant for this analysis. The *weighted IoU* takes more into account the *IoU* of the cells since the *IoU* of the cells prediction is weighted with larger value. This is the reason that it is more relevant for this study once the other metrics capture all classes with same relevance. It is possible to observe that there is no correlation between the number of parameters and the quality of the predictions, since models 8 and 4 are the ones that have more parameters, however they do not have the best prediction results. Model 1, on the other hand, which has one fifth of the number of parameters of models 8 and 4, is the one that showed the best result according to the measurements. In order to have a visual sense of the quality of predictions, Figure 4.5 shows the model prediction results for a specific trap.

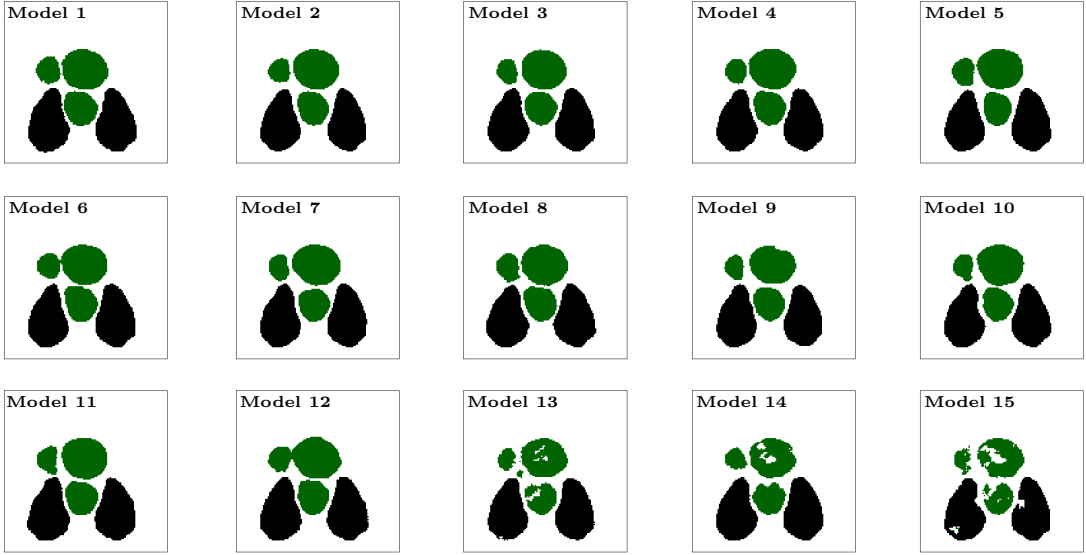

Figure 4.5: Example of prediction results of one specific trap image for all trained models.

One can see in Figure 4.5 that the predictions are relatively accurate. This is why the models do not differ significantly in terms of metrics, with the exception of models 13, 14, and 15, which have the lowest *weighted IoU*.

## 4.5 Influence of input size

This experiment was done to analyze the influence of the input image size on the performance of the network. For that, the *weighted IoU*, *accuracy* and *DICE* were computed as metrics in order to compare the different results. By using Equation 3.3 for $n \in \{6, 7, 8, 9, 10, 11, 12, 13, 14, 15, 16\}$ and the Model 1 with depth $N = 4$, the input size $I_s \in \{96, 112, 128, 144, 160, 176, 192, 208, 224, 240, 256\}$ were tested. The original input images shape without resizing is $195 \times 185$. Figure 4.6 shows the results for the tested input sizes.

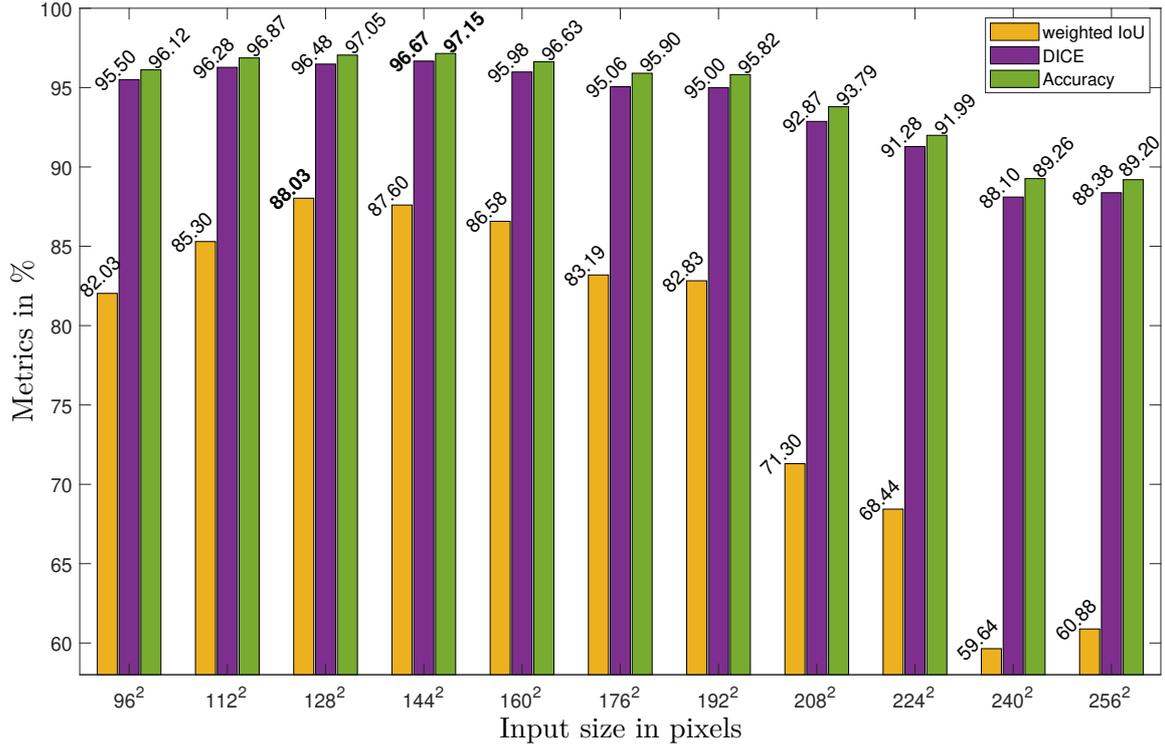

Figure 4.6: Metric results for different input image sizes.

Figure 4.6 shows that the best input size of the neural network is between 144 and 128, for this dataset whose images have shape 195 × 185. It can be seen that it is better to make the size of the input images smaller than bigger, for this bilinear interpolation method. Furthermore, resizing an image to a bigger dimension than the original is not a good approach, once the metrics values are low and there are more computational cost than for small input size for the network to output the class probabilities for every pixel. The input size $I_s = 144$ showed the best results for the *DICE* and *accuracy* metrics, while the best *weighted IoU* was achieved when $I_s = 128$. Therefore, both sizes should be a good choice to use as input size of the networks. Then, $I_s = 128$ was chosen since it is common to use values that are power of 2, and since the smaller the image size, the smaller the computational cost to obtain the network parameters, for example for the convolutional layers once the convolution operation covers the entire images.

## 4.6 Number of image samples

This experiment was done to identify a minimum number of samples per focal plane that results in an accurate prediction. For that, it were made different training sets increasing the number of images per focal plane (brightfield channel) and all training sets were validated on the same validation data, which consists of 70 images. To compare the results, the weighted *IoU*, *DICE* and *accuracy* were computed as metrics. Figure 4.7 shows the computed metric values for the different training sets.

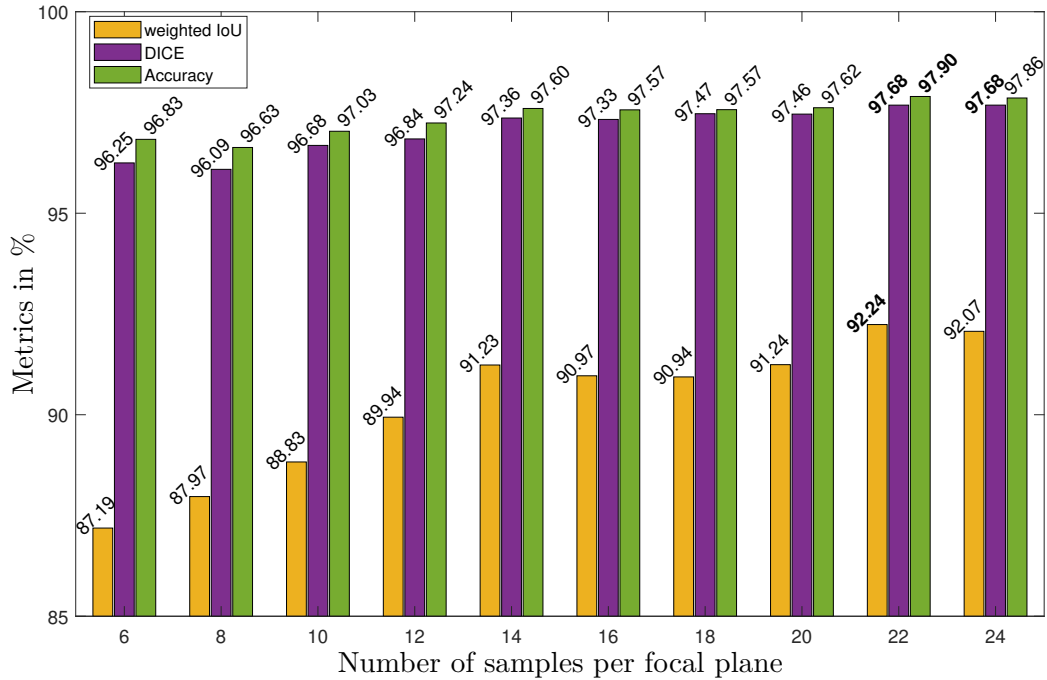

Figure 4.7: Validation metrics result for training sets with different number of samples per focal plane. The best value of each metric is in bold.

It can be concluded from Figure 4.7 that a training set with at least 14 images per focal plane is able to result a network with high evaluation metrics. Furthermore, the best result for all the metrics was for the training set with 22 images per focal plane. Therefore, if a totally new data should be used, for example a new trap shape, 22 samples of this data should be enough for the network to learn the new features and to output an accurate prediction. It should be noted that more samples per focal plane were not tested due to the limitation of the dataset. In this dataset, there were only 190 images whose halos of the focal plane images were similar, of which 70 were used as validation dataset. However, although 22 samples per focal plane showed good metric results, the higher the number of samples used, the better the learning process, considering that the new samples are representative and not just empty traps.

## 4.7 Touching cells results

In order to reduce the number of touching cells, a new class was introduced that represents the region between cells. This new class is called interface class and it was created from the binarization of the weight maps, as described in Section 3.12.2. Furthermore, the weighted cross-entropy loss was also implemented in order to deal with the touching objects problem. The results without any method, with the weighted cross-entropy loss and with the interface class are shown below. Figure 4.8 illustrates two examples of the same trap segmentation without any method, with the weighted cross-entropy loss and with the interface class.

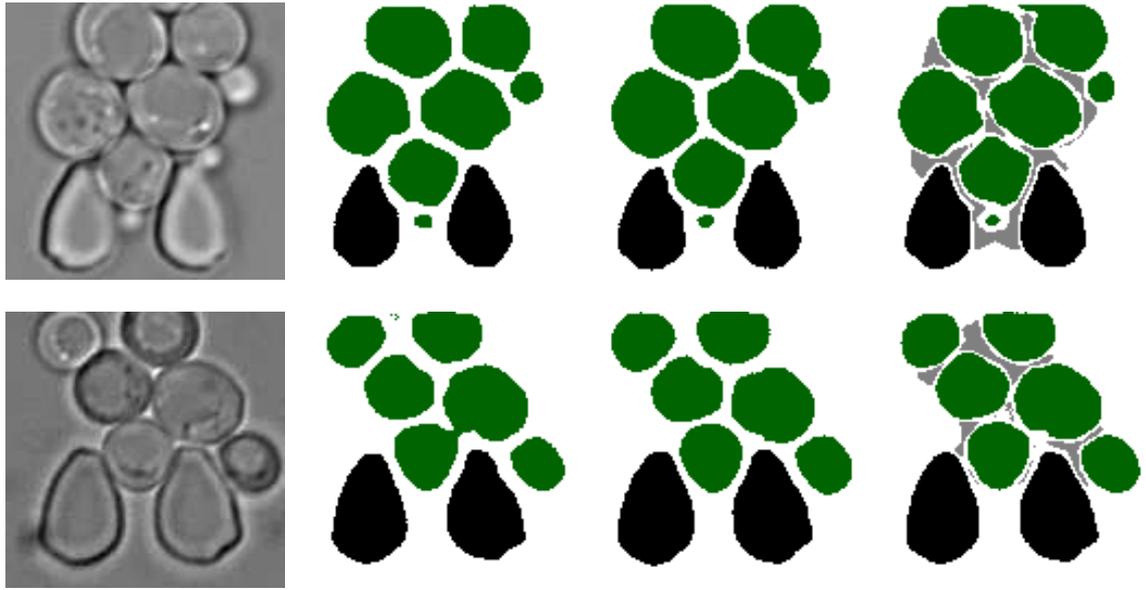

(a) Original trap image.  (b) Prediction without method.  (c) Prediction with weighted loss.  (d) Prediction with interface class.

Figure 4.8: 3 classes prediction examples without any method to deal with touching cells (b), with weighted cross-entropy loss (c) and with the interface class depicted in gray (d).

It can be seen in Figure 4.8 that, for the trap example at the top, the interface class had a better segmentation than the weighted cross-entropy loss method to separate the cells. However, without the approaches, the cells were already not merged. Figure 4.8 also shows an example, at the bottom, where the weighted cross-entropy method aided in the segmentation of separated cells, as well as the interface class method. For this example, both interface method and weighted cross-entropy loss solved the touching cells problem. Table 4.4 contains the mean training epoch time per image sample and mean prediction time per sample in order to analyze how much the methods affect these metrics.

Table 4.4: 3 classes training and prediction times per sample, validation accuracy and *IoU* for the methods applied to the touching cells problem.

| Method | Mean training time/sample (ms) | Mean prediction time/sample (ms) | Validation Acc. | Validation IoU |
| --- | --- | --- | --- | --- |
| weighted loss | 3.2770 | 6.4620 | 0.9705 | 0.9390 |
| interface class | 3.3709 | 6.8621 | 0.9592 | 0.9081 |
| - | 3.2796 | 6.9540 | 0.9705 | 0.9315 |

It can be seen in Table 4.4 that the mean training time per sample and mean prediction time per sample do not differ significantly between the methods. Although the interface class be a more complex problem because there are more classes, this method did not interfere negatively in the computational time of training and prediction per sample. However, its accuracy and *IoU* were slightly lower when compared to values without the interface class and with cross-entropy loss. This is mainly due to the computation of the validation metrics considering the interface class, once this class is not continuously predicted over the background regions.

Performing the same analysis but now with 4 classes, Figure 4.9 shows the segmentations with no method, with the weighted cross-entropy and with the interface class, for the same original trap images of Figure 4.8.

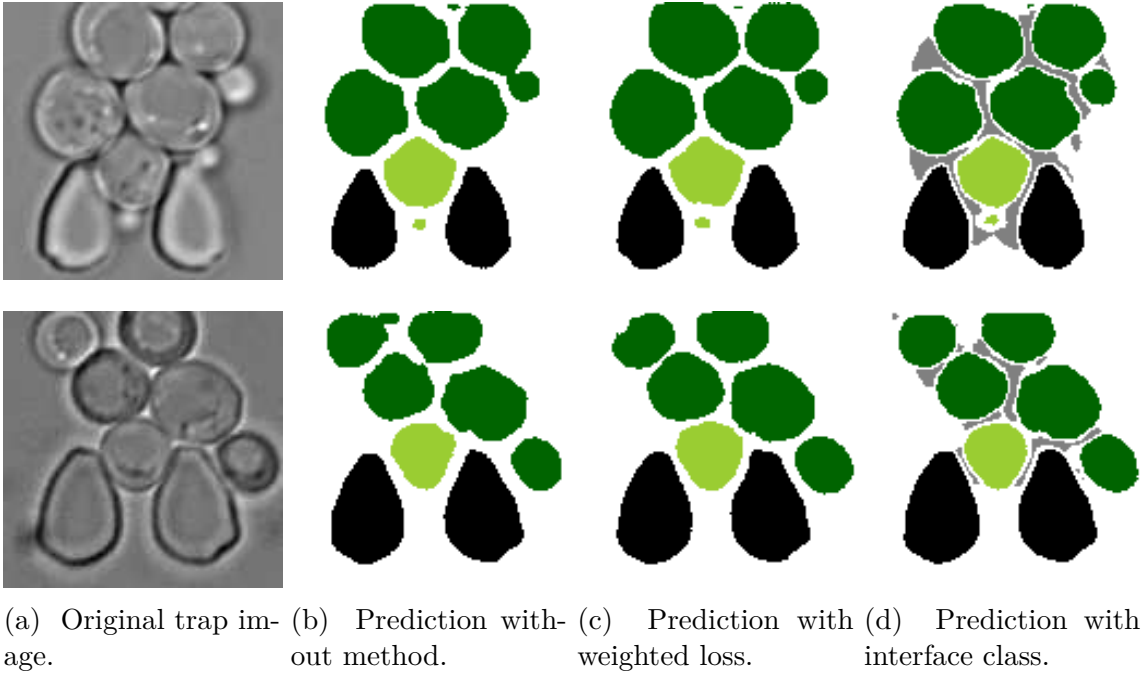

(a) Original trap image.  (b) Prediction without method.  (c) Prediction with weighted loss.  (d) Prediction with interface class.

Figure 4.9: 4 classes prediction examples without any method to deal with touching cells (b), with weighted cross-entropy loss (c) and with the interface class (d).

In the examples of Figure 4.9, both methods applied to the problem of touching cells were able to segment the cells separately, and presented a better segmentation than without any method. Table 4.5 contains the training and prediction times per sample, validation accuracy and $IoU$ for the methods employed in the touching cell problem.

Table 4.5: 4 classes training and prediction times per sample, validation accuracy and $IoU$ for the methods applied to the touching cells problem.

| Method | Mean training time/sample (ms) | Mean prediction time/sample (ms) | Validation Acc. | Validation IoU |
| --- | --- | --- | --- | --- |
| weighted loss | 3.2968 | 6.4848 | 0.9652 | 0.9288 |
| interface class | 3.2587 | 6.9241 | 0.9577 | 0.9098 |
| - | 3.2661 | 6.8020 | 0.9655 | 0.9233 |

Again, the training time per epoch per sample and prediction time do not vary significantly among the methods used. In addition, the presence of the interface class provided the lowest result for accuracy and $IoU$. Therefore, Tables 4.4 and 4.5 show that it is a project decision to choose the number of classes, and the processing time does not significantly influence the segmentation of new images. For this project decision, it should be considered whether only the cell trapped in the trap is significant for the analysis or not. If so, it is not necessary to postprocess to track this target cell, since it is already segmented separately with 4 classes. If all cells should be tracked, it does not matter whether they are segmented with 3 classes or 4, since the segmentations are very close to each other and postprocessing is necessary for tracking independently the cells.

The results with and without the presence of the interface class are shown in the following applied to an experiment. In this case, without the interface class means that the weighted cross-entropy loss was used. The segmentation will depend on whether the neural net was trained with 3 or 4 classes, and also whether the interface class was present during training. In this experiment, microscopic images were obtained in 5 different focal planes for 780 minutes. Each time point image has 24 traps. The results for the case of 3 classes, focal plane $z_0$ and time point 5 are presented in Figure 4.10.

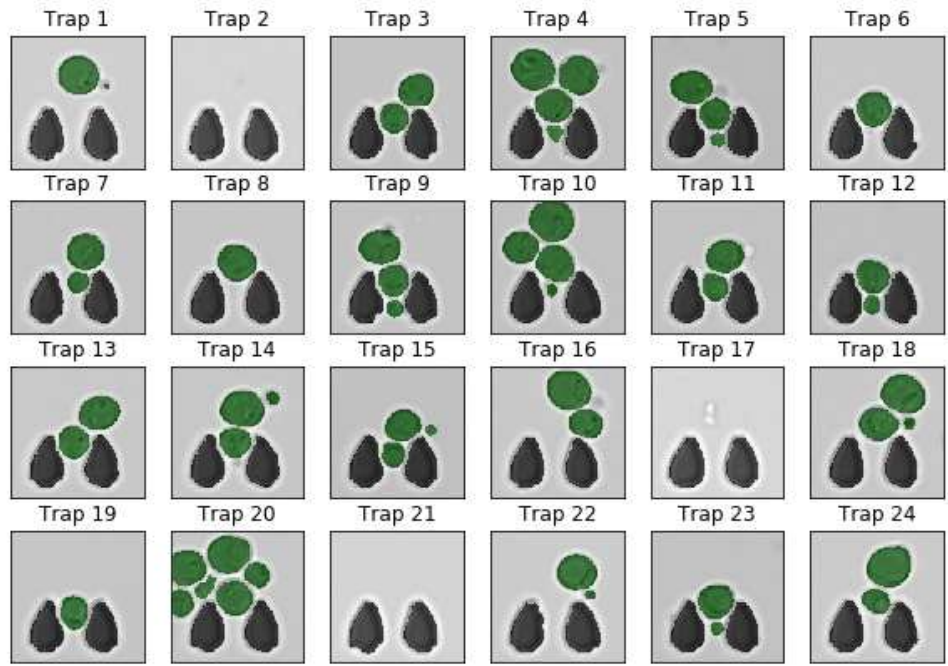

Figure 4.10: Image segmentation in 3 classes for time point 5 and focal plane $z_0$ of an experiment.

The same experiment was tested for the case of 3 classes, with the addition of the new class. For the same focal plane $z_0$ and time point 5 of this experiment, the results can be seen in Figure 4.11.

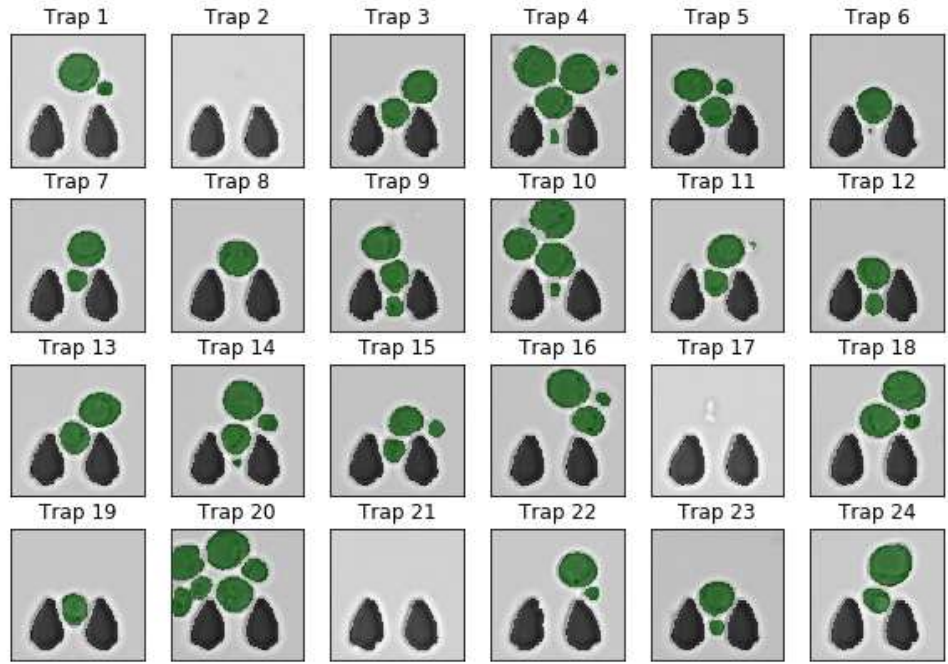

Figure 4.11: Image segmentation in 3 classes for time point 5 and focal plane $z_0$ of an experiment, but with the addition of the interface class representing the region between cells and traps.

It is important to mention that the interface class is not presented with a different color in Figure 4.11 because the new class was converted to background, being set to 0. It is also noted that when comparing Figures 4.10 and 4.11, some cells are correctly segmented with the interface class that are not segmented without the interface class, and vice versa, as is the case, for example, with trap 5. In trap 5, it is noted that without the interface class, a cell at the bottom between traps is detected. However, with the interface class it is not identified, but another cell at the top is segmented. In addition, with the interface class, some cells are predicted which are not identified with the cross-entropy loss, as is the case, for example, with traps 1 and 16. It is also possible to note, however, that in the trap 20 two cells are joined in the prediction with the interface class, which was not what was expected.

The experiment was repeated, but now with 4 classes and with and without the presence of the interface class. The results of the first case, which is without the interface class, are shown in Figure 4.12.

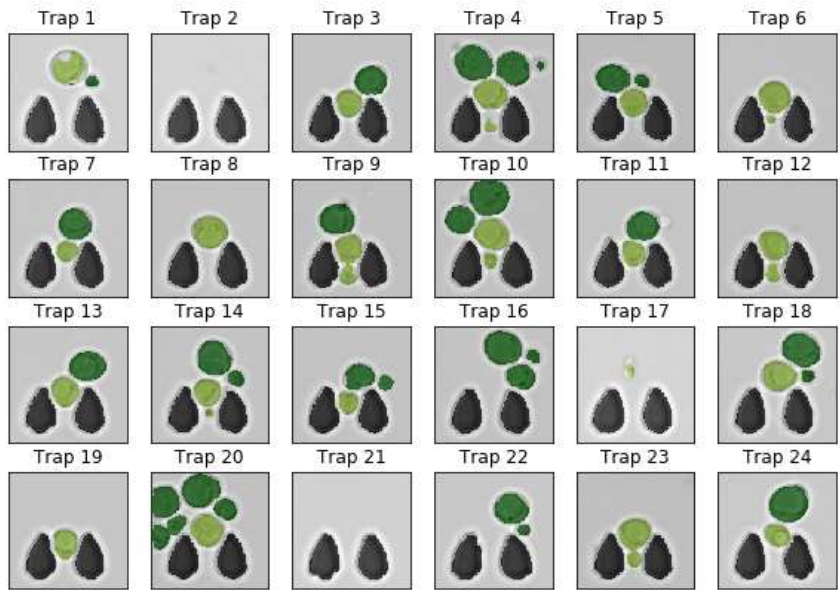

Figure 4.12: Image segmentation in 4 classes for time point 5 and focal plane $z_0$ of an experiment.

In the same way as previously with 3 classes, the interface class was inserted to the case with 4 classes, whose results are shown in Figure 4.13.

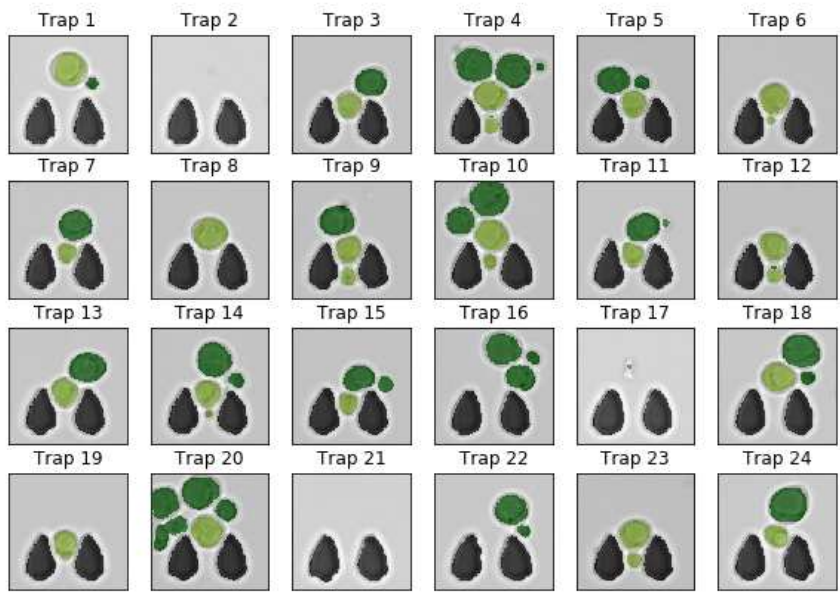

Figure 4.13: Image segmentation in 4 classes for time point 5 and focal plane $z_0$ of an experiment, but with the addition of the interface class.

Comparing Figures 4.12 and 4.13, it is observed that some cells are better segmented with the interface class, as is the case of trap 1. In addition, the new class also helped the neural network in the problem of touching cells, as in trap 23. However, for trap 20, two cells remain connected after segmentation.

## 4.8 Pipeline application

The pipeline application in an experiment consist in, besides segmenting the cells in the microfluidic device, tracking them and calculating their fluorescences in the GFP channel. For this purpose, the images are segmented using the brightfield channel, and these segmentations are subsequently used to calculate the areas of the cells, the respective fluorescences, as well as the fluorescence by area plots. The following results are from a position called 2-Pos_001_004 at focal plane $z_4$, which contains 79 timepoints. There are 19 localized traps, as shown in Figure 4.14.

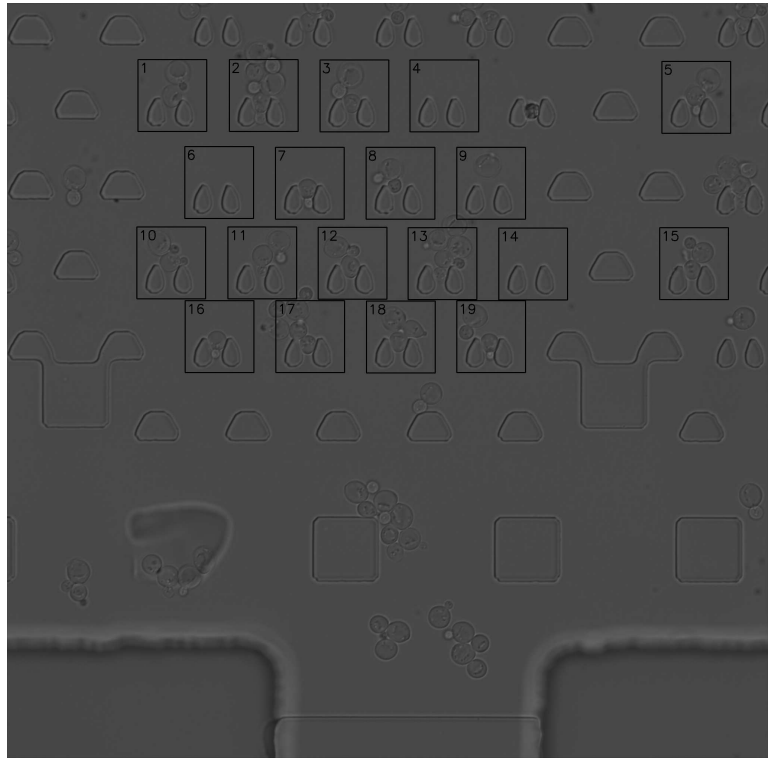

Figure 4.14: Traps of position 2-Pos_001_004 and $z_4$.

Figures 4.17, 4.16 and 4.17 show, respectively, only the fluorescence, the area, and the fluorescence by area over time of all tracked cells in this position.

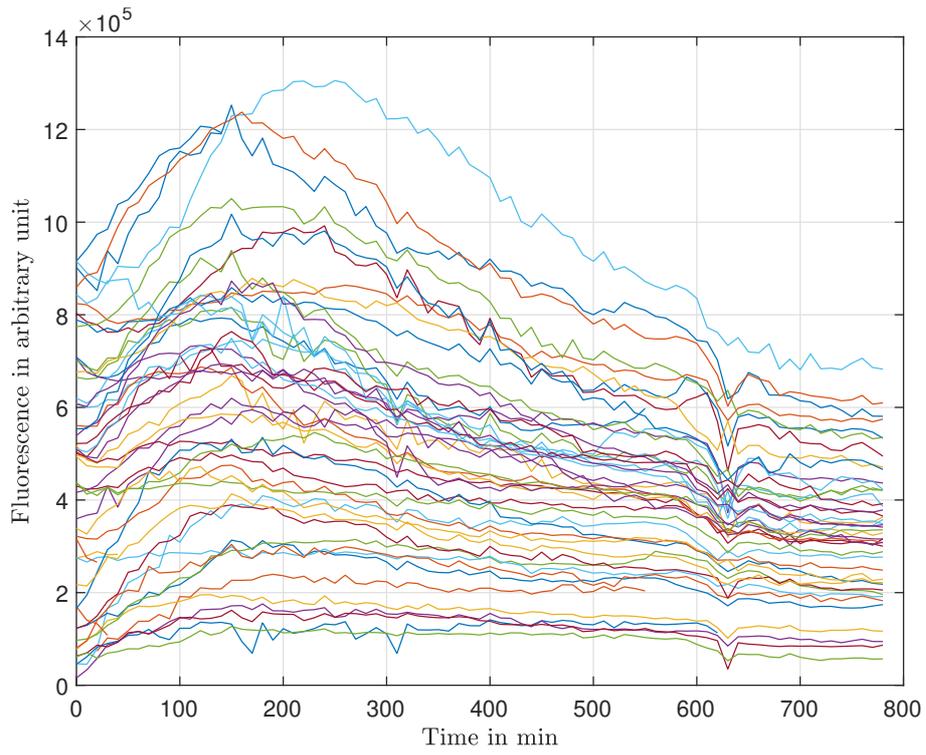

Figure 4.15: Fluorescence of position 2-Pos_001_004 and $z_4$.

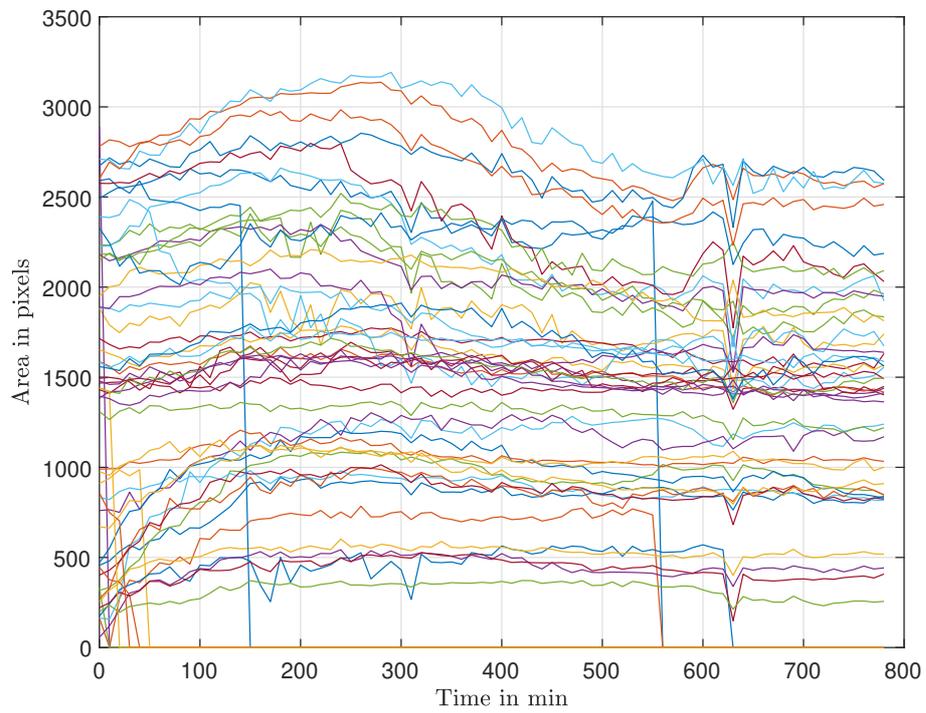

Figure 4.16: Area of position 2-Pos_001_004 and $z_4$.

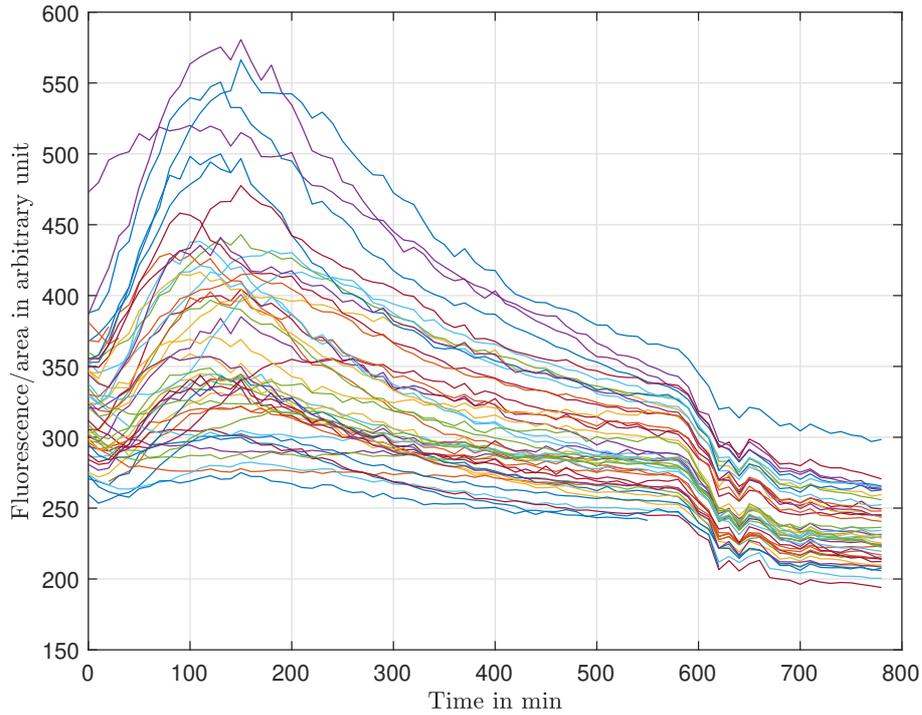

Figure 4.17: Fluorescence/area of position 2-Pos_001_004 and $z_4$.

It can be concluded from Figure 4.15 that the fluorescence of the cells follows the same pattern of intensity over time. There is an increase in the intensity of fluorescence in the first 100 minutes, reaching its maximum value between 100 and 200 minutes, followed by a decay. Figure 4.16 shows that there is a variation in the area of the cells. This variation is due to cell predictions over time, or also due to cell growth or shrinkage throughout the experiment. Finally, Figure 4.15 shows the fluorescence of cells by area, allowing the comparison of fluorescence density by cells, since they can also have different sizes. This curve is the best to be analyzed since it is smoother than just fluorescence, and allows one to have a notion of fluorescence density, regardless of the size of the cells.

Another way to display the results of the experiment is through the Kymograph. This graph allows a 2D analysis of fluorescence over area intensity in time and per cell in a summarized way, where on the x-axis is time, y-axis is the ID of the cells and the color indicates fluorescence over area intensity. Figure 4.18 contains the Kymograph showing the single cell traces.

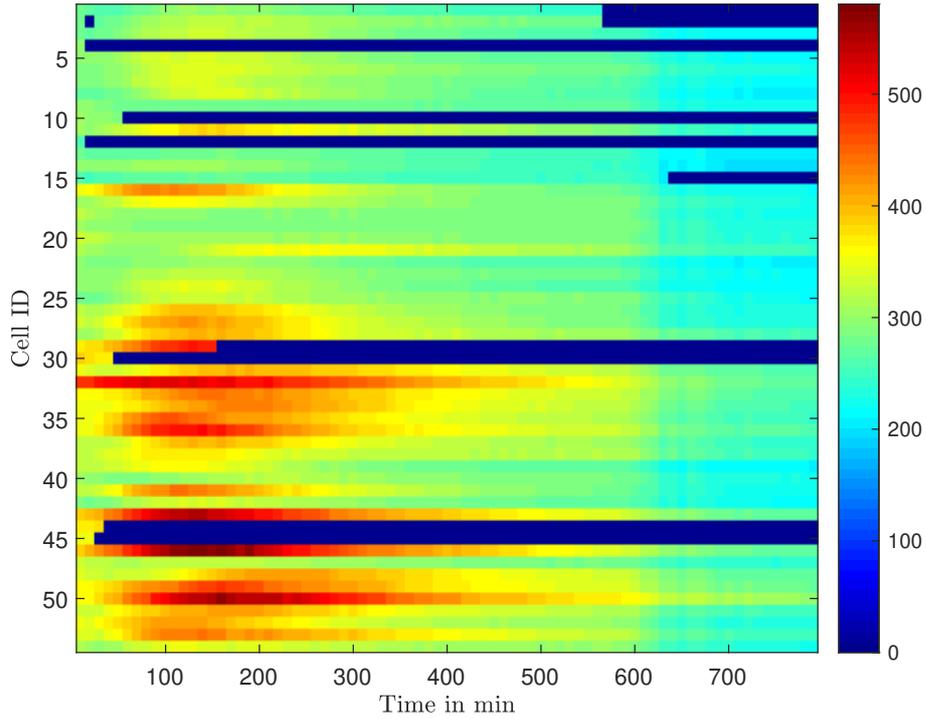

Figure 4.18: Kymograph showing the single cell traces of position 2-Pos_001_004 and $z_4$.

It is important to note that the isolated blue regions in the middle of the images are due to a failure in the prediction of the neural network, where then there was no fluorescence measurement since the contour of the cells was not identified. To avoid zero fluorescence, the average intensity of the time instants posterior and anterior to this time point could be used. However, it was decided not to calculate the mean value in order to visualize where there was a prediction mistake or tracking failure, facilitating a further individual analysis of these cells. Furthermore, the short traces that have zero fluorescence after a certain time might be due to prediction failure where the cell is no longer segmented by the neural network at other timepoints, or it might mean that these cells are daughter cells and have been washed away. If the short traces are not relevant for the analysis of the experiment, it is also possible to set a length threshold to eliminate these cells from the kymograph.

In order to compare the Kymograph of Figure 4.18 with the other focal planes, the summarized kymographs of focal planes $z_0$, $z_1$, $z_2$ and $z_3$ are shown in Figure 4.19.

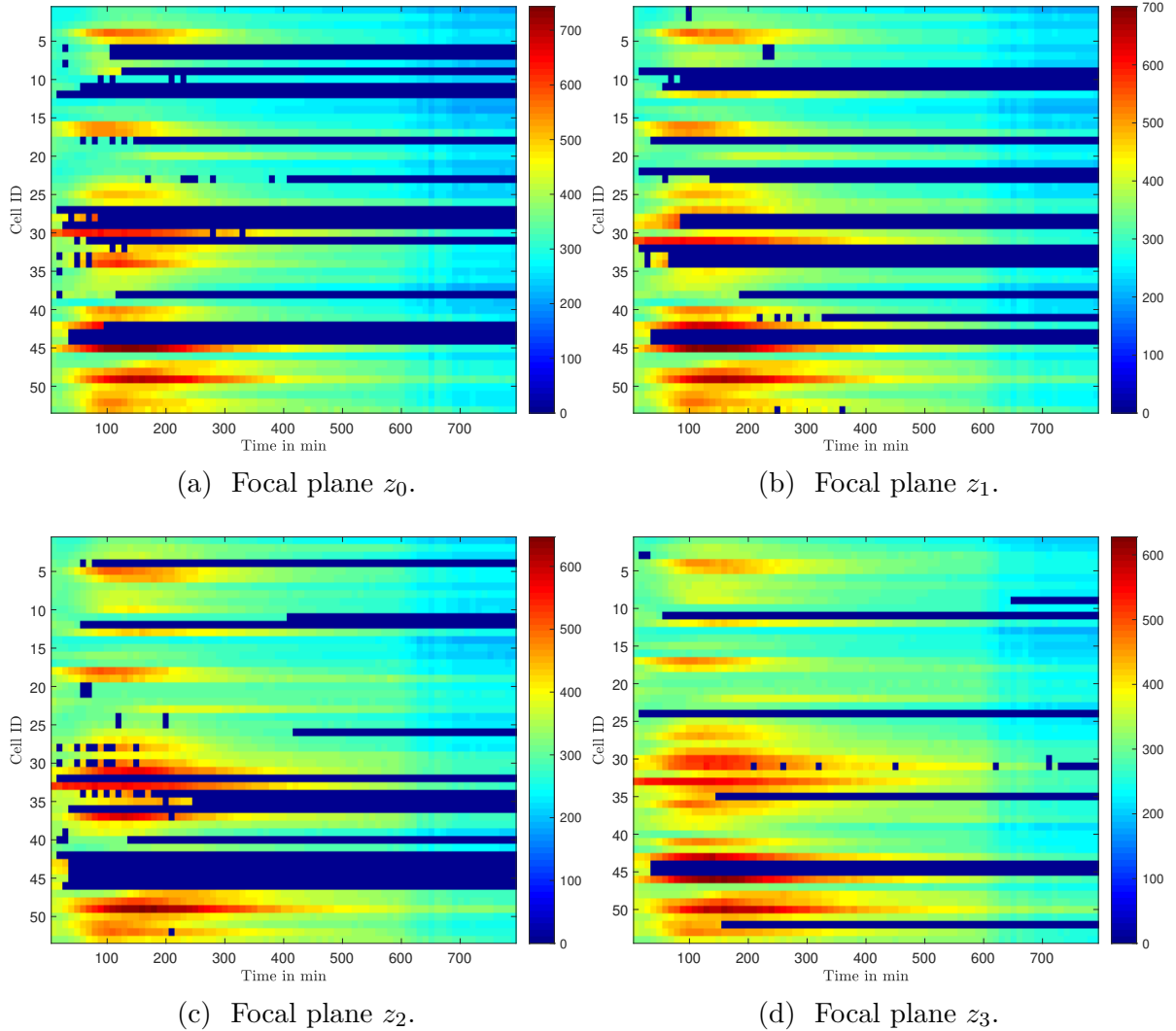

Figure 4.19: Kymographs of position 2-Pos_001_004 for other focal planes.

It can be seen in Figure 4.19 that the Kymographs are quite similar for all focal planes. That was expected once the fluorescence intensity and the cells are the same. What can be different is the prediction of the cells by the neural network, where depending on the focal plane it can be more accurate or it might miss-predict some cells, that is, missing cells. The most clear example where the predictions were not satisfactory is for focal plane $z_2$ shown in Figure 4.19c. In this focal plane, the borders of the cells were not well defined and the network could not segment correctly. However, it is possible to use the information of every focal plane such that they complement each other.

If the behavior of a specific cell or trap must be studied carefully, the analysis of the entire position may be chaotic and may not provide useful information. However, cells and traps have ID's such that they can be analyzed individually. As an example, trap number 12, showed in Figure 4.20 of the same position of the experiment was analyzed individually.

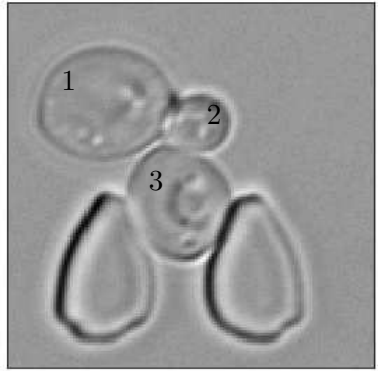

Figure 4.20: Trap 12 of position 2-Pos_001_004 with cell ID's.

The fluorescence, area, fluorescence over area and Kymograph of this trap 12 are found, respectively, in Figures 4.21, 4.22, 4.23 and 4.23.

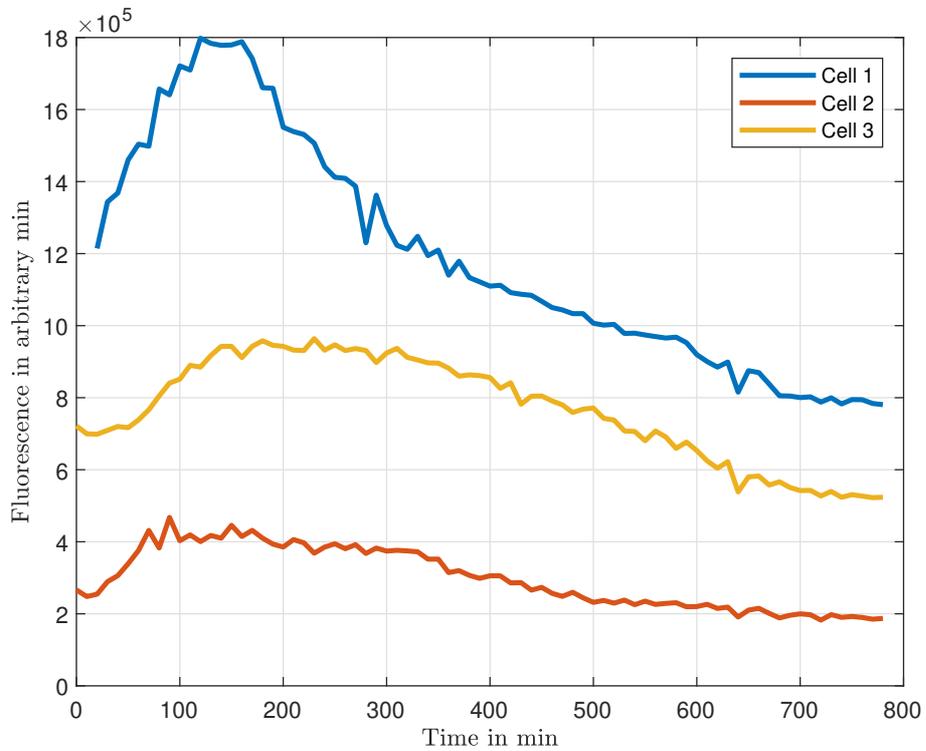

Figure 4.21: Fluorescence of trap 12 from position 2-Pos_001_004.

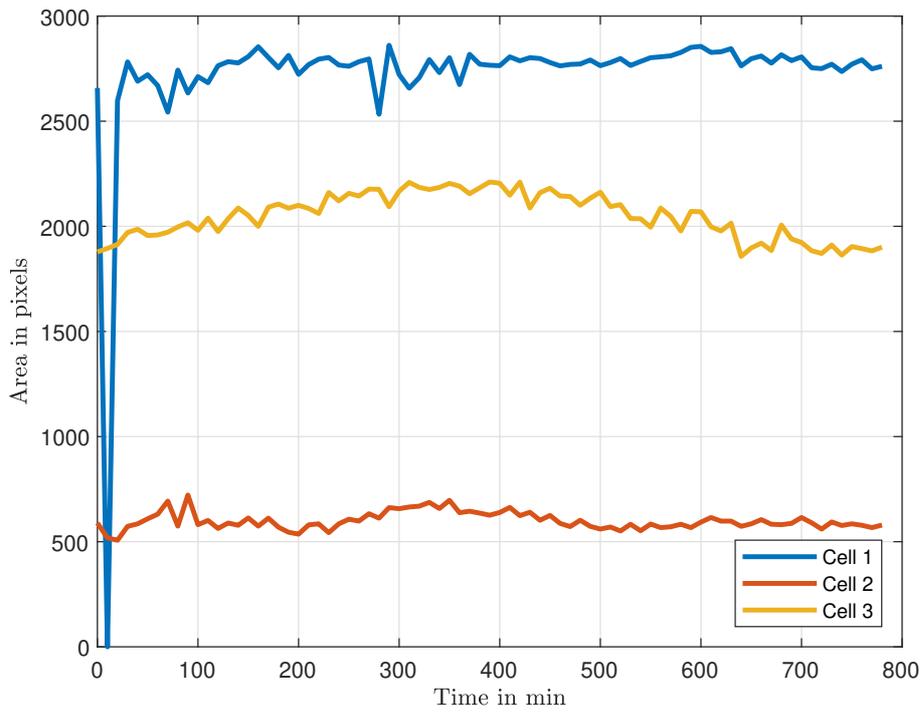

Figure 4.22: Area of trap 12 from position 2-Pos_001_004.

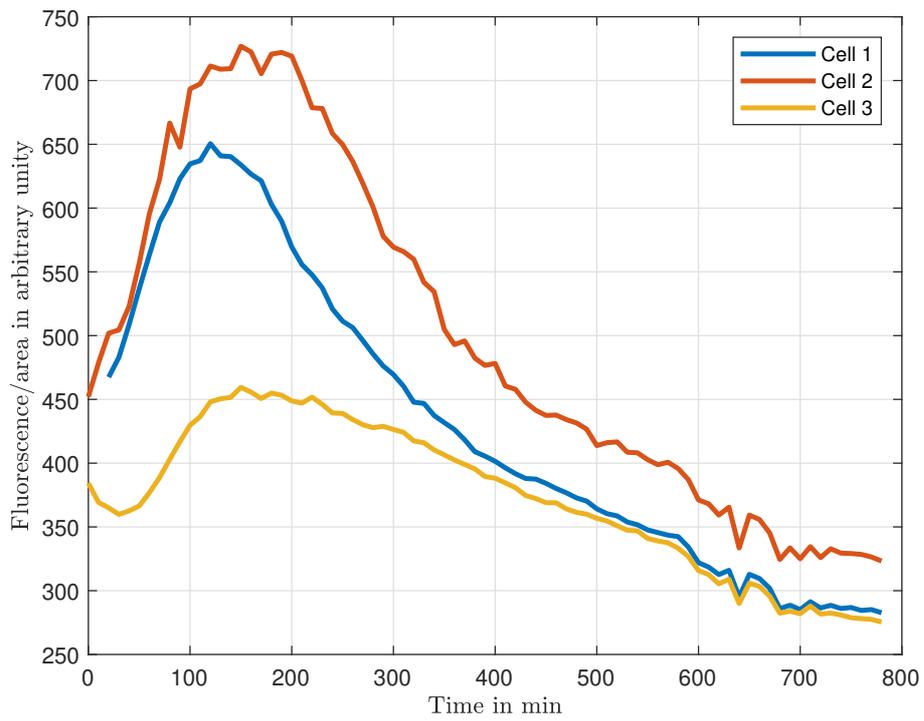

Figure 4.23: Fluorescence/area of trap 12 from position 2-Pos_001_004.

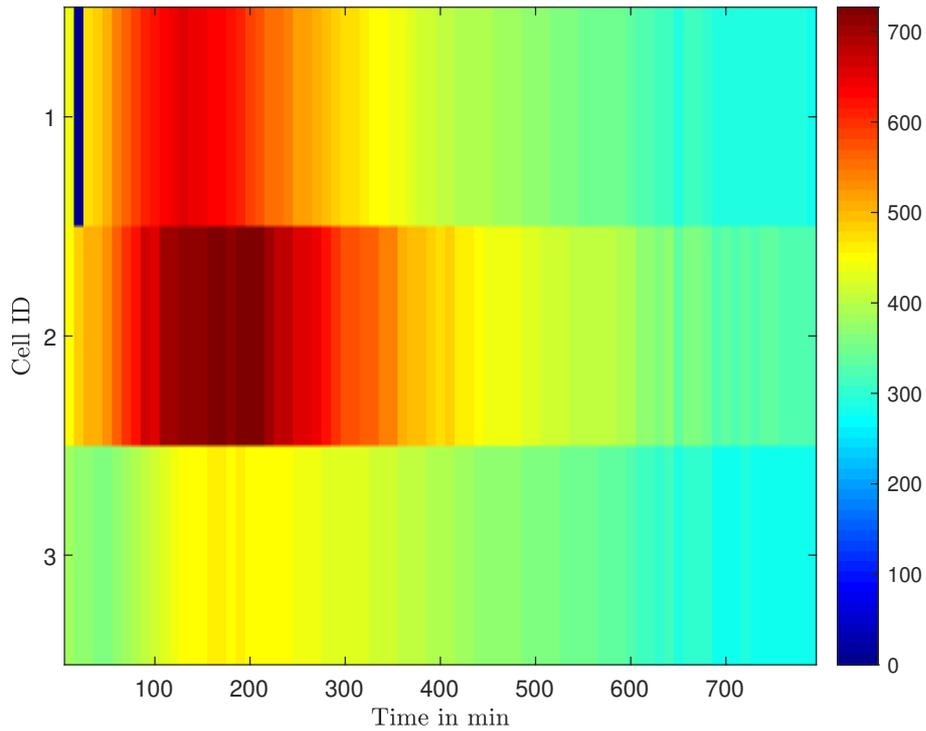

Figure 4.24: Kymograph of trap 12 from position position 2-Pos_001_004.

In this case of data only from trap 12, the data is cleaner and easier to analyze the individual cells than in the results with the entire position. It is also observed the coherence of the cell ID's comparing Figures 4.20 and Figure 4.22, once the cell with the biggest area is the cell 1, and the cell with smallest area is the cell 2.

Besides the analysis of only one position, the simultaneous study of several positions of an experiment is also possible. Sometimes it is interesting to analyze this case in order to obtain a general idea of the experiment. Figures 4.25 and 4.26 show the results of fluorescence over area, and the kymograph of all the cells contained in 4 positions, which are called 2-Pos_001_002, 2-Pos_001_003, 2-Pos_001_004, and 2-Pos_003_003.

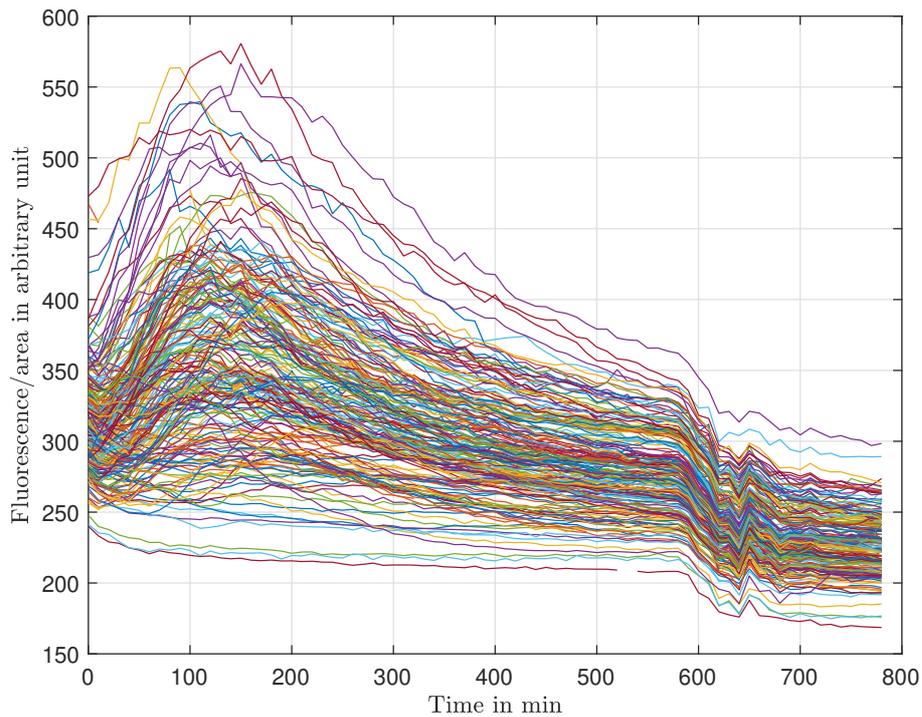

Figure 4.25: Fluorescence/area of 4 positions of an experiment.

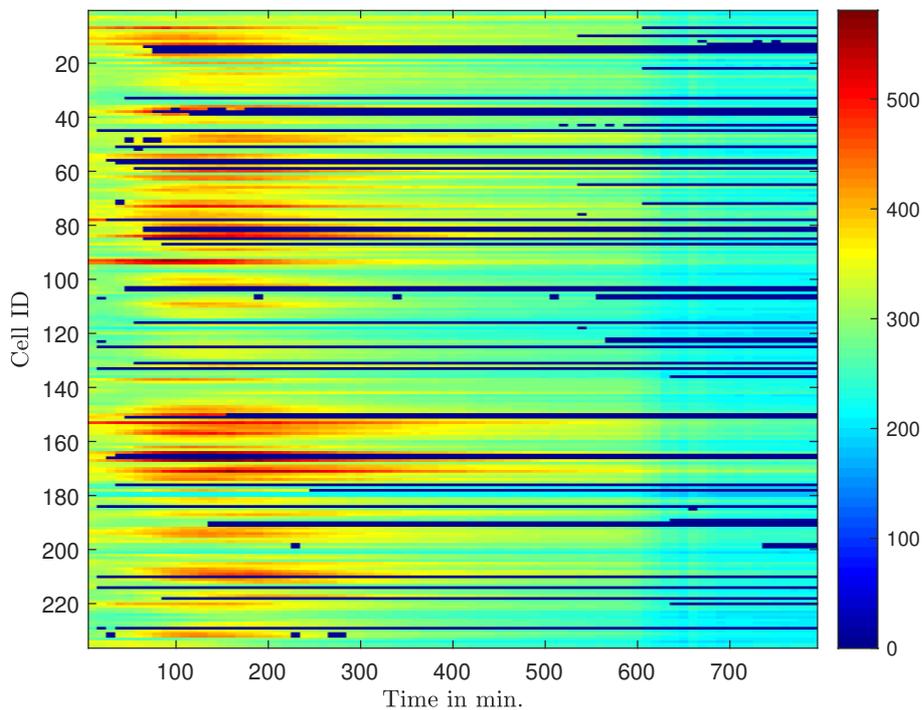

Figure 4.26: Kymograph of 4 positions of an experiment.

It is observed that, between 600 and 700 minutes, all positions show correlation in the behavior of fluorescence over area curves. This occurred due to an oscillation of the light emitted by the microscope, which interfered the measurements of all positions. This analysis with many positions at the same time helps to identify errors during the experiment, while also allowing to verify that the fluorescence curves of cells, despite different positions, have the same pattern of curves.

One example of experiment error is shown in Figure 4.27. In this experiment, it can be seen that after approximately 240 minutes the microscope lost totally the focus. As a consequence, the cells not even the traps were registered in the images.

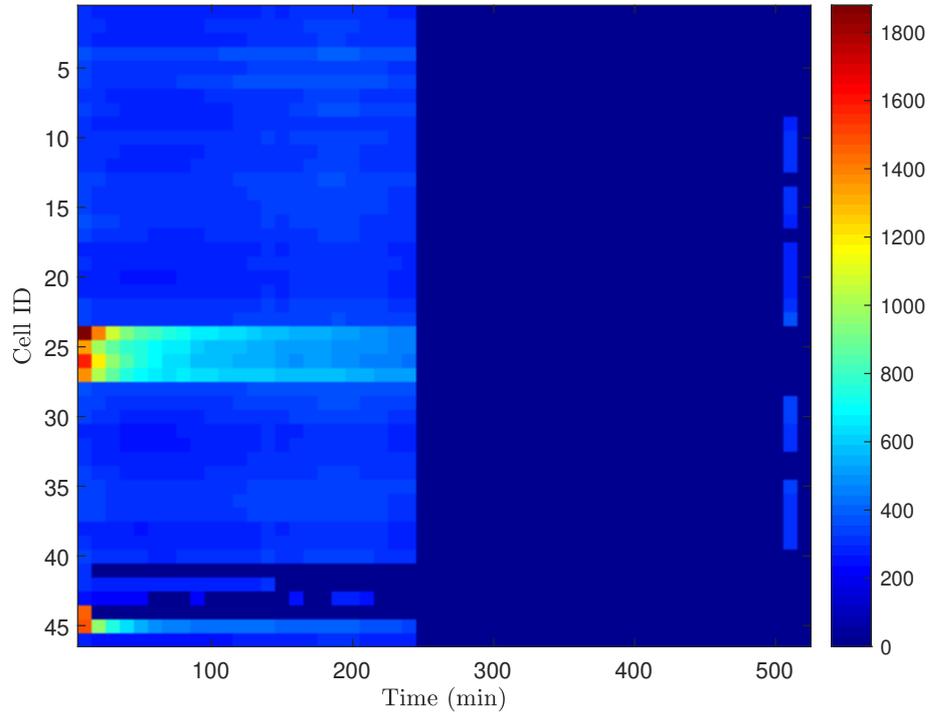

Figure 4.27: Kymograph of an experiment where the microscope went out of focus.

This experiment can be better understood in Figure 4.28, where the 6 cells with high fluorescence in the first 3 timepoints are highlighted in the GFP image. In addition, Figure 4.28c shows the experiment at timepoint 24, when the microscope lost the focus.

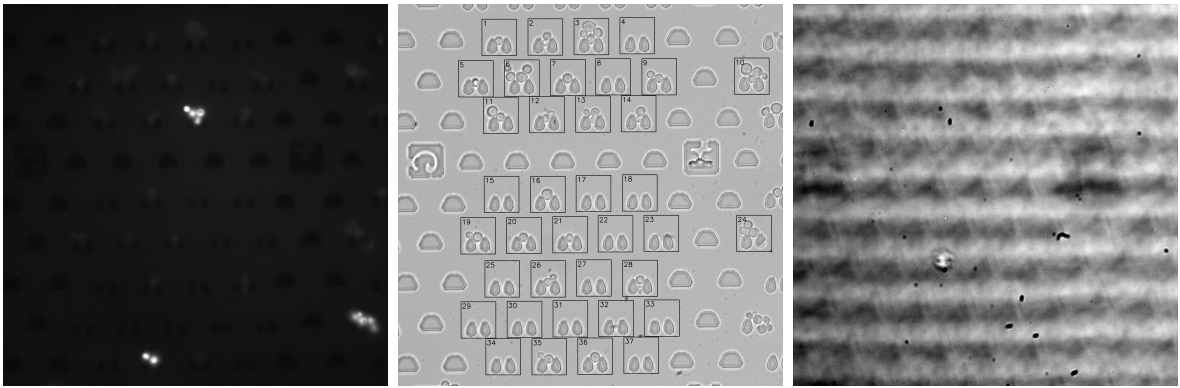

(a) GFP image at timepoint 0.

(b) Brightfield image at timepoint 0 with located traps.

(c) Brightfield image at timepoint 24.

Figure 4.28: GFP and brightfield images of experiment that went out of focus.

## 4.9 Processing time analysis

The processing time of each step of the experiment was measured separately for a temporal analysis. Afterwards, the time of trap localization and shift correction are compared, as well as prediction time and extraction of information from the experiment.

The first analysis is about trap localization and shift correction. Table 4.6 shows the mean computational time of trap localization at the reference timepoint 0, and the mean cross-correlation time to find the correction pixels at the other time points.

Table 4.6: Mean computational time of localizing the traps at timepoint 0 and of finding the shift correction for other timepoint.

| Mean time of traps localization (s) | Mean time of shift correction (s) |
|---|---|
| 4.8666 | 0.2282 |

It can be concluded from Table 4.6 that it is advantageous to use shift correction in the other timepoints since this time is on average much lower than the time of localizing the traps in the reference image.

The next temporal measurement is made for preprocessing and postprocessing, predicting the neural network, for watershed setting different labels for each cell present in traps and for cell tracking. As the number of traps at each position of the experiment may vary, the measured values then indicate the mean processing time per trap image. The measured values are shown in Table 4.7.

Table 4.7: Mean processing time of preprocessing, postprocessing, prediction, watershed and tracking steps by trap image.

| Pre-processing (ms) | Prediction (ms) | Post-Processing (ms) | Watershed (ms) | Tracking (ms) |
|---|---|---|---|---|
| 0.2285 | 5.7064 | 0.6822 | 1.9289 | 3.7770 |

Table 4.7 shows that the longest image processing time is given by the prediction of the neural network. The measured experiment contains 24 traps in each timepoint image, that is, 24 samples. Thus, the average time of prediction of each big image at one timepoint is 136, 9536 ms. Although this is the longest time at this stage, it is still relatively much shorter than the trap localization time and shift correction.

Finally, the total time for processing a big image, with all traps at a timepoint, is also measured, as indicated in Table 4.8.

Table 4.8: Mean processing time for processing all computations in one timepoint image.

| Get all data (s) | Total time (s) timepoint 0 | Total time (s) timepoint >0 |
|---|---|---|
| 0.3080 | 7.4328 | 0.5684 |

In this case, get all data means to calculate the fluorescence and all useful information of one entire timepoint image. Besides, total time means the time since to read the big images and make the trap localization and shift correction until get all data information. Comparing Tables 4.6 and 4.8, it is observed that 40.14% of the total processing time of a timepoint is due to shift correction to align the timepoints with the reference. For the reference timepoint 0, the total processing time was measured as 7.4328 seconds. This was also due to the longer processing time to localize trap positions and due to the first communication between GPU and CPU and to access the cuDNN library, which took a mean measured time of 2.38378 seconds. This first communication is rather slow but affects only the first timepoint. Without considering the localization and correction of traps, the mean processing time per timepoint image is less than half a second (308 ms), which corresponds to 50.18% of the total processing time of a timepoint. This indicates that the bottleneck of the pipeline is not the image segmentation performed by the neural network, but in the preprocessing of the images to localize the traps at all timepoints and tracking.

### 4.9.1 Pipelines comparison

The previous pipeline available in BCS had many parameters without variables in the Python code, making it difficult to generalize to an experiment by varying, for example, the size of the input images or the model to be chosen by the pipeline. Among other factors, the previous pipeline was not optimally implemented, since the algorithm was highly hardcoded, making it difficult to change some test parameters. In addition, it was tested the processing time of the previous pipeline with the current one to check if the algorithms of the pipeline steps were optimized. Table 4.9 contains the mean processing time for a timepoint image by using the CPU, where the previous pipeline was originally implemented, and by using the GPU in order to a fair comparison.

Table 4.9: Mean processing time of the previous and current pipelines by using CPU and GPU.

| Processing unit | Previous pipeline total time (s) timepoint 0 | Previous pipeline total time (s) timepoint >0 | Current pipeline total time (s) timepoint 0 | Previous pipeline total time (s) timepoint >0 |
|---|---|---|---|---|
| CPU | 10.4087 | 6.8365 | - | - |
| GPU | 9.4343 | 3.7433 | 0.5684 | 7.4328 |

It can be seen from Table 4.9 that the previous pipeline is much slower than the pipeline developed in this thesis. To process the complete timepoint image, the current pipeline needs 0.5684 seconds, while the previous one adapted to GPU needs 3.7433 seconds to process the same information. In other words, the new pipeline requires only 15.18% of the processing time of the previous pipeline to process the same information. The time point 0 is still relatively high for both pipelines. However, the average processing time of roughly 2.4 seconds is due to GPU communication to access the cuDNN library. Even disregarding this time, it is possible to see that the total processing time has been reduced in the current pipeline due to the new tracking implementation, the optimized implementation of the predicted masks labels as postprocessing step and the structured and organized way in which the data is processed. Therefore, it is concluded that there has been an improvement of more than 50% with regard to processing time considering how fast the pipeline performs for timepoints different from the initial of the pipeline. This result is very significant for a study of real time application of the pipeline in an experiment, which was not possible with the previous pipeline.

### 4.9.2 Quasi real-time experiment capability

Considering the work method of the microscope to generate the images, it is possible to verify if it is possible to apply the pipeline in real time in an experiment. It was seen by means of experiments that, for an experiment with 5 focal planes, it takes approximately 4 seconds for brightfield images to be collected in a position of the microfluidic device in a timepoint. That is, the first brightfield image of the next position will be generated after 4 seconds. This means that that microscope requires approximately 1 second to generate together one brightfield and one GFP image. This time is a fair reference since it also includes the channel change time to collect the GFP images and the displacement of the microscope to the next position of the microfluidic device. For timepoints other than the initial one, the total processing time to segment each brightfield image and calculate the information of every single cell is roughly 0.6 seconds. This means that the segmentation and processing this image is faster than the measurement by the microscope. In the previous pipeline this was not possible, since the processing of each image was on average 4 times slower than the measurement image time by the microscope, as shown in Table 4.9. Therefore, an online capability of the pipeline has been enbled with the new pipeline. In other words, a quasi real time processing was achieved.

To improve the processing time, the first thing to do is to invest in the localization of traps, since this step contributes the most to the total processing time of an experiment. With this improvement, the experiment can be performed in quasi real time with more time slack, once the initial timepoint was the bottleneck of the pipeline. Finally, the current pipeline was applied to a complete position of the microfluidic device, with 79 timepoints and 5 focal planes. This position represents 6.2GB of data. The processing of this data was performed in exactly 237.5148 seconds, i.e., 4 minutes.

# Chapter 5

# Conclusions

Based on the convolutional neural network with U-Net architecture, the microscopic images of yeasts were then segmented and applied to the pipeline that returned the area and fluorescence information of each individual cell over time. Several neural networks were tested and it was found that the original U-Net architecture can be reduced for this application in the BCS group. It contributes to speed up the segmentation time compared to the bigger network, which is relevant for a quasi real time application. Furthermore, an accurate network was built through studies on the effects of some hyperparameters and on a representative training dataset.

For batch size, the larger the batch size chosen, the larger the number of epochs should be for the model to reach the same validation metric as a model trained with smaller batch size. In addition, it is also concluded that the neural network learns to segment traps and cells relatively quickly, since with less than 30 epochs the neural network is able to achieve high segmentation performance for the dataset used. However, more epochs can be used during training because the Keras library allows saving the model with the lowest validation loss, which ensures that the saved model does not have high performance only for training data, i.e., there was no overfitting.

The original images obtained by microscopy have a size $2048 \times 2048$ pixels. A trap present in this image, increased by an upper region to cover the cells, has an average size $185 \times 195$. It was experimentally determined that an input size of $128 \times 128$, for the available dataset, allows the model to segment the images with high performance.

Another parameter now related to the dataset is the minimum number of images needed for the neural network to learn new halo characteristics from focal planes. It was determined that 22 sample images of each new halo characteristic, with the use of augmented images, is sufficient for the neural network to learn how to extract and predict the new characteristic.

Regarding the analysis of different U-Net models, it is concluded that a neural network with fewer parameters is sufficient to segment the biomedical images of yeasts. This answers affirmatively to the initial question "Can the original U-Net architecture be reduced to a smaller architecture for this application?". The original U-Net has 31.030.723 learnable parameters, while the model 1 used showed better performance with only 5.137.635, that is, there was a reduction of 83.33% in the number of parameters of the original model. In addition, other even smaller models could have been used, such as model 11 with 1.861.827 learnable parameters. However, model 1 was chosen since it presented the highest performance according to the adopted metrics.

The pipeline was able to segment the images and track individual cells, saving fluorescence and area information for each individual cell. These results were summarized in two main graphs, namely the fluorescence curve by area over time and the so-called

Kymograph. In addition, the current pipeline showed improvements in performance when compared to the previous one available in the BCS group. The current pipeline is for instance available in a private GitLab, but with the possibility of being open to the public.

Finally, the current pipeline can be applied in a quasi real time experiment, considering the aspects raised in Section 4.9.2. This also answers affirmatively to the question "Is it feasible to process an experiment in real time in such a way that the fluorescence intensity of individual cells is measured during an experiment over time?" since it is feasible.

## 5.1 Outlook

During the course of this project and after the analysis of the results, new issues arose that can be implemented in future work in order to improve the algorithms and the pipeline, which are listed hereafter.

- The use of a neural network to localize traps. This would be a common object detection and localization task with bounding box, and it might improve the processing time of the initial timepoint 0, since this was the bottleneck of the total processing of an experiment.

- Adapt the pipeline to read data from different saved file format. Previously, the images generated by the microscope were in *.tif* format. However, a new approach is used in the BCS group that saves the experiment as a *.nd*2 file. This file format requires a different interaction with the saved images, but it is still possible to adapt the pipeline to process this type of format.

- The use of a neural network to track individual cells. Although cell tracking by image processing algorithms is relatively fast, a neural network can also be used for this purpose, as used in [48].

- Expand dataset to other trap shapes so that the neural net can be invariant to the trap shape.

# Appendices

# Appendix A

# Looking inside more layers of the network

Figures A.1 to A.6 show the output images of a specific image sample for more convolutional layers of the U-Net, from decoder and encoder path. The yellow represents high activation value, and purple represents zero activation. The blue scale in between yellow and purple represents intermediate activation values

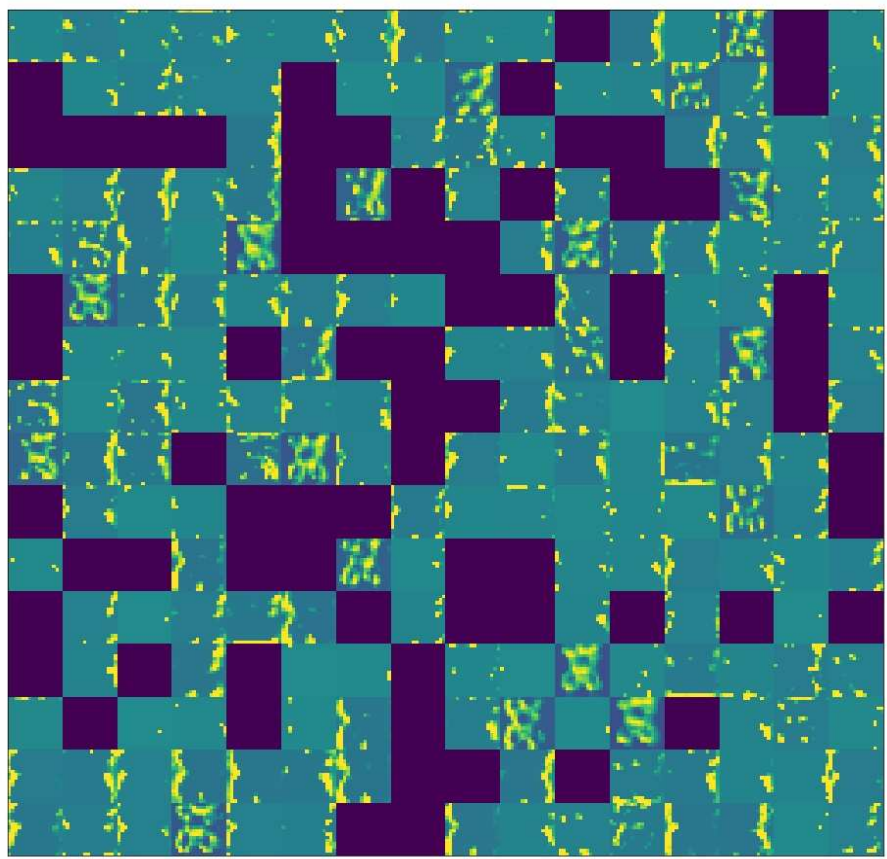

Figure A.1: Output images of the 256 filters after the second convolutional layer of the fourth encoder block.

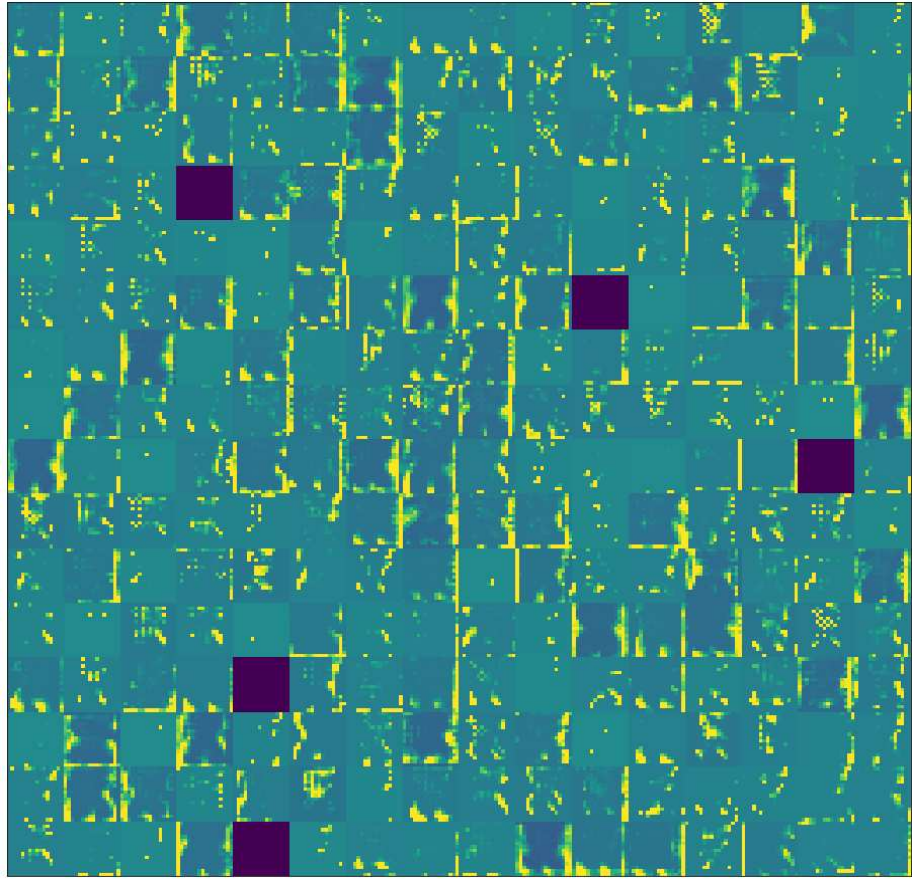

Figure A.2: Output images of the 256 filters after the second convolutional layer of the first decoder block.

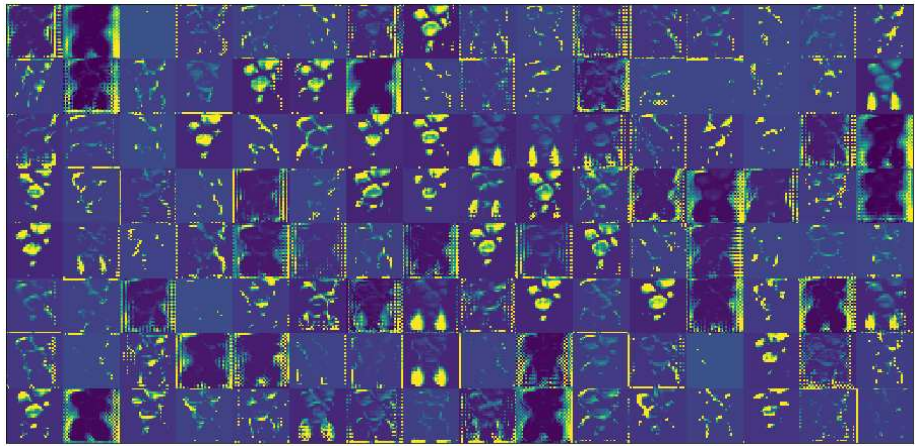

Figure A.3: Output images of the 128 filters after the second convolutional layer of the second decoder block.

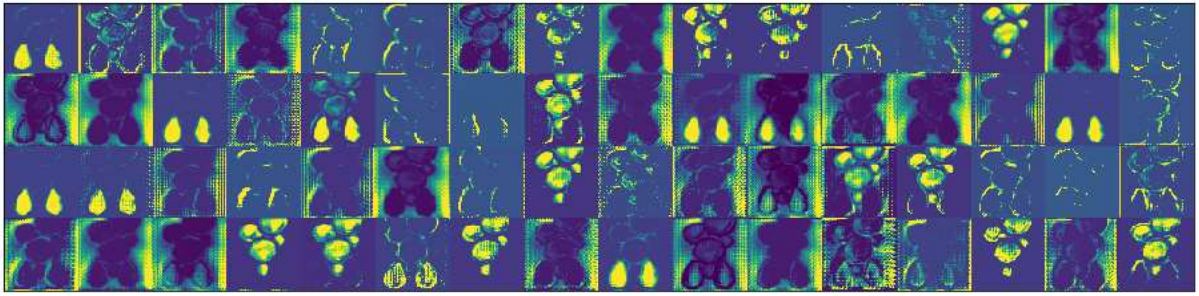

Figure A.4: Output images of the 64 filters after the second convolutional layer of the third decoder block.

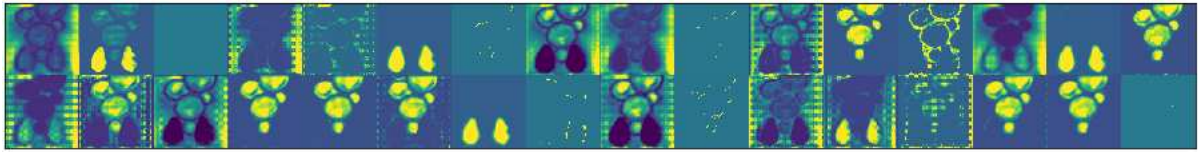

Figure A.5: Output images of the 32 filters after the second convolutional layer of the fourth decoder block.

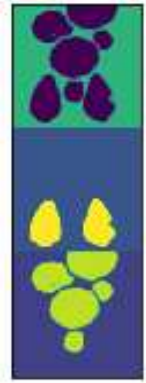

Figure A.6: Output images of the last $1 \times 1$ convolutional layer that outputs the segmentation map.

# Appendix B

# Pipeline application in a different experiment

The experiment is called Pos28 from Jascha Diemer. Figure B.1 shows trap localization of first timepoint (tp).

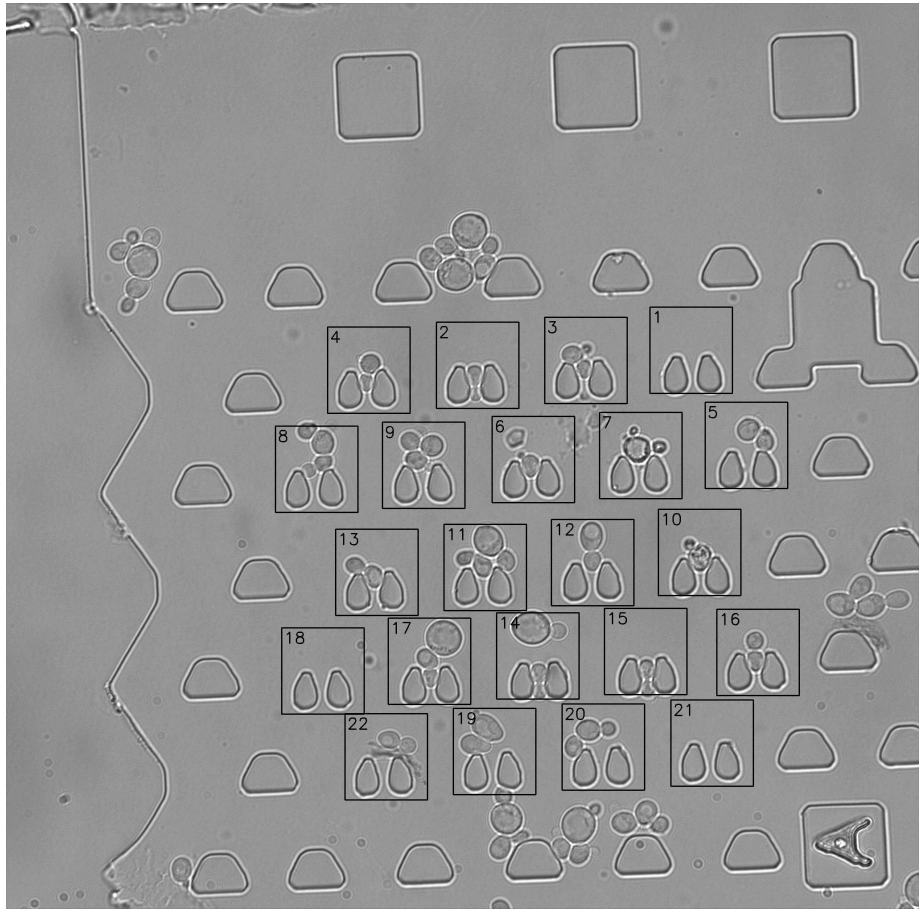

Figure B.1: Brightfield 0 at timepoint 0 of Jascha Diemers's experiment, Pos28.

Figure B.2 shows the predictions in the first 30 timepoints of trap 4.

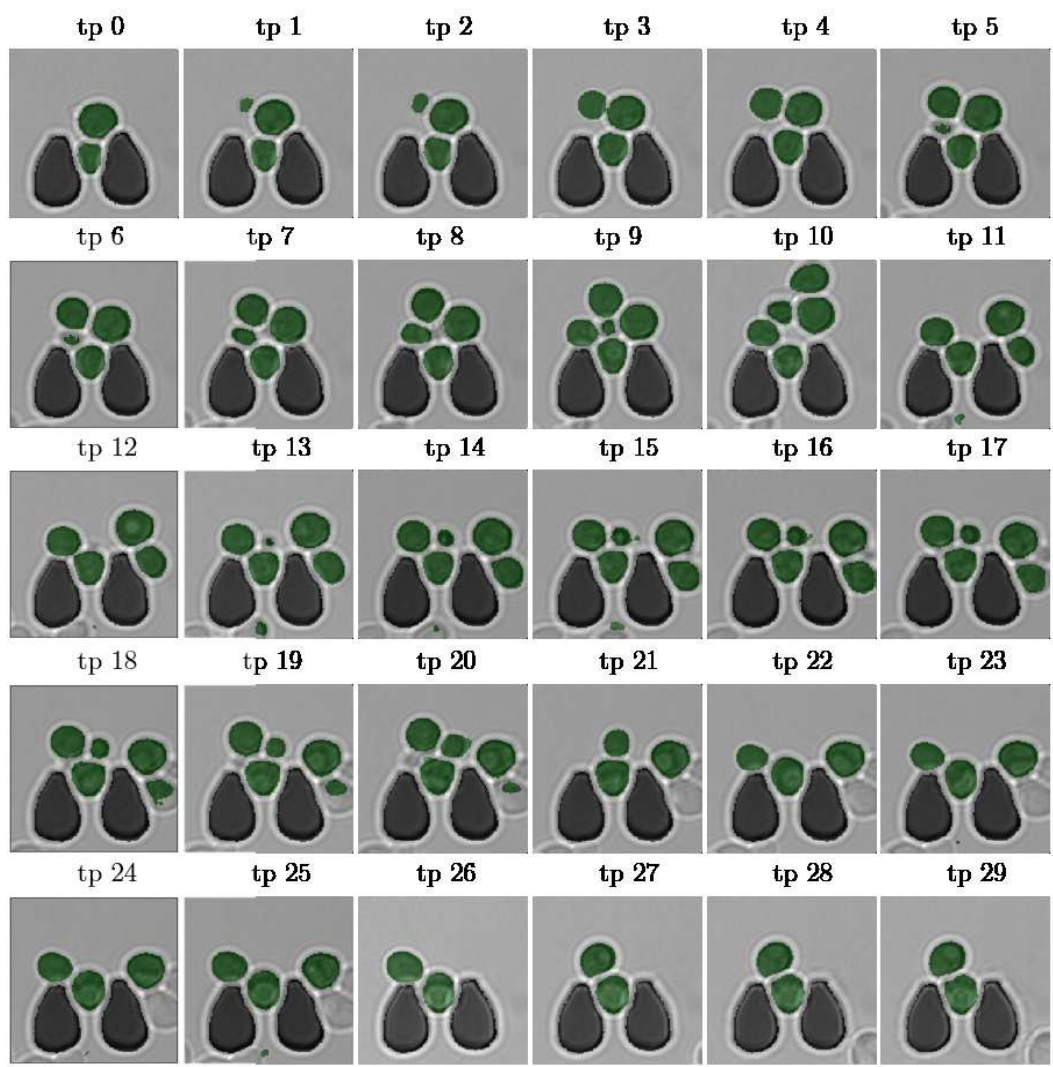

Figure B.2: Predictions of trap 4 in 30 timepoints.

Figure B.3 shows the fluorescence, area, fluorescence over area plots, as well as the kymograph of trap 4.

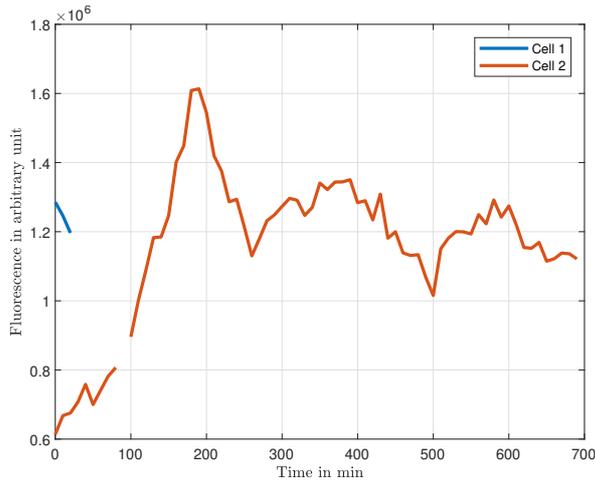
(a) Fluorescence of trap 4.

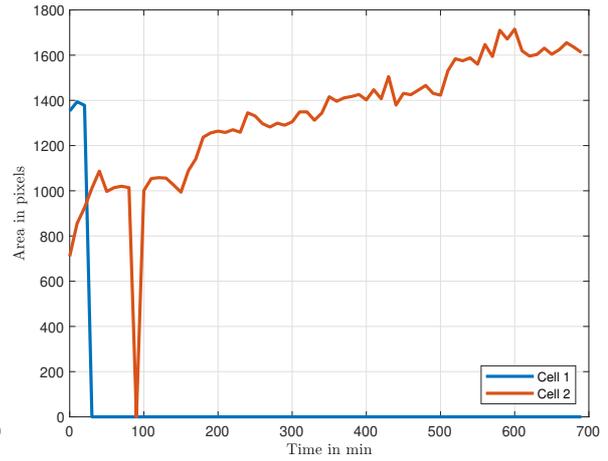
(b) Area of cells in trap 4.

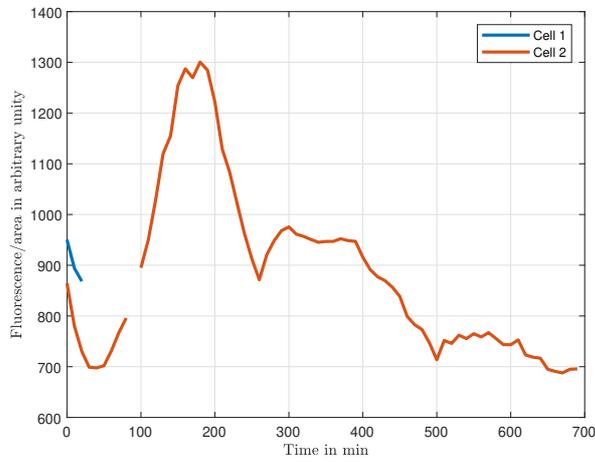
(c) Fluorescence over area of cells in trap 4.

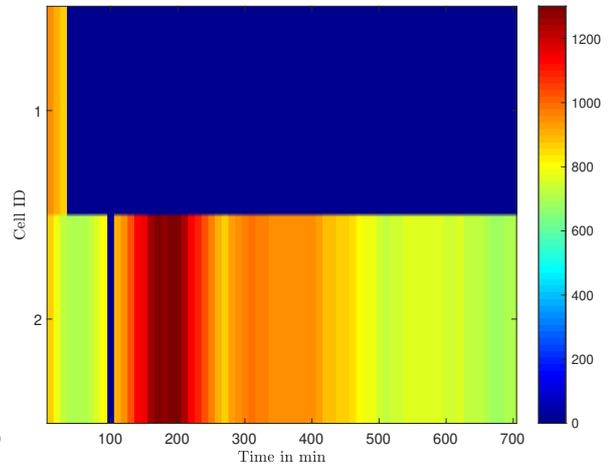
(d) Kymograph of trap 4.

Figure B.3: Area, fluorescence, fluorescence/area and kymograph of trap 4.

The tracking failed to create ID's for the new cells and to follow large cell movements. Only the trapped cell was tracked. However, the predictions of the convolutional neural network look accurate.

# Appendix C

# Different models applied to an experiment

The position is called 2-Pos_001_004 from ExperimentsAH2018, 18/02/2018, H2E_60x_2. The results of Figure C.1 are from the model trained with 3 classes without weighted cross-entropy loss and without interface class.

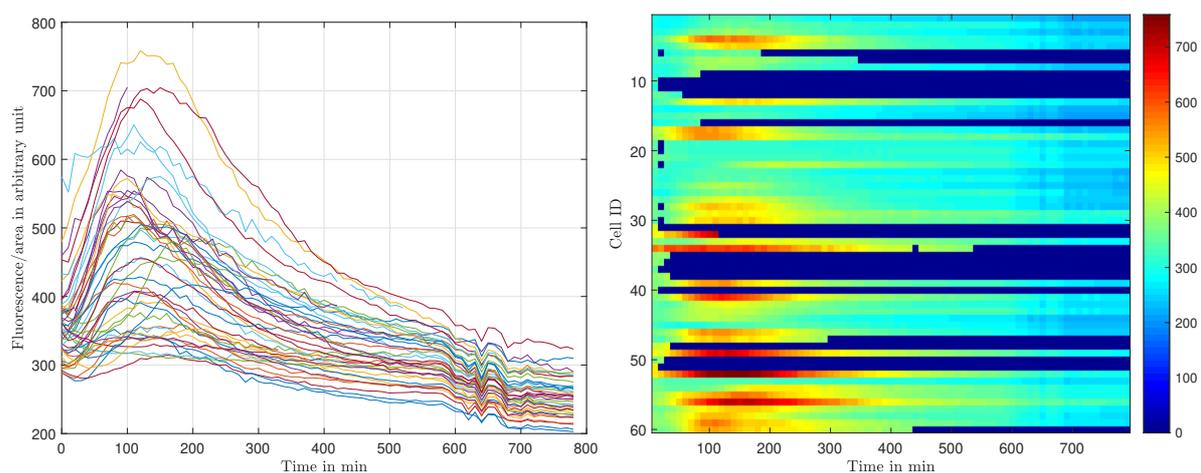

Figure C.1: Fluorescence over area and kymograph of model trained with 3 classes and no method to deal with touching cells.

Figure C.2 shows the results of the model trained with 3 classes with the interface class.

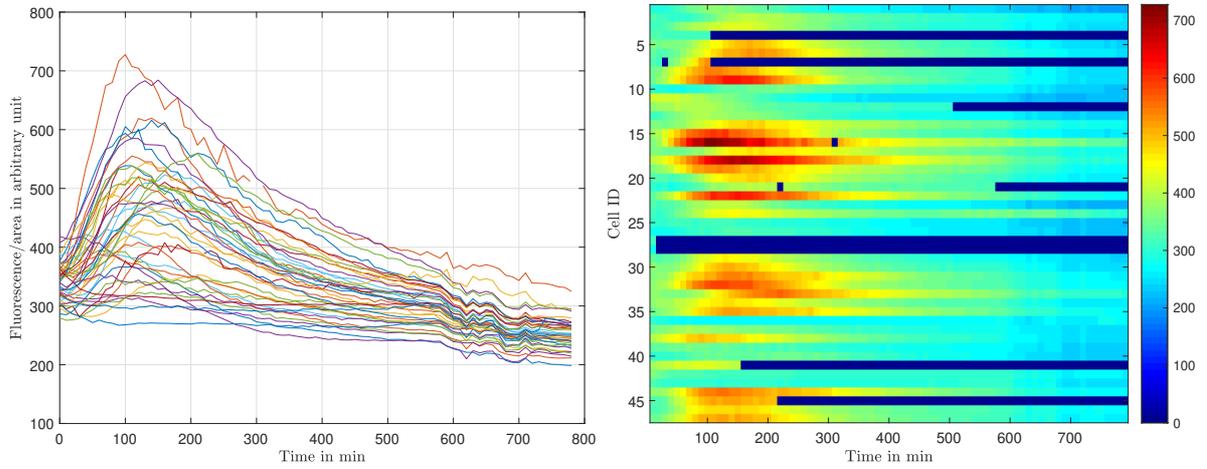

Figure C.2: Fluorescence over area and kymograph of model trained with 3 classes and interface class.

Figure C.3 shows the results of the model trained with 4 classes and with the weighted cross-entropy loss.

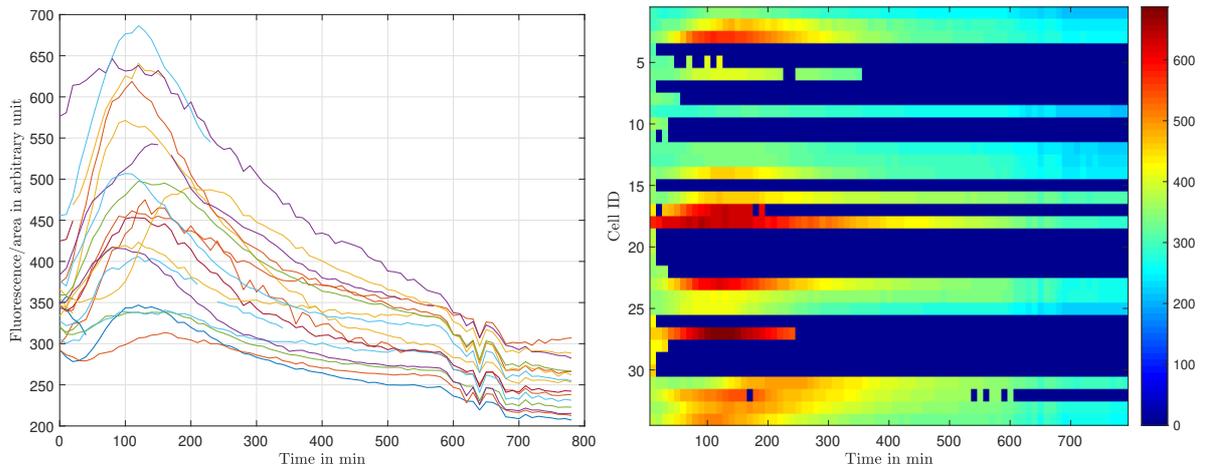

Figure C.3: Fluorescence over area and kymograph of model trained with 4 classes and with weighted cross-entropy loss.

Figure C.4 shows the results of the model trained with 4 classes and with the interface class.

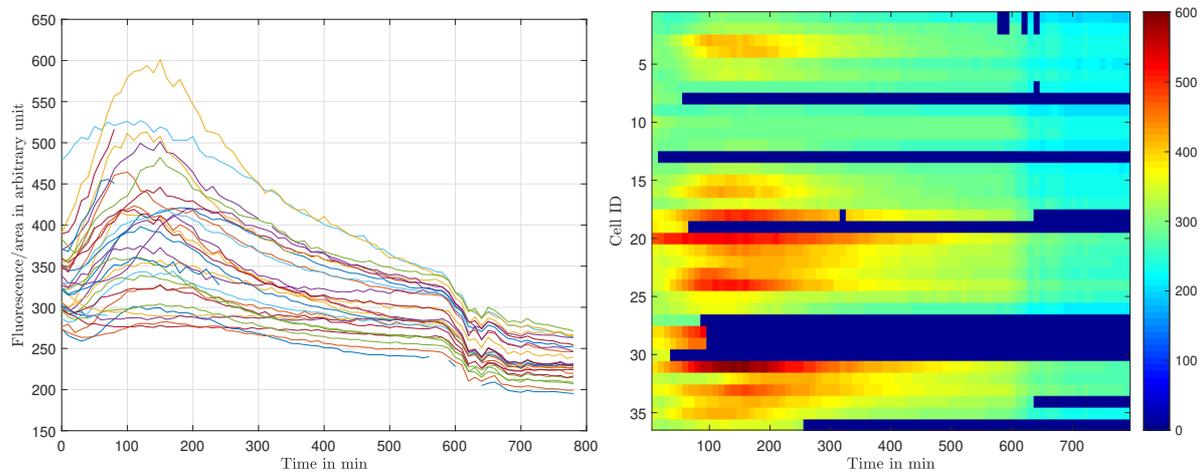

Figure C.4: Fluorescence over area and kymograph of model trained with 4 classes and with interface class.